\documentclass{emulateapj}
\usepackage{apjfonts}
\usepackage{amsmath}

\newcommand{\pdfgraphics}{\ifpdf\DeclareGraphicsExtensions{.pdf,.jpg}\fi}

\newcommand \microjy{$\mu$Jy}

\newcommand \spitzer{\textit{Spitzer}}
\newcommand \chandra{\textit{Chandra}}
\newcommand \xmm{\textit{XMM}}

\def\gsim{\mathrel{\rlap{\lower4pt\hbox{\hskip1pt$\sim$}} \raise1pt\hbox{$>$}}}
\def\lsim{\mathrel{\rlap{\lower4pt\hbox{\hskip1pt$\sim$}} \raise1pt\hbox{$<$}}}

\slugcomment{Received 2011 August 4; accepted 2012 January 18}
\shorttitle{IRAC AGN Selection}
\shortauthors{DONLEY ET~AL}

\begin{document}
\pdfgraphics

\title{Identifying Luminous AGN in Deep Surveys: Revised IRAC Selection Criteria}

\author{J. L. Donley\altaffilmark{1}, A. M. Koekemoer\altaffilmark{1}, M. Brusa\altaffilmark{2}, P. Capak\altaffilmark{3}, C. N. Cardamone\altaffilmark{4}, F. Civano\altaffilmark{5}, O. Ilbert\altaffilmark{6}, C. D. Impey\altaffilmark{7}, J. S. Kartaltepe\altaffilmark{8}, T. Miyaji\altaffilmark{9,10}, M. Salvato\altaffilmark{2}, D. B. Sanders\altaffilmark{11},  J. R. Trump\altaffilmark{12}, G. Zamorani\altaffilmark{13}}

\altaffiltext{1}{Space Telescope Science Institute, 3700 San Martin Drive, Baltimore, MD 21218; Giacconi Fellow; donley@stsci.edu}
\altaffiltext{2}{Max Planck Institut f\"{u}r extraterrestrische Physik, Giessenbachstrasse 1, D-85748 Garching bei M\"{u}nchen, Germany}
\altaffiltext{3}{Spitzer Science Center, California Institute of Technology, MC 220-6, 1200 East California Boulevard, Pasadena, CA 91125, USA}
\altaffiltext{4}{Department of Physics, Massachusetts Institute of Technology, 77 Massachusetts Ave., Cambridge, MA 02139, USA}
\altaffiltext{5}{Harvard Smithsonian Center for Astrophysics, 60 Garden St., Cambridge, MA 02138, USA}
\altaffiltext{6}{Laboratoire d'Astrophysique de Marseille (UMR 6110), CNRS-Universit\'{e} de Provence, 38, rue Fr\'{e}d\'{e}ric Joliot-Curie, 13388 Marseille Cedex 13, France}
\altaffiltext{7}{Steward Observatory, University of Arizona, 933 North Cherry Avenue, Tucson, AZ 85721, USA}
\altaffiltext{8}{National Optical Astronomy Observatory, 950 North Cherry Avenue, Tucson, AZ 85719, USA}
\altaffiltext{9}{Instituto de Astronom\'ia, Universidad Nacional Aut\'onoma de M\'exico, Ensenada, Baja California, M\'exico (mailing address: PO Box 439027, San Diego, CA 92143-9027, USA)}
\altaffiltext{10}{University of California, San Diego, Center for Astrophysics and Space Sciences, 9500 Gilman Drive, La Jolla, CA 92093-0424, USA}
\altaffiltext{11}{Institute for Astronomy, University of Hawaii, 2680 Woodlawn Drive, Honolulu, HI 96822, USA}
\altaffiltext{12}{UCO/Lick, University of California, Santa Cruz, CA 95064, USA}
\altaffiltext{13}{INAF-Osservatorio Astronomico di Bologna, via Ranzani 1, I-40127, Bologna, Italy}

\begin{abstract}
\spitzer\ IRAC selection is a powerful tool for identifying luminous AGN.  For deep IRAC data, however, the AGN selection wedges currently in use are heavily contaminated by star-forming galaxies, especially at high redshift.  Using the large samples of luminous AGN and high-redshift star-forming galaxies in COSMOS, we redefine the AGN selection criteria for use in deep IRAC surveys.  The new IRAC criteria are designed to be both highly complete and reliable, and incorporate the best aspects of the current AGN selection wedges and of infrared power-law selection while excluding high redshift star-forming galaxies selected via the BzK, DRG, LBG, and SMG criteria. At QSO-luminosities of log $L_{\rm 2-10 keV} $(ergs~s$^{-1}$)$\ge 44$, the new IRAC criteria recover $75\%$ of the hard X-ray and IRAC-detected \xmm-COSMOS sample, yet only 38\% of the IRAC AGN candidates have X-ray counterparts, a fraction that rises to 52\% in regions with \chandra\ exposures of 50-160 ks.  X-ray stacking of the individually X-ray non-detected AGN candidates leads to a hard X-ray signal indicative of heavily obscured to mildly Compton-thick obscuration (log~$N_{\rm H}$ (cm$^{-2}) = 23.5 \pm 0.4$).  While IRAC selection recovers a substantial fraction of luminous unobscured and obscured AGN, it is incomplete to low-luminosity and host-dominated AGN.

\end{abstract}


\section{Introduction}

While supermassive black hole (SMBH) growth and galaxy formation were once assumed to proceed independently of one another, a new picture is emerging in which common triggering and/or feedback mechanisms drive the formation and evolution of both SMBHs and their hosts. To determine which processes (e.g., secular evolution or major galaxy mergers) are primarily responsible for moving a galaxy onto the present-day M-$\sigma$ relation, however, we first require a complete census of luminous AGN activity.  

In X-rays, the typical AGN outshines even the most actively star-forming galaxy, and as such, deep X-ray surveys provide the most reliable means of AGN selection.  However, while X-rays penetrate low to moderate columns of obscuring dust and gas, 2-10 keV X-ray surveys miss a significant fraction of moderately obscured AGN ($\sim 25$\% at $N_{\rm H} = 10^{23}$~cm$^{-2}$) and nearly all Compton-thick AGN \citep[$N_{\rm H} > 10^{24}$~cm$^{-2}$,][]{treister04,ballantyne06,tozzi06}.  From fits to the cosmic X-ray background, \cite{gilli07} predict that both moderately obscured and Compton-thick AGN are as numerous as unobscured AGN at high luminosity (log $L_{\rm 0.5-2 keV} $(ergs~s$^{-1}$)$> 43.5$), and are four times as numerous as unobscured AGN at low luminosity (log $L_{\rm 0.5-2 keV} $(ergs~s$^{-1}$)$< 43.5$).  The obscured and Compton-thick AGN missed in deep X-ray surveys therefore serve not only as important probes of SMBH/galaxy co-evolution, but likely constitute a significant fraction of the total AGN population at all luminosities.

To identify obscured AGN not recovered by X-ray surveys, studies have turned to the mid-infrared (MIR).  Not only does the MIR emission from AGN-heated dust trace the reprocessed radiation absorbed in other wavebands, but it is itself relatively insensitive to intervening obscuration.  MIR selection therefore identifies many heavily obscured AGN, nearly half of which are missed in deep X-ray surveys \citep{donley08}. While many MIR-based selection criteria are therefore designed to specifically target heavily obscured AGN \citep[e.g.,][]{daddi07agn,fiore08,fiore09}, \spitzer\ IRAC selection is sensitive to the hot dust signature present in $\gsim 80-95\%$ of luminous AGN regardless of obscuration \citep{hao10,hao11}.  As such, IRAC selection is a potentially powerful technique not only for identifying the heavily obscured AGN missed in the X-ray, but for selecting luminous obscured \textit{and} unobscured AGN when deep X-ray data are unavailable.

The IRAC color-color cuts most commonly used for AGN selection were defined by \cite{lacy04,lacy07} and \cite{stern05} using shallow IRAC data to which additional flux cuts at 8~\micron, 24~\micron\ or R-band served to reject all but the brightest sources ($S_{\rm 8\mu m} \ge 1$~mJy, \cite{lacy04}; $R < 21.5$ and $S_{\rm 8.0\mu m} \ge 76$ \microjy, \cite{stern05}; $S_{\rm 24\mu m} \gsim 5$~mJy, \cite{lacy07}).  While these initial color cuts therefore effectively select luminous AGN in samples containing only AGN and bright, low-redshift star-forming galaxies \citep[see also][]{sajina05}, they extend into regions of IRAC color space populated by moderate to high-redshift ($z\gsim 0.5$) star-forming galaxies in the deep IRAC surveys now available across many cosmological fields \citep[e.g.,][]{barmby06,donley07,donley08,cardamone08,yun08,brusa09,park10}. IRAC power-law selection, which identifies only the most robust of the IRAC color-selected AGN, has therefore been adopted to minimize contamination by normal galaxies \citep{aah06,donley07,donley08}. This technique, however, depends on both the IRAC photometry and the often-underestimated photometric errors, adding a degree of complexity not present in simple color-color cuts.

Our understanding of the MIR source population has increased substantially in recent years, thanks in part to the \spitzer\ IRS spectra now available for large samples of both local and high redshift sources, and in part to the availability of deep IRAC data in multi-wavelength survey fields.  It is therefore time to revisit the IRAC selection of AGN and redefine the selection criteria for use in deep survey fields.  To do so, we focus on the 2~deg$^2$ Cosmic Evolution Survey \cite[COSMOS,][]{scoville07, koekemoer07}. Unlike the deeper Great Observatories Origins Deep Survey (GOODS) fields, the combined size and depth of COSMOS provides large samples of both luminous AGN and normal IRAC-detected star-forming galaxies out to $z\sim 3$.  Furthermore, intensive spectroscopic follow-up campaigns and photometric redshifts tuned to both AGN and normal galaxies give near complete redshift constraints on both the AGN and star-forming populations \citep{lilly07,trump09,salvato09,ilbert09}.

This paper is organized as follows. In \S2, we provide an overview of IRAC AGN selection and discuss which AGN should reliably be selected by this method.  We present the relevant COSMOS datasets in \S3, and we discuss in \S4 the IRAC power-law selected AGN in COSMOS.  In \S5, we present the properties of the COSMOS \xmm\ sample, and in \S6, we summarize the \xmm\ sample's trends in IRAC color space.  We then investigate the properties of the full IRAC sample in \S7 and of high-redshift star-forming galaxies in \S8.  In \S9, we present the revised selection criteria, and we summarize the results in \S10.  Throughout the paper, we assume the following cosmology: ($\Omega_{\rm m}$,$\Omega_{\rm \Lambda},H_0$)=(0.27, 0.73, 70.5~km~s$^{-1}$~Mpc$^{-1}$), and we quote all magnitudes in the AB system unless otherwise noted.

\section{IRAC AGN Selection: an Overview}

The premise behind IRAC AGN selection is illustrated in Figure 1, where we plot composite AGN$+$starburst SEDs constructed using the QSO1 and M82 templates of \cite{polletta08}. The SEDs of normal star-forming galaxies display a prominent dip between the 1.6 \micron\ stellar bump and the long-wavelength emission from star-formation heated dust ($T_{\rm dust} \sim 25-50$~K).  Dust near an AGN's central engine, however, can reach a sublimation temperature of $T_{\rm dust}\sim 1000-1500$~K and thus radiate into the near-infrared (NIR).  If the AGN is sufficiently luminous compared to its host galaxy, the superposition of black body emission from the AGN-heated dust will fill in the dip in the galaxy's SED and produce a red, power-law like thermal continuum across the IRAC bands.  As shown in Figure 1, while the UV-optical SEDs of obscured and unobscured AGN bear little resemblance to one another, this characteristic MIR spectral shape should remain so long as the obscuring medium is optically thin at NIR-MIR wavelengths.

\begin{figure}
\epsscale{1.15}
\plotone{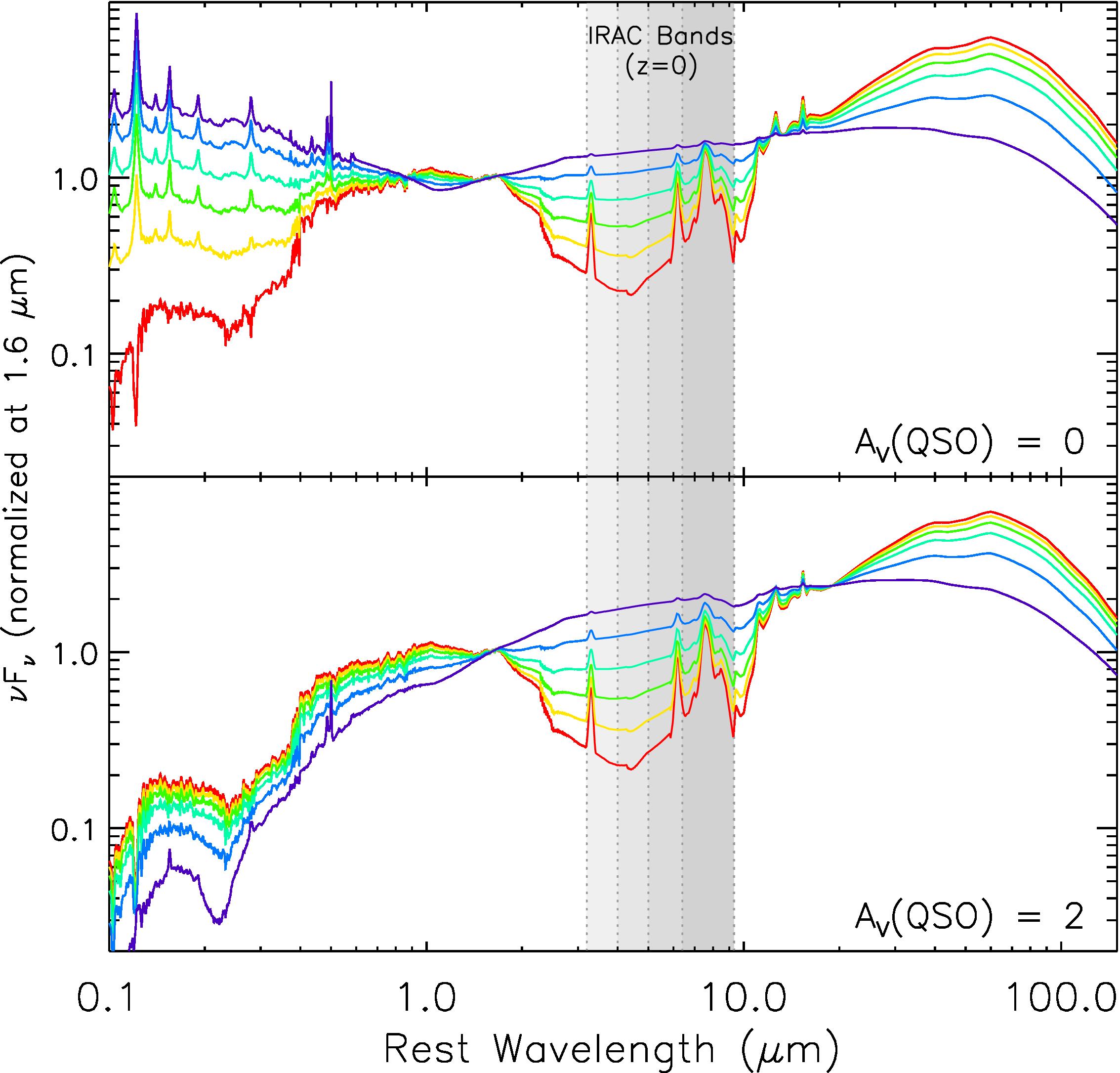}
\caption{Composite SEDs constructed using the QSO1 and M82 templates of \cite{polletta08}, scaled to give 1-10 \micron\ AGN contributions of 0\% (red in the online version) to 95\% (purple in the online version).  The final SEDs have been normalized at 1.6 \micron.  In the lower panel, we apply an extinction of $A_{\rm V}=2$ to the QSO1 SED using the \cite{draine03} extinction law. The four IRAC bands at $z=0$ are shaded.  While luminous unobscured and obscured AGN have very different UV-optical SEDs, luminous AGN should display a red MIR power-law SED regardless of obscuration.}
\end{figure}

The optical thickness of the obscuring medium at NIR-MIR wavelengths depends on the structure of the torus.  A smooth and geometrically thick torus will obscure even the MIR emission from Type 2 AGN, leading to a large offset in MIR luminosity between Type 1 and Type 2 AGN \citep{pier93}.  While no studies have observed the degree of MIR anisotropy predicted by smooth torus models, Type 2 AGN do appear to be $\sim 3-5$ fainter in the MIR than Type 1 AGN when the radio continuum luminosity is used to constrain the AGN's intrinsic luminosity \citep{heckman95,buchanan06,haas08,leipski10}.  However, no offset in MIR luminosity is observed when the AGN luminosity is instead normalized by the absorption-corrected X-ray luminosity and when the AGN's contribution to the MIR emission is isolated either spatially or spectrally \citep{lutz04,gandhi09}.  This suggests that the radio-based comparison may be biased, at least in part, by the contribution from beamed synchrotron emission in Type 1 AGN \citep{cleary07}.

\begin{figure*}
$\begin{array}{c}
\includegraphics[angle=90,scale=0.72]{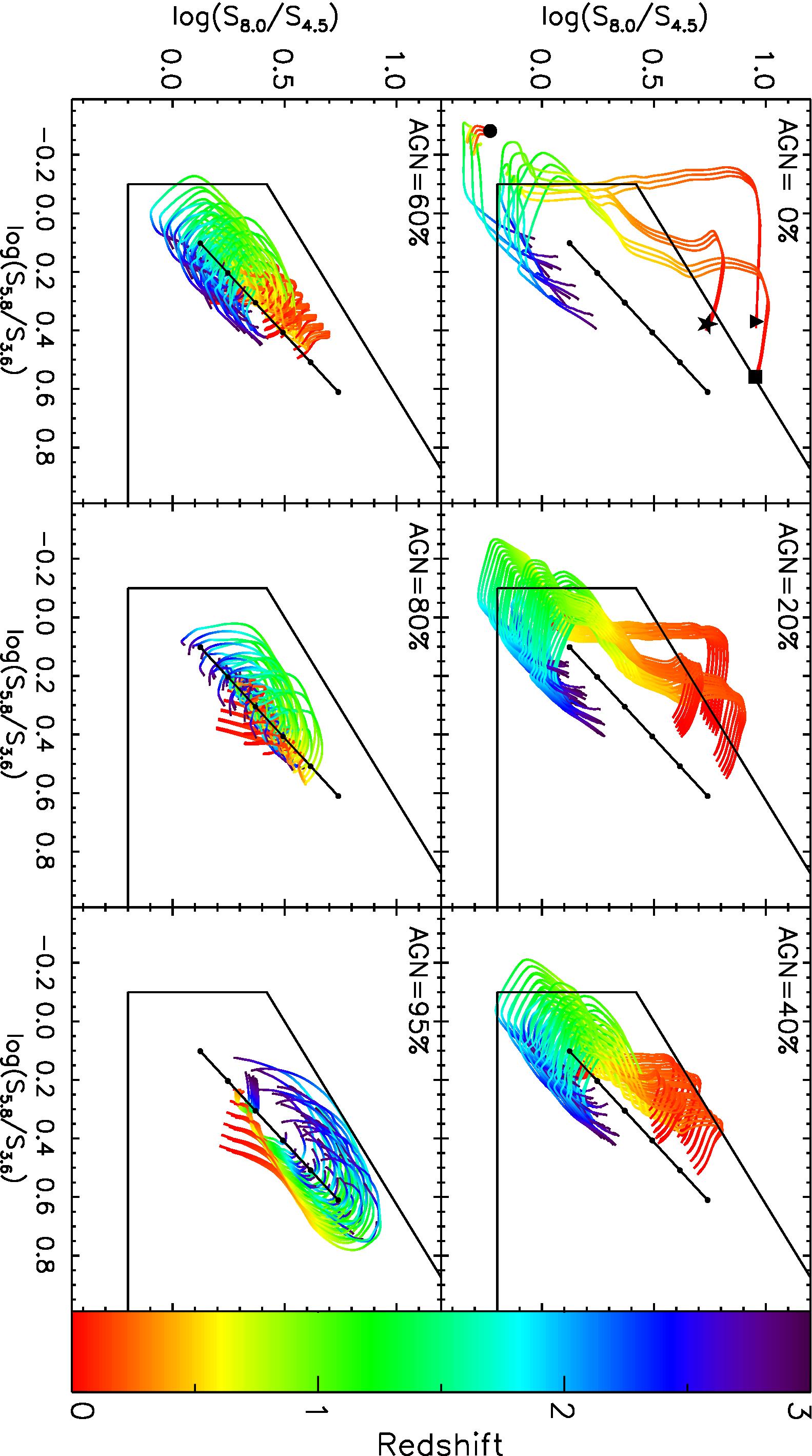} 
\end{array}$
\caption{Predicted $z=0-3$ IRAC colors of AGN/galaxy composite SEDs in \cite{lacy04,lacy07} color space, where the AGN fraction is defined between 1 and 10 \micron.  The star-forming templates represent the ULIRG IRAS 22491 \citep[square,][]{polletta08}, the starburst M82 \citep[star,][]{polletta08}, a normal star-forming spiral galaxy \citep[triangle,][]{dale02}, and an elliptical galaxy \citep[circle,][]{polletta08}, where large symbols mark each family of purely star-forming templates at $z=0$. The AGN template is the QSO1 template of \cite{polletta08}.  Additional extinctions of $A_{\rm V} = 0-2$ and $A_{\rm V} = 0-20$ are applied to the star-forming and AGN components, respectively. The wedge is the AGN selection region of \cite{lacy07}, and the line represents the power-law locus from $\alpha = -0.5$ (lower left) to $\alpha = -3.0$ (upper right).  While the templates of purely star-forming galaxies avoid the power-law locus, they enter the current AGN selection region at both low and high redshift.  As the AGN's contribution to the MIR emission increases, the SEDs move inwards and redwards toward the power-law locus.}
\end{figure*}

\begin{figure*}
$\begin{array}{c}
\includegraphics[angle=90,scale=0.72]{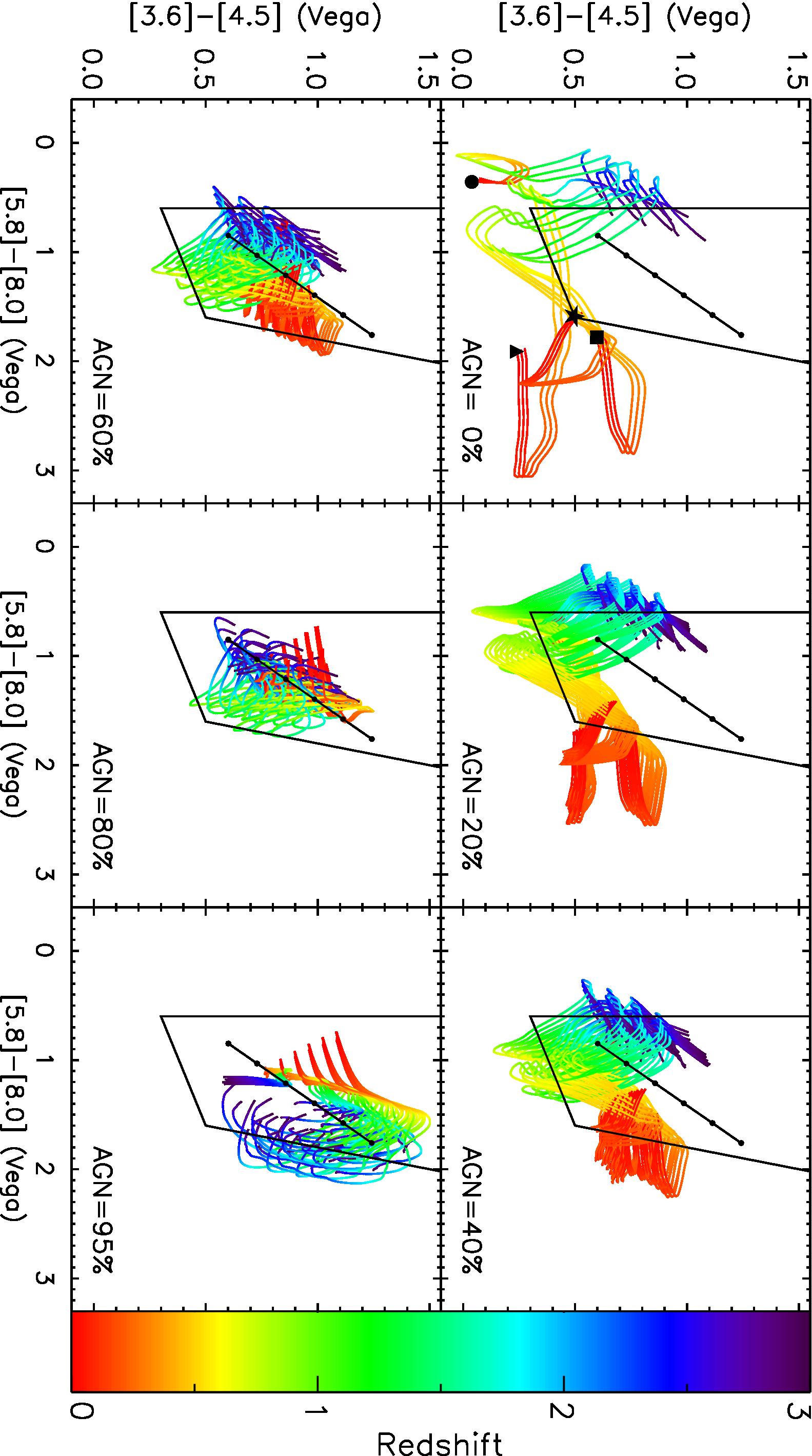} 
\end{array}$
\caption{Same as Figure 2, but for \cite{stern05} color space. The wedge is the AGN selection region of \cite{stern05}.}
\end{figure*}

While the observed degree of MIR anisotropy therefore remains an open question, a low optical depth is a natural consequence of clumpy torus models, which predict only weak NIR-MIR anisotropies that essentially disappear by $\lambda \sim 12$\micron\ \citep[e.g.][]{nenkova08a,nenkova08b}. Even at 1 \micron, the clumpy torus model of \cite{nenkova08b} predicts an edge to pole flux ratio of $\lsim 5$ for a cloud optical depth of $\tau_{\rm V} = 60$. For samples of Type 1 and Type 2 AGN averaged across all viewing angles, we might therefore expect only small systematic offsets in the observed MIR emission, and only at the lowest rest-frame wavelengths probed by the IRAC bands. We will investigate this issue in more detail in Sections 6 and 9.   

\vspace*{0.5cm}
\subsection{IRAC colors of AGN/star-forming composites}

To illustrate the expected IRAC colors of galaxies with varying AGN contributions, we plot in Figures 2 and 3 the $0<z<3$ IRAC colors of composite SEDs constructed from the QSO1, elliptical (ell2), starburst (M82), and ULIRG (IRAS 22491) templates of \cite{polletta08} and the star-forming template of \cite{dale02}, plotted in both \cite{lacy04,lacy07} and \cite{stern05} color space.  To allow for the effects of reddening, we apply additional extinctions of $A_{\rm V} = 0-2$ and $A_{\rm V} = 0-20$ to the star-forming and AGN components, respectively, using the \cite{draine03} extinction curve \cite[e.g.,][]{hickox07,hou11}.

We also show in Figures 2 and 3 the AGN selection wedges of \cite{lacy07} and \cite{stern05} along with the IRAC power-law locus (the line on which a source with a perfect IRAC power-law SED would fall), where circles denote power-law slopes of $\alpha = -0.5$ to  $-3.0$ ($f_{\rm \nu} \propto \nu^{\alpha}$).  While the power-law locus itself extends to bluer slopes, luminous AGN are expected to display red slopes of $\alpha \le -0.5$ \citep{aah06,donley07}.

In both \cite{lacy04,lacy07} and \cite{stern05} color-space, purely star-forming templates generally avoid the region surrounding the power-law locus, at least out to $z\sim3$ (the potential for contamination by higher redshift galaxies will be discussed in \S8).  As the AGN becomes increasingly dominant, the IRAC colors then shift inwards and redwards towards the power-law locus.  Generally speaking, the more luminous the AGN, the redder the IRAC colors.  A source's precise location in IRAC color space, however, will depend not only on the relative AGN/host contributions, but on its redshift, the reddening of its host and AGN components, and the host galaxy type.

\begin{deluxetable*}{llccc}
\centering
\tabletypesize{\footnotesize}
\tablecaption{Pure Starbursts with IRS Spectra}
\tablehead{
\colhead{Number}             &
\colhead{Mean $z$}           &
\colhead{Sample}             &
\colhead{Fields}             &
\colhead{Selection}          
}
\startdata
16   & $0.06 \pm 0.04$    & \cite{houck07}        & NDWFS          & $f_{\rm 24} > 10$ mJy                   \\ 
182  & $0.13 \pm 0.10$    & \cite{sargsyan09}     & SWIRE, NDWFS   & z$<$0.5 starbursts                     \\  
22   & $0.71 \pm 0.06$    & \cite{fu10}           & COSMOS         & $f_{\rm 24} > 0.7$ mJy  , $z\sim 0.7$   \\
24   & $0.74 \pm 0.22$    & \cite{dasyra09}       & SFLS           & $f_{\rm 24} > 0.9$ mJy                  \\
4    & $0.75 \pm 0.29$    & \cite{brand0870}      & NDWFS          & 70 \micron\ selected, optically-faint  \\
5    & $0.86 \pm 0.14$    & \cite{farrah09}       & SWIRE          & 70 \micron\ selected, optically-faint  \\
11   & $1.63 \pm 0.12$    & \cite{farrah08}       & SWIRE          & IRAC bump selected                     \\
6    & $1.75 \pm 0.60$    & \cite{pope08smg}      & GOODS-N        & SMGs                                   \\
4    & $1.82 \pm 0.10$    & \cite{weedman06agnsb} & SWIRE          & IRAC bump selected                     \\
6    & $2.05 \pm 0.48$    & \cite{huang09}        & EGS            & IRAC bump selected                     \\
2    & $2.59 \pm 0.63$    & \cite{menendez09}     & SWIRE          & SMGs                                   \\
\enddata
\end{deluxetable*}

\begin{figure*}
$\begin{array}{cc}
\includegraphics[angle=0,scale=0.46]{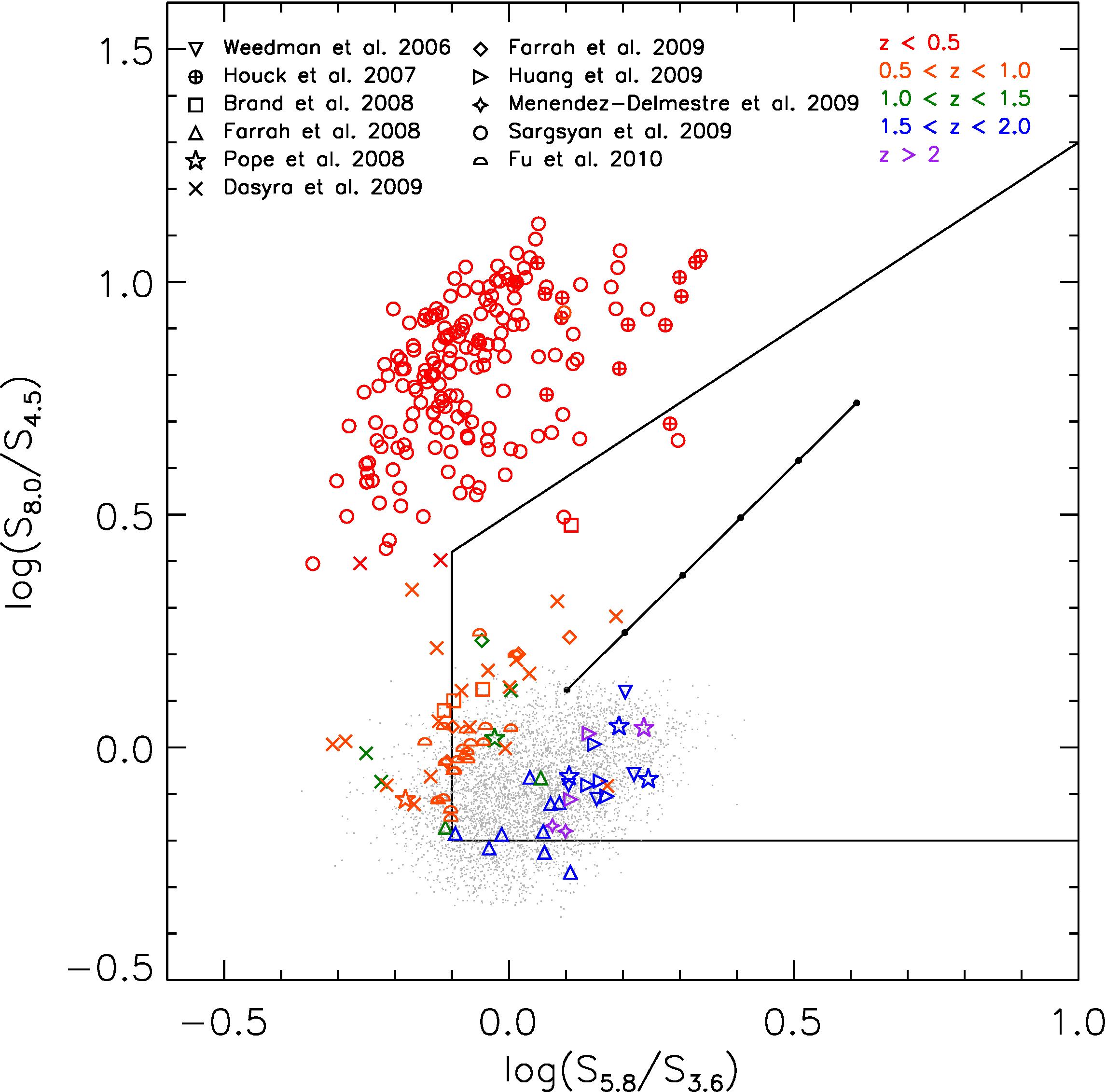} &
\includegraphics[angle=0,scale=0.46]{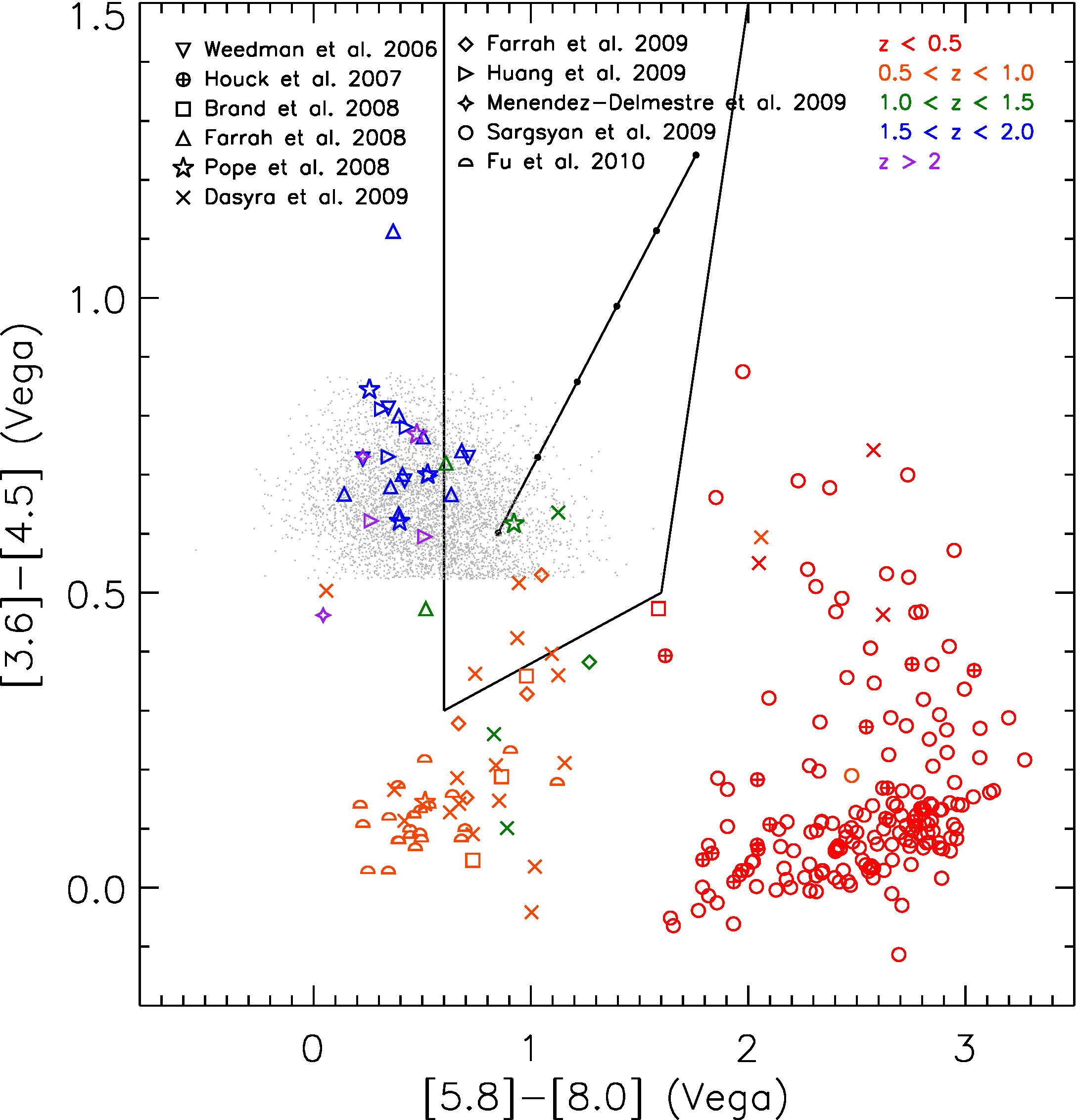} 
\end{array}$
\caption{IRAC colors of pure starbursts, as determined via \spitzer\ IRS spectroscopy (large symbols).  The small grey points represent sources that have not been observed with IRS, but that meet the $1.5 < z < 3$ star-forming IRAC criteria of \cite{huang09}. The wedges show the current AGN selection regions of \cite[][left]{lacy07} and \cite[][right]{stern05}, and the line shows the power-law locus from $\alpha = -0.5$ to $\alpha = -3.0$.  A significant fraction of purely star-forming galaxies lie within the current AGN selection regions, particularly at high redshift.}
\end{figure*}

\subsection{IRAC colors of purely star-forming galaxies}

While the purely star-forming templates shown in Figures 2 and 3 lie outside of the power-law region in IRAC color space, they enter the \cite{lacy04,lacy07} and \cite{stern05} AGN selection wedges at both low and high redshift \citep{barmby06,donley07,donley08}.    Thanks to recent \spitzer\ IRS spectroscopic campaigns in fields with deep IRAC data, however, we need no longer rely solely on templates to constrain the IRAC colors of purely star-forming galaxies.  

The IRAC colors of \spitzer\ IRS sources classified as pure starbursts are plotted in Figure 4 (see Table 1 for details).   While each IRS study uses slightly different criteria for identifying purely star-forming galaxies, we standardize this definition by requiring that EW(6.2 \micron)$ \ge 0.4$ when the 6.2 \micron\ PAH feature lies within the observed bandpass \citep{weedman09} and EW(11.3 \micron)$ > 0.8$ when it does not \citep{dasyra09}.  We exclude any source flagged as having blended IRAC photometry,  and plot for comparison the colors of sources that have not been observed with IRS, but that meet the $1.5 < z < 3$ star-forming IRAC criteria of \cite{huang09}.

The IRAC colors of the IRS star-forming galaxies are generally consistent with the predictions from Figures 2 and 3.  In \cite{lacy04,lacy07} color space, low redshift galaxies lie predominantly above the AGN wedge, though some fall just within the AGN selection region.  At $0.5<z<1.5$, they fall both to the left of and within the AGN wedge, and at $z>1.5$, the vast majority of star-forming galaxies lie within the AGN selection region.  In \cite{stern05} color space, most low and moderate redshift star-forming galaxies fall below the selection region, and many of the high-redshift galaxies lie to the left of the AGN wedge.  However, a significant fraction of the \cite{huang09} $1.5 < z < 3$ star-forming candidates scatter into the AGN selection region, as do several moderate and high-redshift IRS-selected starbursts.

With the exception of a $z=0.73$ ULIRG from \citet[][shown by a cross near the $\alpha \sim -1.0$ power-law locus]{dasyra09}, the star-forming galaxies lie well away from the power-law locus, especially at $\alpha \le -1.0$ and in \cite{lacy04} color space.  The IRS spectrum of this source displays strong PAH emission with EW(11.3\micron)$ = 1.43 \pm 0.07$ \citep{dasyra09}, and a low $S/N$ optical spectrum shows strong [OII], H$\beta$, and [OIII] emission but no obvious high-ionization lines \citep{papovich06}.  At $z=0.73$, however, neither the optical nor MIR spectral coverage extend into the 1-5 \micron\ rest-frame regime sensitive to AGN heated dust.  Given the available data, it is therefore unclear whether this source's power-law like IRAC SED is due to a contribution from AGN-heated dust, or whether it is a true star-forming interloper.  Nevertheless, it is clear from Figures 2, 3, and 4 that while the current AGN selection regions inadequately separate AGN from moderate- to high-redshift star-forming galaxies, luminous AGN should occupy a well-defined region of color space with minimal star-forming contamination.  To test these predictions and better constrain the region of IRAC color space that can reliably be used for AGN selection, we turn to COSMOS.

\section{COSMOS Data and Sample Selection}

The 2~deg$^2$ COSMOS field is characterized by both deep and wide coverage across much of the electromagnetic spectrum.  We focus here on the \textit{XMM-Newton} and \spitzer\ IRAC datasets that cover the full survey area \citep{hasinger07,sanders07}, as well as the deeper \chandra\ data available over the central 0.9~deg$^2$ \citep{elvis09}.  We also make use of the intermediate and broad-band optical-NIR data used to construct photometric redshifts for the X-ray--detected and normal galaxy populations \citep[][Capak et al. 2011, in prep.]{capak07,salvato09,ilbert09,salvato11}.  

\subsection{IRAC}

The S-COSMOS IRAC data cover the full 2.3 deg$^{2}$ COSMOS field to 1200s depth, reaching $5\sigma$ sensitivities of 0.9, 1.7, 11.3, and 14.6~\microjy\ in the 3.6, 4.5, 5.8, and 8.0 \micron\ bands, respectively \citep{sanders07}.  The IRAC sample used here contains the 26251 non-flagged IRAC sources that lie in the 1.73 deg$^{2}$ region of COSMOS with high-quality unmasked optical-NIR data \citep{ilbert09,ilbert10}, have an optical/NIR counterpart within a 2\arcsec\ search radius, are not flagged as stars by \cite{ilbert09} or \cite{mccracken10}, have a photometric or spectroscopic redshift, and meet the $5\sigma$ sensitivity limits of \cite{sanders07} in each of the four IRAC bands.  The aperture-corrected IRAC photometry was calculated in 1.9\arcsec\ apertures using SExtractor \citep{sanders07}.

\subsection{\xmm}
Our primary sample of X-ray selected AGN is drawn from the $\sim 40$~ks \xmm-COSMOS survey \citep{hasinger07,cappelluti09,brusa10}.  While the \chandra\ data of \cite{elvis09} reach greater depths over the central 0.9 deg$^2$ and will be used later in the paper, the larger area of the \xmm\ survey better samples the population of luminous yet low space density AGN, resulting in roughly equal numbers of Seyfert galaxies (log~$L_{\rm x}$(ergs~s$^{-1}$)$ \le 44$) and QSOs (log~$L_{\rm x}$(ergs~s$^{-1}$)$ \ge 44$).  

The full \xmm-COSMOS sample of \cite{brusa10} contains 1797 sources, 93\% of which have a spectroscopic or photometric redshift indicative of an extragalactic source. (Of the remaining sources, 5.5\% are Galactic stars, and 1.5\% lack a redshift estimate).  We restrict our \xmm\ sample to the sources that lie within the 1.73 deg$^2$ region of COSMOS with deep, uniform, and unmasked optical-NIR data \citep{ilbert09}, have a reliable optical/NIR counterpart in \cite{brusa10}, are not flagged as stars in \cite{brusa10}, \cite{ilbert09}, or \cite{mccracken10}, have a photometric or spectroscopic redshift, and have high quality IRAC counterparts that meet or exceed the COSMOS $5\sigma$ sensitivity in each of the four IRAC bands \citep{sanders07}.  Applying all but the IRAC detection criterion gives a sample of 1183 sources, 12\% of which are excluded in our final sample of 1039 \xmm\ sources with high significance IRAC counterparts. 

Of the \xmm\ sources, 63\% have a spectroscopic redshift as compiled by \cite{brusa10} and the remaining 37\% have an AGN-specific photometric redshift from \cite{salvato09}. To constrain the IRAC properties of the \xmm\ sample, we use the IRAC flux densities from \cite{ilbert09} \cite[as given by][]{brusa10}, which have been re-extracted at the positions of the \xmm\ sources to minimize the effects of blending.  To permit comparison with the 3$\arcsec$ optical-NIR aperture photometry, \cite{ilbert09} multiplied all of the IRAC flux densities by 0.75.  Here, we undo this correction to recover the total IRAC flux densities. 

\subsection{\chandra}

The COSMOS \chandra\ coverage reaches 160~ks depth over 0.5~deg$^2$ and 80~ks depth over an additional 0.4~deg$^2$ \citep{elvis09}.  The full catalog of \chandra\ sources contains 1761 sources, 1010 of which have optical/NIR and IRAC counterparts from \cite{elvis09} and Civano et al. (2011, in prep.) that meet the criteria described above. Of these, 56\% are also \xmm\ sources, and 61\% have spectroscopic redshifts.  For the remaining \chandra\ sources, we adopt the AGN-specific photometric redshifts of \cite{salvato11}. 

\subsection{Optical-NIR Photometry and Redshifts}

Optical and NIR photometry are taken from the intermediate and broad-band catalogs of \cite{capak07}, \cite{ilbert09}, \cite{mccracken10} and \cite{capak11}. We apply the band-specific systematic offsets given by \cite{ilbert09}, which correct for errors in the zero-points and minimize the systematic scatter in the photometric redshifts, and we aperture correct the photometry using the band-independent auto-offset aperture corrections. 

High confidence spectroscopic redshifts and spectral classifications were compiled from the Sloan Digital Sky Survey \citep[SDSS,][]{kauffmann03,adelman06}, the Magellan/IMACS and MMT observing campaigns of \cite{prescott06}, \cite{trump07}, and \cite{trump09}, zCOSMOS \citep{lilly07}, and subsequent Keck runs.  Photometric redshifts for non X-ray sources are taken from \cite{ilbert09}, whereas AGN-specific photometric redshifts for the \xmm\ and \chandra\ samples were drawn from \cite{salvato09} and \cite{salvato11}, respectively. 

\section{Power-law Selection: A Starting Point}

IRAC power-law selection identifies sources that lie near the power-law locus in color space, and therefore selects only the most secure AGN-dominated candidates from the IRAC AGN wedges \citep[e.g.,][]{aah06,donley07,donley08,park10}.  For a source to be selected as a power-law AGN, a line of slope $\alpha \le -0.5$ must provide a good fit to its logarithmic IRAC photometry ($f_{\rm \nu} \propto \nu^{\alpha}$, or log$f_{\rm \nu} = \alpha$log$\nu + b$).  The goodness of fit is measured by the $\chi^2$ probability, $P_{\rm \chi} $, with a limit of $P_{\rm \chi} \ge 0.1$ \citep{donley07,donley08}.  

IRAC power-law selection therefore depends not only on the IRAC flux densities used for color-color selection, but also on the IRAC errors typically derived using SExtractor.   Unfortunately, SExtractor tends to underestimate photometric uncertainties, as it does not account for the correlated sky noise present in mosaiced data \cite[e.g.,][]{gawiser06,barmby08}.  Furthermore, the cataloged measurement errors do not account for the IRAC calibration uncertainty, which is at best 3\% and is at worst 10\% when the sub-pixel response and the array-location-dependent changes in pixel solid angle and spectral response are not taken into account \citep{reach05}. The SExtractor errors can therefore be as low as $\Delta f_\nu/f_\nu$ = 0.1\% for our IRAC sample, with the brightest IRAC sources exhibiting the most underestimated errors.   

Using such highly underestimated errors, it becomes nearly impossible to obtain a statistically acceptable fit to a given SED, even when the SED appears to closely match the photometry.  As an example, consider the 25 \xmm\ sources that lie in the immediate vicinity (within $0.025$~dex) of the $\alpha = -1.0$ power-law locus. Using the flux densities and errors from the official COSMOS IRAC catalog, the power-law criterion is met by only 30\% of the fainter half of the sample and by none of the brighter half of the sample.  

This effect is further illustrated in Figure 5, where we plot the IRAC colors of the COSMOS \xmm\ and IRAC samples and identify the sources selected as power-law galaxies using the cataloged errors and those selected as power-law galaxies when additional uncertainties of 3\%, 5\%, 10\%, and 15\% are added to the cataloged errors. To quantify the dependency of power-law selection on the IRAC errors, we define a box around the power-law locus in \cite{lacy04,lacy07} color space that encloses nearly all of the \xmm\ sources with red ($\alpha \le -0.5$) IRAC colors (see Figure 5).  If we were to select IRAC power-law galaxies using the cataloged errors, \textit{we would recover only 5\% of the \xmm\ sources in the power-law box}.  Adding additional uncertainties of 3\%, 5\%, 10\% and 15\% to the IRAC errors increases this fraction to 41\%, 69\%, 96\%, and 98\%, respectively.     

\begin{figure*} 
$\begin{array}{cccc}
\includegraphics[angle=0,scale=0.45]{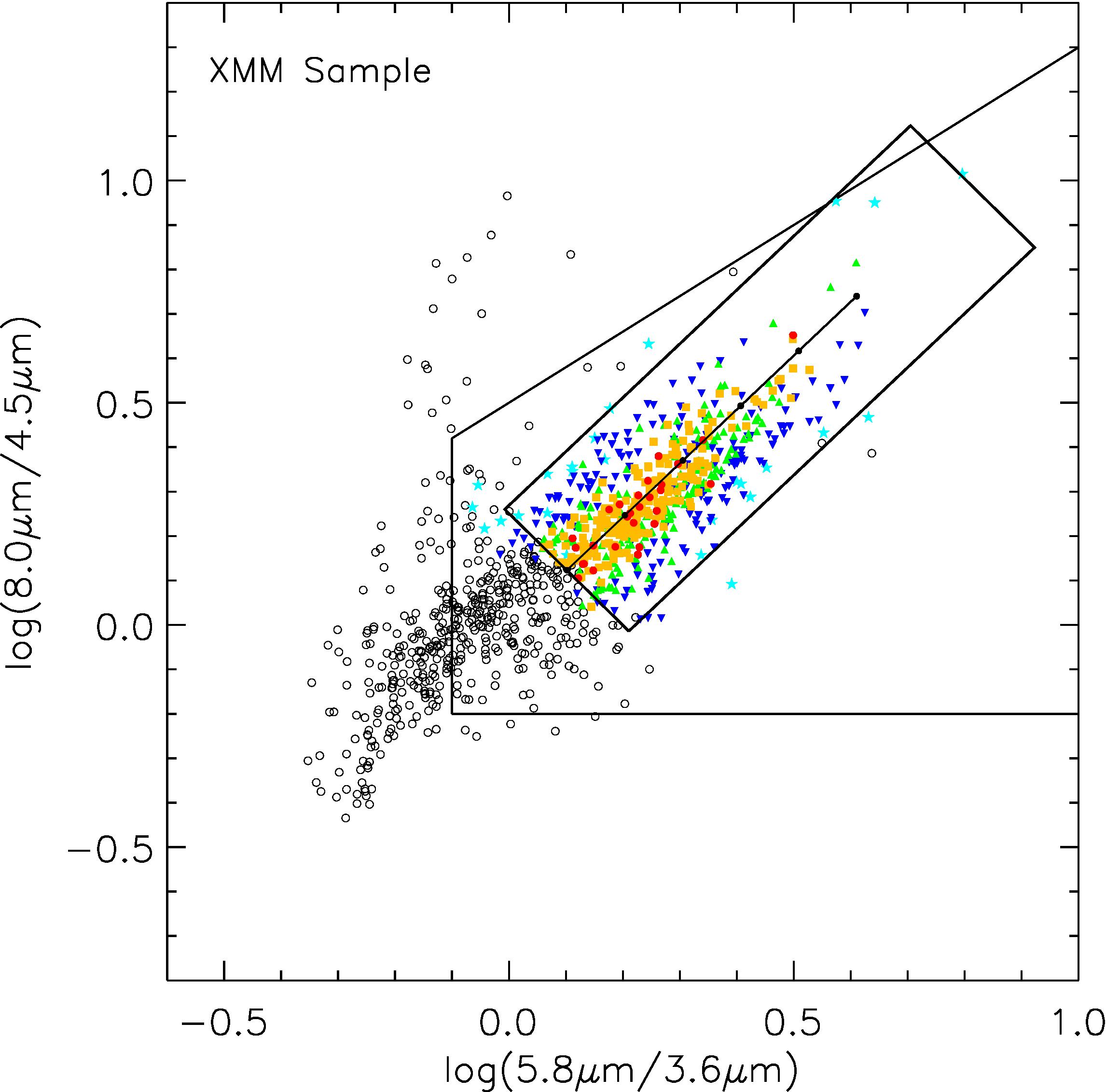} &
\includegraphics[angle=0,scale=0.45]{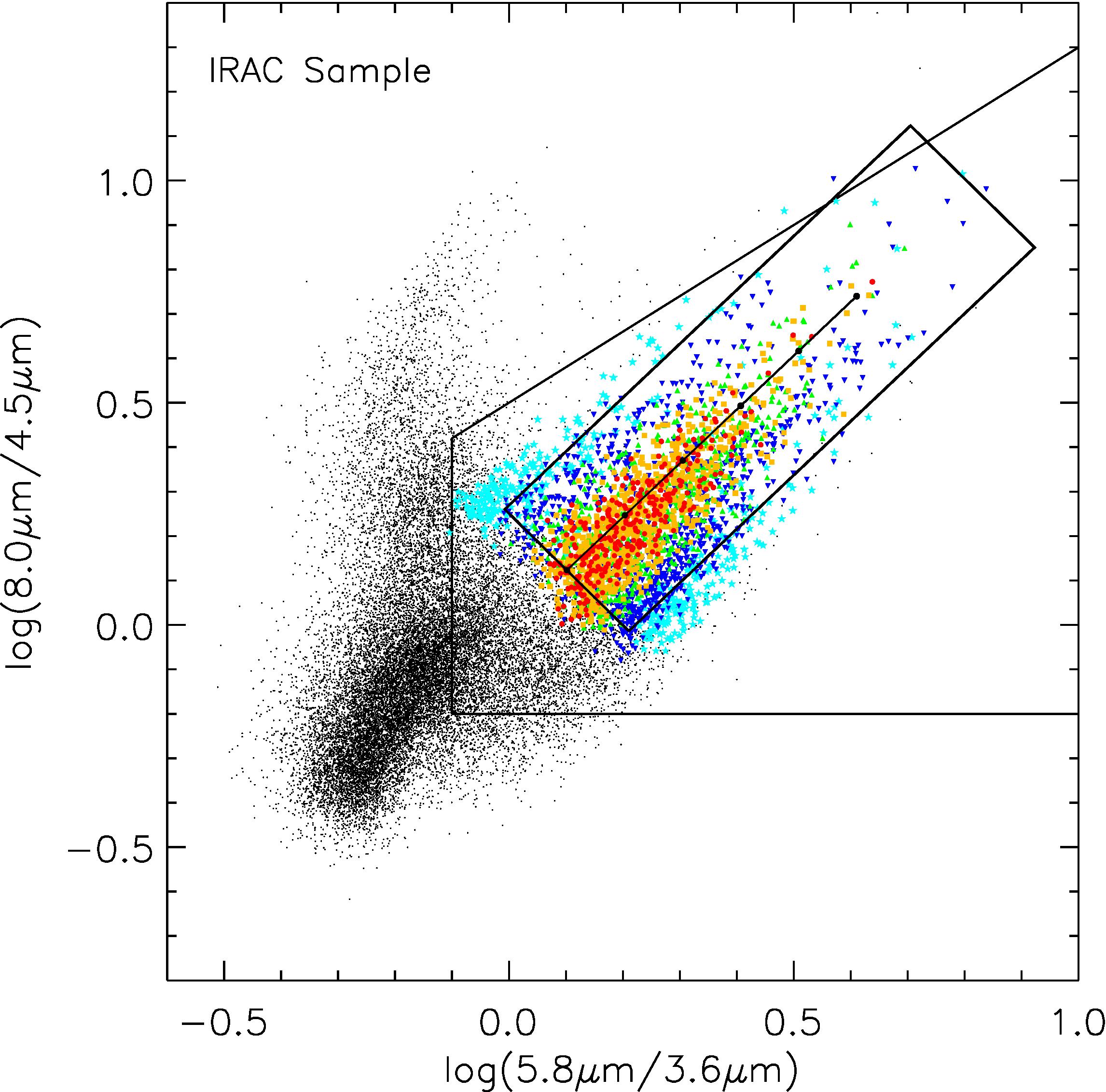} \\
\includegraphics[angle=0,scale=0.45]{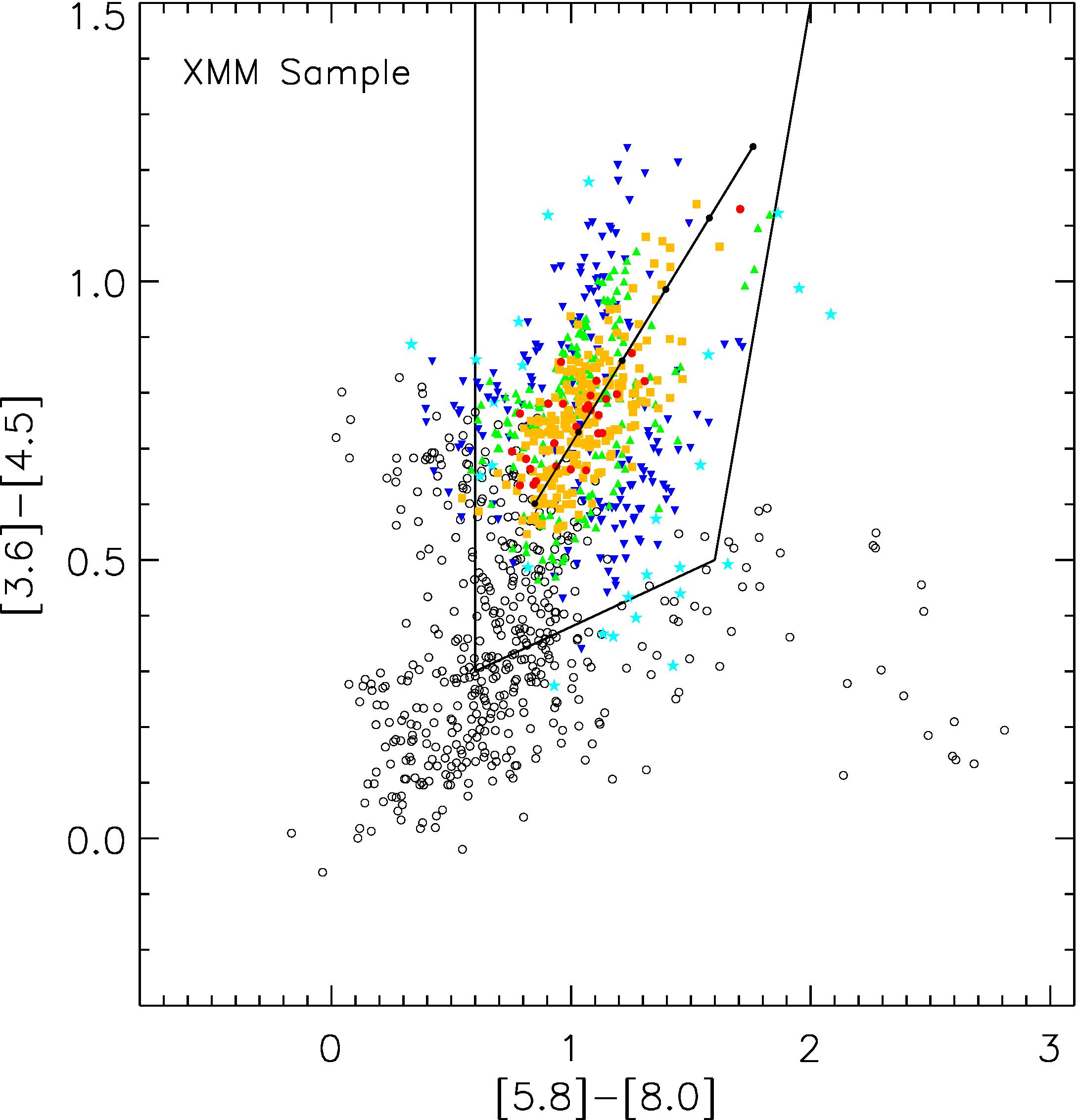} &
\includegraphics[angle=0,scale=0.45]{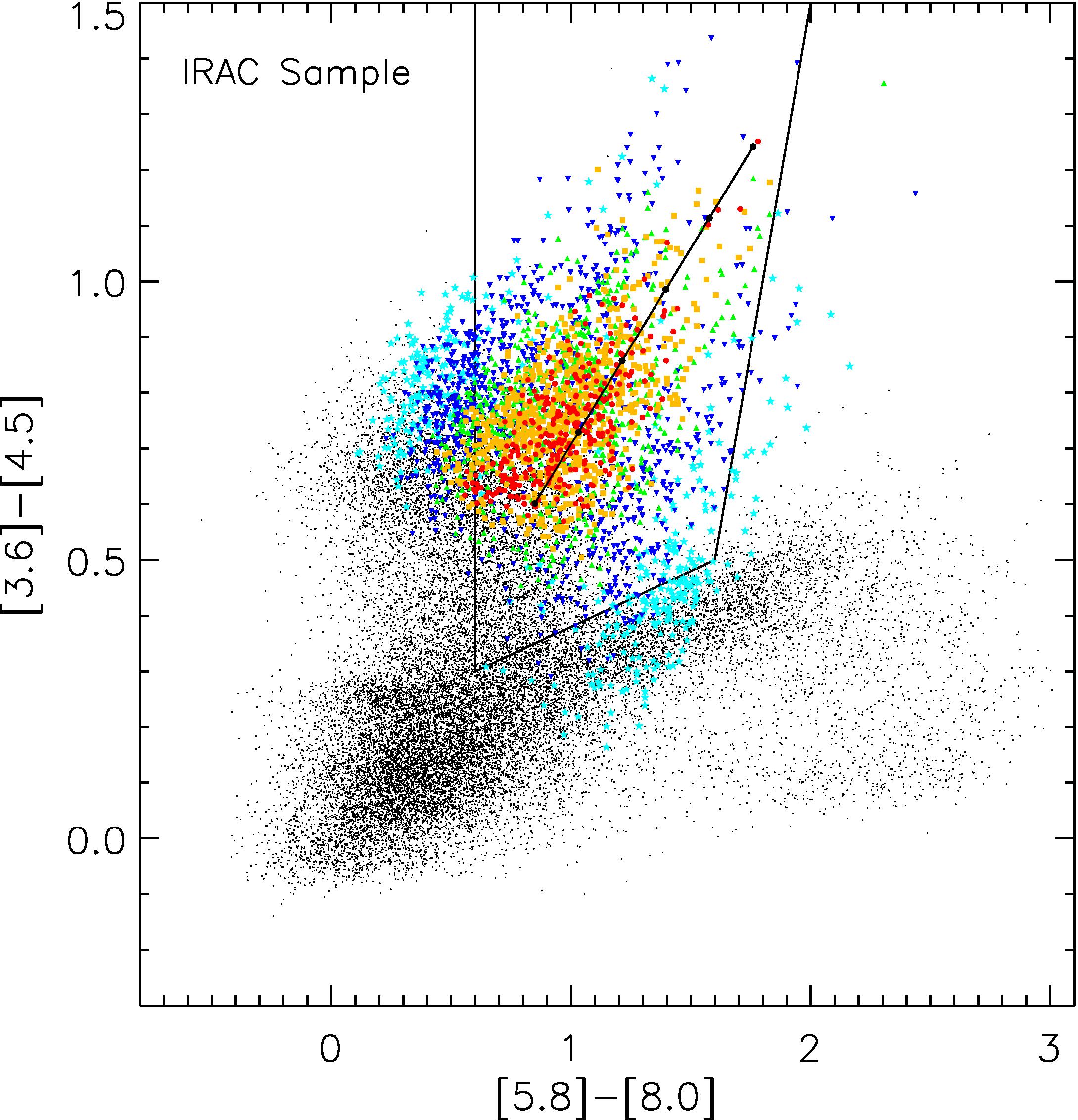} \\
\end{array}$
\epsscale{1.1}
\caption{IRAC color-color diagrams for the \xmm\ (left) and IRAC (right) samples in COSMOS, where the top and bottom plots show \cite{lacy04,lacy07} and \cite{stern05} color-space, respectively.  Solid circles (red in the online version) indicate those sources that would be selected as $\alpha \le -0.5$ power-law galaxies using the IRAC flux densities and errors from the official COSMOS catalog, and squares (gold in the online version) indicate the additional power-law galaxies selected when the IRAC calibration uncertainty of 3\% is added to the IRAC errors \citep{reach05}.  Likewise, (green) triangles, (blue) upside-down triangles, and (cyan) stars indicate the additional power-law galaxies selected when uncertainties of 5\%, 10\%, and 15\% are added to the cataloged errors, respectively. To select complete samples of AGN that lie near the power-law locus in color space but that may exhibit small deviations from a perfect power-law SED, a 10\% error on the flux must be assumed.}
\end{figure*}

Because the sources in the full IRAC sample tend to be fainter than those in the \xmm\ sample, they display more scatter in IRAC color space.  Nonetheless, we observe a similar trend in power-law selection: 12\%, 40\%, 62\%, 91\%, and 96\% of the sources in the power-law box are selected as power-law galaxies using the cataloged errors and additional uncertainties of 3\%, 5\%, 10\% and 15\%, respectively. 

Adjustments to the SExtractor-derived IRAC errors are clearly required to identify complete samples of AGN that lie near the power-law locus in color space. In previous work, we have assumed an overall IRAC calibration uncertainty of 10\% to recover the majority of X-ray sources with red IRAC colors \citep{aah06,donley07,donley08}.  While this likely overestimates the true errors on the flux, this assumption facilitates the selection of sources with small deviations from a perfect power-law SED.  As shown in Figure 5, however, these red, power-law dominated AGN occupy a well-defined region of \cite{lacy04,lacy07} IRAC color space.  Power-law galaxies could therefore be selected on the basis of their IRAC colors alone, thus eliminating the dependence of power-law selection on poorly constrained photometric errors and assumed uncertainties. 

In contrast, power-law galaxies cannot be cleanly identified using color-color cuts in \cite{stern05} color-space.  This increased scatter arises from the use of adjacent bands (3.6/4.5~\micron\ and 5.8/8.0~\micron), which minimizes the wavelength baseline, increases the sensitivity to small variations in the SEDs, and groups together the two low-sensitivity IRAC channels \citep[5.8 and 8.0 \micron, see also][]{donley07, donley08, richards09}.  We therefore focus below on the \cite{lacy04,lacy07} representation of IRAC color space, although we will continue to show relevant plots in both representations to motivate the need for new selection criteria. 

We take as a starting point for these revised IRAC color-color cuts the $\alpha \le -0.5$ power-law box defined above. Not only does this region enclose nearly all \xmm\ sources with red IRAC colors, but it also tightly encloses the AGN-dominated templates shown in Figure 2.  While this power-law box should therefore be highly complete, at least to luminous AGN that dominate the light from their hosts, high-redshift star-forming galaxies may enter the bluest portions of the power-law box (see Figure 4).  In the sections that follow, we therefore use the COSMOS \xmm, \chandra, and IRAC samples, as well as samples of high-redshift galaxies, to further refine this revised AGN selection region and to quantify its completeness and reliability. 

\section{\xmm\ sample: Properties}

To constrain the efficiency and completeness of IRAC selection and test for any trends in AGN properties across IRAC color space, we first determine the intrinsic obscuration and X-ray luminosity of the COSMOS \xmm\ sample. 

\vspace*{0.5cm}
\subsection{AGN Type: Optical and X-ray Classification}

AGN can be divided into Type 1 (unobscured) or Type 2 (obscured) classes using a variety of criteria, including optical spectral classification, optical SED fitting, IR to optical flux ratios, X-ray spectral fitting, and X-ray hardness or flux ratios \citep[for a discussion of the agreement between classification schemes, see][]{trouille09}.  Here, we focus on two methods.  For the first, we adopt the approach of \cite{brusa10} and use the optical spectra classification when available and the best-fit broad-band SED type from \cite{salvato09} otherwise, and refer to the resulting classes as ``optical Type 1'' and ``optical Type 2''.  Of the \xmm\ sources with Type 1 and Type 2 spectra, 85-90\% are also best-fit by Type 1 and Type 2 SEDs, respectively, indicating that these two variations on optical classification are well-matched. Using this approach, we find an optical Type 2 fraction of 58\% for the \xmm\ sample.  

For the second approach, we turn to the X-ray and use each source's redshift, hard to soft X-ray flux ratio, and an assumed intrinsic X-ray photon index of $\Gamma = 1.8$ \citep{tozzi06} to estimate its intrinsic column density, $N_{\rm H}$.  We then group sources into ``X-ray Type 1'' and ``X-ray Type 2'' classes using a cut of $N_{\rm H} = 10^{22}$~cm$^{-2}$.   For the brightest sources in our sample, we compare our estimated column with that derived from X-ray spectral fitting \citep{mainieri07}. For the 24 sources with measurable columns from both techniques, we find a typical absolute scatter of log~$N_{\rm H}$ (cm$^{-2}) = 0.28$ and a mean offset of $\Delta$log~$N_{\rm H}$(cm$^{-2}$)$ = 0.02$. The agreement between our simple flux ratio method and the full spectral fitting is therefore generally good. 

Using the flux-ratio-derived column density, we measure X-ray types for $\sim 3/4$ of our sample and find an X-ray Type 2 fraction of 52\%. Of the remaining sources, 90\% are detected in only the soft band but have upper limits on $N_{\rm H}$ that are potentially consistent with the Type 2 class (due to the comparably low sensitivity of the \chandra\ hard band), and 10\% are detected in only the hard band but have lower limits on $N_{\rm H}$ that are potentially consistent with the Type 1 class.  This leads to ambiguous classifications.

Comparing the X-ray and optical types, we find that $65$\% of the optically-classified AGN have matching X-ray types when the latter is available.  While this fraction is far from 100\%, $\sim20-30\%$ of AGN have mismatched optical and X-ray types even when classifications are made using deep X-ray data and optical spectra, an apparent discrepancy whose proposed explanations include spectral variability, host galaxy dilution, and disappearance of the broad-line region in AGN accreting at $L/L_{\rm Edd} \le 0.01$ \citep[e.g.,][]{tozzi06,page06,caccianiga07,trouille09,trump09,trump11a,trump11b}. To further complicate matters, observational scatter in the measured X-ray fluxes, when combined with the strict dividing line in column density between X-ray Type 1 and Type 2 sources, leads to discrepant X-ray types for 15-30\% of the sources in our sample detected by both \chandra\ and \xmm.

\subsection{Intrinsic X-ray Luminosity}

For the $70\%$ of \xmm\ sources in our sample with a hard-band detection, we estimate the intrinsic (e.g. absorption-corrected) rest-frame 2-10 keV luminosities using the column densities estimated above and the observed 2-10 keV fluxes.  While we can place only lower limits on the columns of the 13\% of sources that lack soft-band detections (with a median value of log~$N_{\rm H}$(cm$^{-2}$)$\ge22.7)$, their estimated intrinsic luminosities would increase by an average of only $\sim 0.4$ dex if all such sources were Compton-thick with log~$N_{\rm H}$(cm$^{-2}$)$=24$ (and by only $\sim 0.1$ dex if log~$N_{\rm H}$(cm$^{-2}$)$=23.5$).  In contrast, because of the comparably low sensitivity of the hard X-ray band, the majority (77\%) of sources detected only in the soft band have column density upper limits in excess of log~$N_{\rm H}$(cm$^{-2}$)$=22$, and 24\% have upper limits greater than log~$N_{\rm H}$(cm$^{-2}$)$=23$ (assuming an intrinsic X-ray spectrum with $\Gamma = 1.8$).  Such high columns can translate into order of magnitude corrections to the intrinsic luminosity estimated from the observed soft-band flux, so we do not attempt to estimate intrinsic luminosities for the 30\% of \xmm\ sources detected only in the soft band. 

All of the \xmm\ sources with a hard band detection have both optical and X-ray classifications, with optical and X-ray Type 2 fractions of 57\% and 58\%, respectively, and 68\% have spectroscopic redshifts.  For sources with a measurable column or those with only a hard-band detection, we estimate the absorption-corrected rest-frame luminosity by again assuming an intrinsic $\Gamma$ of 1.8. For sources with no measurable column (i.e., those that are softer than our assumed $\Gamma$ of 1.8), we use the Galactic absorption-corrected hard to soft flux ratio to calculate the observed photon index, $\Gamma$, which we then use to determine the rest-frame luminosity.

\subsection{Broad-band SEDs}

\begin{figure*}
$\begin{array}{cccc}
\includegraphics[angle=0,scale=0.45]{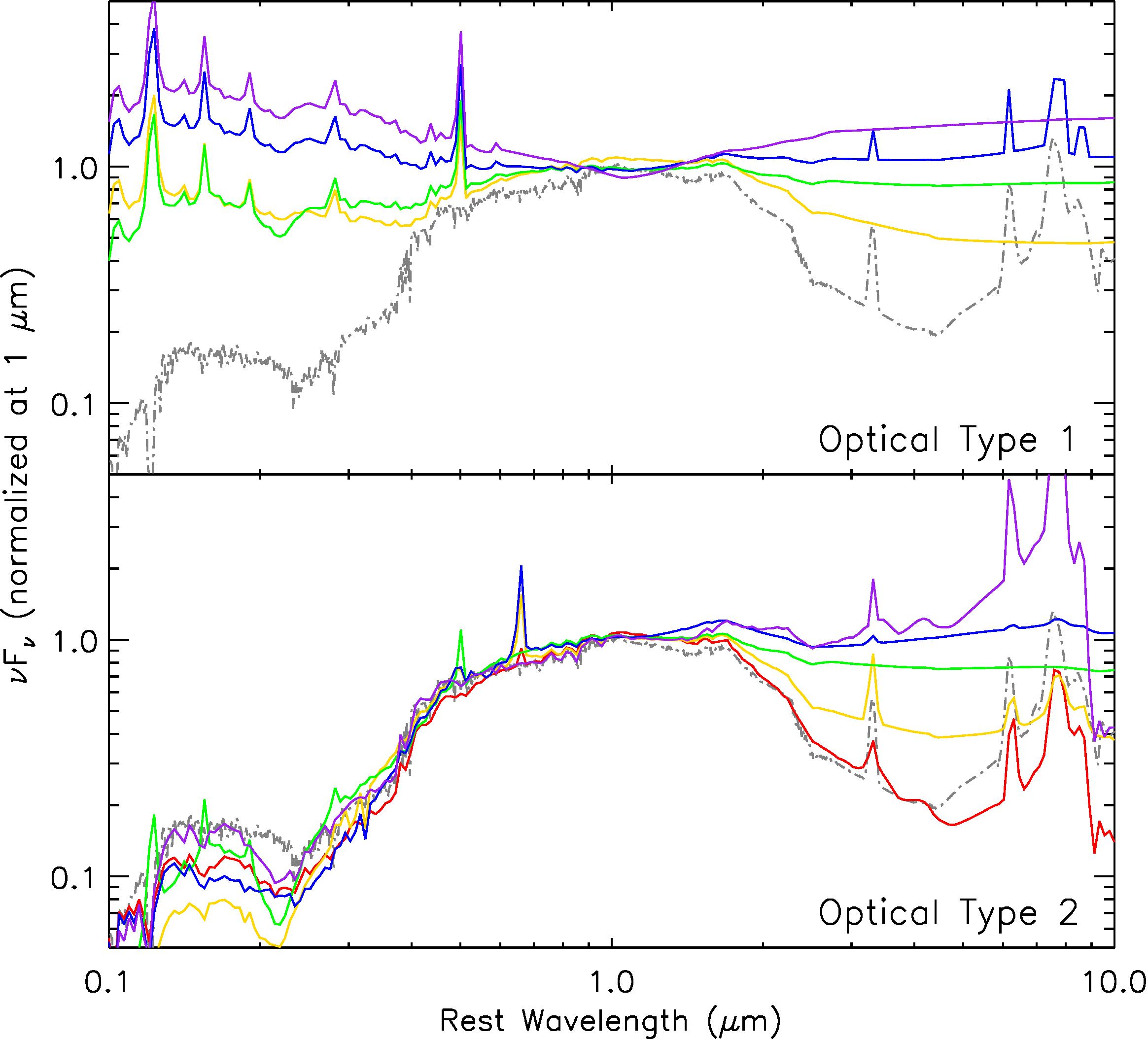} &
\includegraphics[angle=0,scale=0.45]{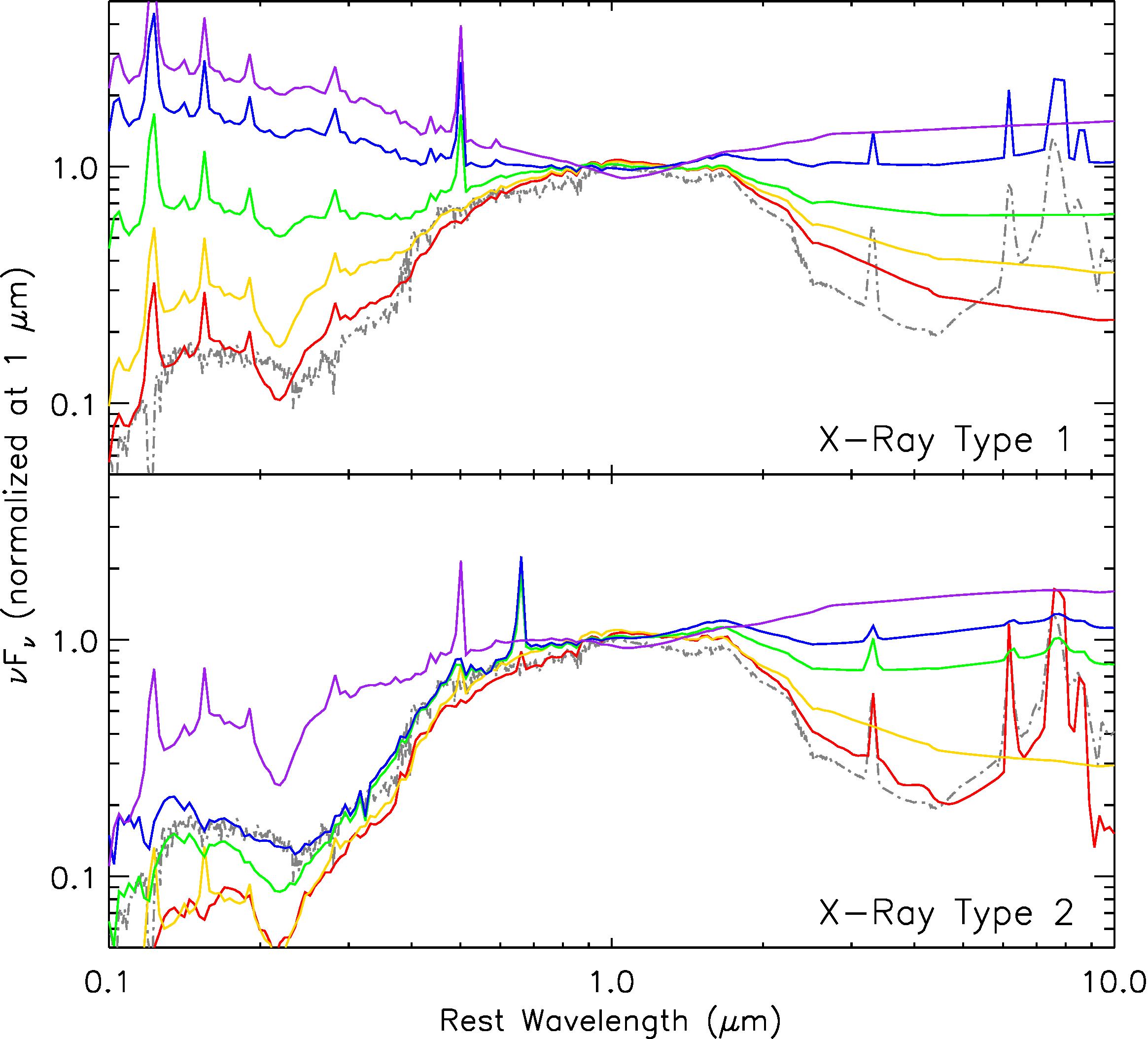} \\
\end{array}$
\caption{Template combinations (see \S5.3) that best fit the median SEDs of the optical Type 1 and Type 2 AGN (left) and X-ray Type 1 and Type 2 AGN (right) in the \xmm\ sample.  From light to dark (or red to purple in the online version), the SEDs represent the following luminosity bins: log $L_{\rm x}$ $<$ 43.0 (red), 43.0 $<$ log $L_{\rm x}$ $<$ 43.5 (yellow), 43.5 $<$ log $L_{\rm x}$ $<$ 44.0 (green), 44.0 $<$ log $L_{\rm x}$ $<$ 44.5 (blue), log $L_{\rm x}$ $>$ 44.5 (purple).  The grey dash-dotted SED is that of the starburst M82.  MIR spectral shape is relatively independent of the X-ray or optical type, and is primarily a function of AGN luminosity, with more luminous AGN displaying redder MIR SEDs. The differences in the SEDs of X-ray and optically classified AGN can be explained by the selection effects and biases discussed in \S5.3.1.}
\end{figure*}

Before discussing the IRAC colors of the \xmm\ sample, we examine the median SEDs of sources binned by X-ray or optical type and X-ray luminosity.  Doing so sheds light on the nature of AGN of different classes, and also illustrates the inherent differences and biases in the classification schemes which will affect our later interpretation. 

We plot in Figure 6 the best-fit templates to the median SEDs of the X-ray and optical Type 1 and Type 2 populations, as a function of intrinsic X-ray luminosity.  To create these templates, we shift each individual SED into the rest-frame and normalize it to 1 \micron\ (the approximate location of both the inflection point seen in luminous AGN and the midpoint of the stellar bump prominent in lower-luminosity AGN).  We then calculate the median SED in bins of log $\Delta \lambda ($\AA$) = 0.1$ for all bins containing at least 10 points.  This procedure gives a low-resolution median SED, albeit one with occasional gaps in wavelength coverage.  To provide a clearer representation of the median SEDs, we then fit each median SED with all possible combinations of one of six AGN templates (Sey1.8, Sey2, QSO1, BQSO1, TQSO1, QSO2) and one of 14 star-forming templates (comprising all elliptical and spiral templates as well as the starburst/ULIRG templates of N6090, M82, Arp220, and I22491) from \cite{polletta08}, choosing as the final fit the template combination with the lowest reduced chi-squared.  To account for obscuration, we allow for independent reddening of the AGN and star-forming templates using the \cite{draine03} extinction law.  

The SEDs shown in Figure 6 should be interpreted as rough guides to the overall behavior of the \xmm\ sources.  In general, they are far more highly sampled in the UV/optical than in the MIR, and they do not reflect the intrinsic variation seen in any given class of AGN (for more detailed SED analyses, see \cite{salvato09}, \cite{hao10}, \cite{lusso10}, \cite{trump11b}, Lusso et al. (2011, in prep.), and Elvis et al. (2011, in prep.)). That said, they clearly illustrate the redder MIR colors of more luminous AGN (see also Figures 1, 2, and 3) and confirm that this trend is relatively insensitive both to obscuration and to the classification scheme (X-ray vs. optical).

\subsubsection{Differences between X-ray and Optical Classification Schemes}

While the MIR SEDs plotted in Figure 6 are relatively insensitive to the obscuration scheme, X-ray and optical AGN classification returns samples with different UV-optical SEDs.  This apparent discrepancy can be attributed in part to the physically distinct samples identified by the two criteria, and in part to selection biases. For a source to be classified as ``optical Type 1'', it must display either broad optical emission lines or have an SED best fit by an unobscured quasar template relatively flat in $\nu F_{\rm \nu}$.  All remaining sources, including those with narrow optical emission lines, those with no emission lines yet X-ray luminosities exceeding $2 \times 10^{42}$~ergs~s$^{-1}$  \citep[e.g., optically dull AGN,][]{moran02,comastri02}, and those lacking optical spectra yet best-fit by a galaxy-dominated template are classified as ``optical Type 2''.  In the absence of high-quality spectra, low-luminosity Type 1 AGN (e.g., Seyfert 1's) with host-dominated SEDs are therefore likely to be misclassified as optical Type 2 AGN \citep[see also discussion in][]{brusa10}. 

However, the vast majority (88\%) of low luminosity (log~$L_{\rm 2-10 keV}$(ergs~s$^{-1}$)$ < 43.5$) X-ray Type 1 AGN in our \xmm\ sample have robust optical spectra, yet 77\% lack broad optical emission lines.  While we cannot rule out the possibility that the host galaxies of these low-luminosity AGN have diluted their broad emission lines \cite[e.g.,][]{page06,caccianiga07,civano07}, \cite{trump11b} show that the X-ray unobscured narrow-line AGN in \xmm-COSMOS have low accretion rates of $L/L_{\rm Edd} < 10^{-2}$, and may therefore be fueled by radiatively inefficient accretion flows incapable of supporting a broad-line region.  The X-ray Type 1, yet optical Type 2, classification may therefore correctly represent the intrinsic nature of these low-luminosity AGN, whose seemingly discrepant properties give rise to many of the observed differences in Figure 6.

X-ray classification is less sensitive to accretion rate and host galaxy properties, but is subject to a redshift-dependent bias.  At low redshift, the soft X-ray band (0.5-2 keV) is sensitive to relatively low obscuring columns.  At high redshift, however, low to moderate column densities become poorly constrained as the observed X-ray bands sample the higher energy rest-frame emission less subject to intrinsic obscuration. Small observational uncertainties in X-ray flux therefore translate to large uncertainties in column density, and because low columns are far more poorly constrained than high columns, Type 1 AGN are far more likely to be misclassified as Type 2 AGN than vice versa.  We illustrate and quantify this well-known effect in more detail in the Appendix. This tendency to misclassify high-redshift (e.g., high-luminosity) X-ray Type 1 AGN as X-ray Type 2 AGN likely accounts for the blue UV-optical continuum of the most luminous X-ray Type 2 SED in Figure 6.

\section{\xmm\ Sample: Trends in IRAC Color Space}

\begin{figure*}
$\begin{array}{cc}
\includegraphics[angle=0,height=3.6in]{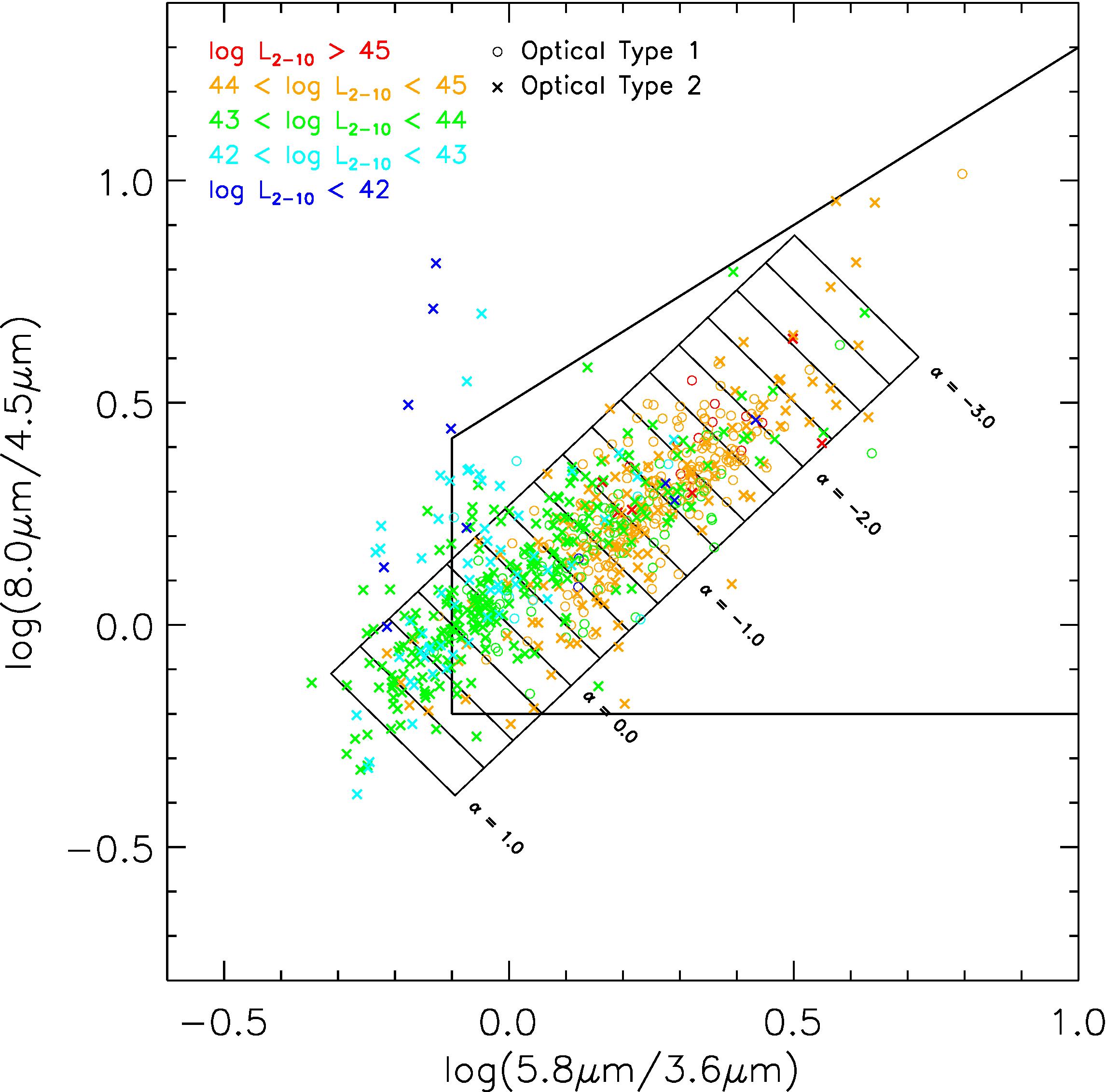} &
\includegraphics[angle=0,height=3.6in]{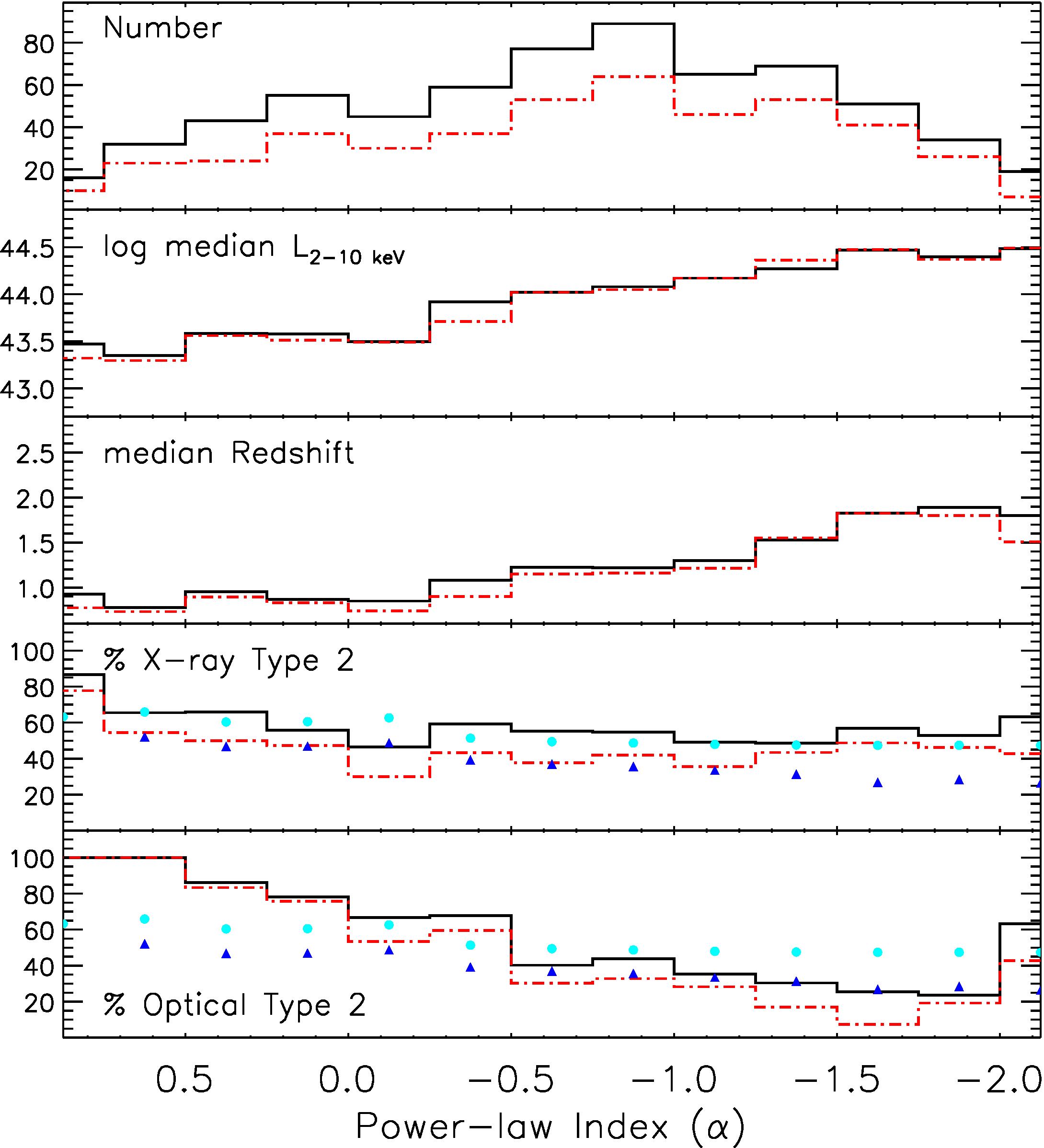} \\
\end{array}$
\caption{left panel: IRAC color-color diagram for hard-band--detected \xmm\ sources, as a function of X-ray luminosity and optical type. The power-law box has been extended from $\alpha = 1.0$ to -3.0. right panel: Properties of the \xmm\ sources in each of the bins in $\alpha$ containing at least 10 sources ($\alpha \ge -2.25$).  The trends for sources with spectroscopic redshifts are shown by dash-dotted histograms (red in the online version). The filled circles and triangles give the expected Type 2 AGN fractions from \cite{gilli07} and \cite{hasinger08}, respectively, based on the median X-ray luminosity in each bin.  AGN with redder IRAC colors (more negative power-law indices) tend to have higher X-ray luminosities, lie at higher redshift, and be somewhat less obscured.}
\end{figure*}

To test for trends in IRAC color space as a function of luminosity, redshift, and AGN type, we plot in Figure 7a the IRAC colors of all \xmm\ sources detected in hard X-rays.  We then extend the power-law box defined in \S4 from $\alpha = -3.0$ to $\alpha = 1.0$, enclosing 92\% of the \xmm\ sources, and plot in Figure 7b the median luminosity, redshift, and X-ray and optical types of the \xmm\ sources in bins of power-law slope. 

\subsection{Trends in X-ray Luminosity and Redshift}
A clear trend is seen in X-ray luminosity, with the more luminous X-ray sources displaying redder IRAC colors, as expected \citep[see also][]{cardamone08,brusa09,eckart10,park10,trump11b}. While there is no cut in power-law slope that cleanly distinguishes QSOs (log~$L_{\rm 2-10 keV}$(ergs~s$^{-1}$)$ > 44$) from less luminous Seyfert galaxies (log~$L_{\rm 2-10 keV}$(ergs~s$^{-1}) < 44$), a cut of $\alpha = -0.78$ best separates the two populations, with equal fractions (72\%) of high-luminosity AGN falling redward of the cut and low-luminosity AGN falling blueward of the cut.

The median redshift also increases for sources with redder IRAC colors, from $z\sim1$ at $\alpha = 1.0$ to $z\sim2$ at $\alpha = -2.25$.   This correlation is driven primarily by the observed trend in luminosity, as only the most luminous sources (which tend to have redder IRAC colors) can be detected out to high redshift.  Between $z=1$ and $z=2$, the SED templates shown in Figure 2 move roughly perpendicularly to the power-law locus, at least for AGN fractions $\le 80\%$.  Only for templates with AGN fractions of  $95\%$ and moderate $A_{\rm V, AGN}\sim 2-8$ do we see a trend towards redder IRAC colors between $z=1$ and $z=2$, although this shift of $\Delta \alpha \sim 0.25-0.4$ is too weak to account for the observed correlation between redshift and IRAC color. 

\subsection{Trends in AGN Type}
While the X-ray Type 2 fraction is essentially independent of $\alpha$, the optical Type 2 fraction drops smoothly from $\alpha = 1$ to $\alpha \sim -1.6$ before rising again at the reddest IRAC colors.  These seemingly inconsistent trends remain, and are even enhanced, if we consider only those sources with spectroscopic redshifts (shown in red).  

Because high luminosity AGN tend to be less obscured than their low luminosity counterparts, at least in X-ray selected samples, we might expect sources with redder IRAC colors to have a lower Type 2 fraction \citep[e.g.][]{ueda03,gilli07,hasinger08}. We plot in the right panel of Figure 7 the expected Type 2 fractions of \cite{gilli07} and \cite{hasinger08} at the median X-ray luminosity of each bin.  While a slight decrease in the Type 2 fraction would be expected over the observed range in luminosity, the trend we observe for the optically-classified AGN is much stronger than predicted.

These observed trends, however, reflect the biases and selection effects discussed in \S5.3.  At blue IRAC colors, the excess of low-luminosity, optical Type 2 AGN can be attributed to the population of unobscured narrow-line AGN in \xmm-COSMOS \citep{trump11a}.  Not only are these weak accretors likely to lack a broad-line region, but their anomalously blue IRAC colors also suggest that they, like many low-luminosity, low accretion rate radio galaxies, may lack a hot dust torus \citep{trump11a,ogle06}.  Likewise, the excess of X-ray Type 2 AGN at red IRAC colors can be attributed to the overestimation of X-ray column densities at high redshift (see \S5.3 and the Appendix). 

Given the nature of these biases, an AGN's intrinsic obscuration is best represented by the X-ray classification at low luminosity/redshift, and by the optical classification at high luminosity/redshift.  If we therefore assume the X-ray derived Type 2 fraction at blue slopes of $\alpha \gsim -0.5$, and the optically-derived Type 2 fraction at red slopes of $\alpha \lsim -0.5$, we find a distribution that favors unobscured AGN at red IRAC colors but whose evolution is not nearly as pronounced as the optical classification alone would suggest.  This trend is likely driven primarily by the anti-correlation between luminosity and obscuration, and while this new hybrid estimate of the obscured AGN fraction may display a somewhat steeper evolution than those of \cite{gilli07} and \cite{hasinger08}, a direct comparison is complicated by the scatter in X-ray luminosity in a given power-law bin, our necessary exclusion of sources with ambiguous X-ray types, as well as the expected incompleteness of the \xmm\ sample to the obscured AGN that are the target of MIR selection.

\begin{figure*}
$\begin{array}{llllll}
\includegraphics[angle=0,height=2.2in]{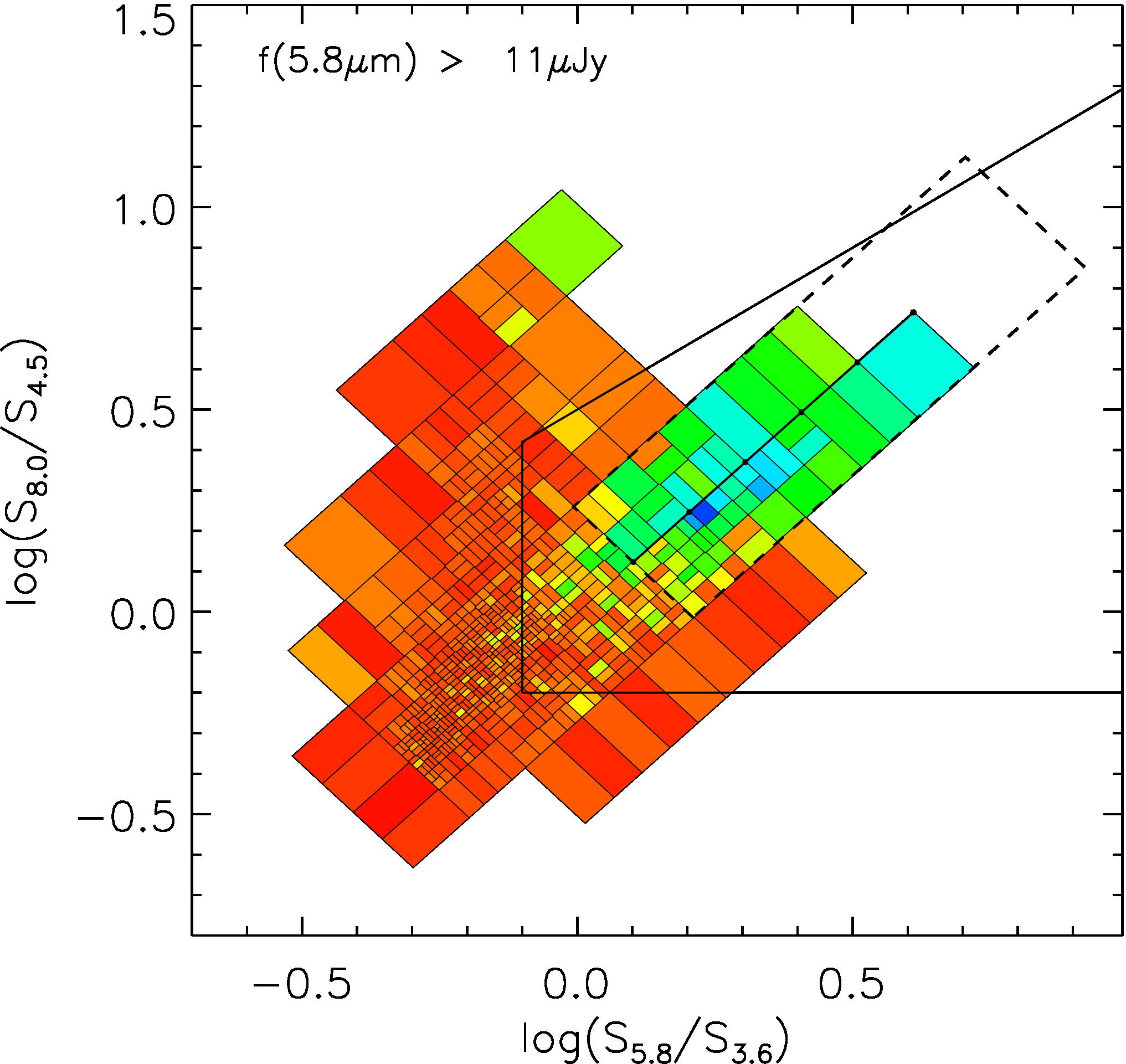}&
\includegraphics[angle=0,height=2.2in]{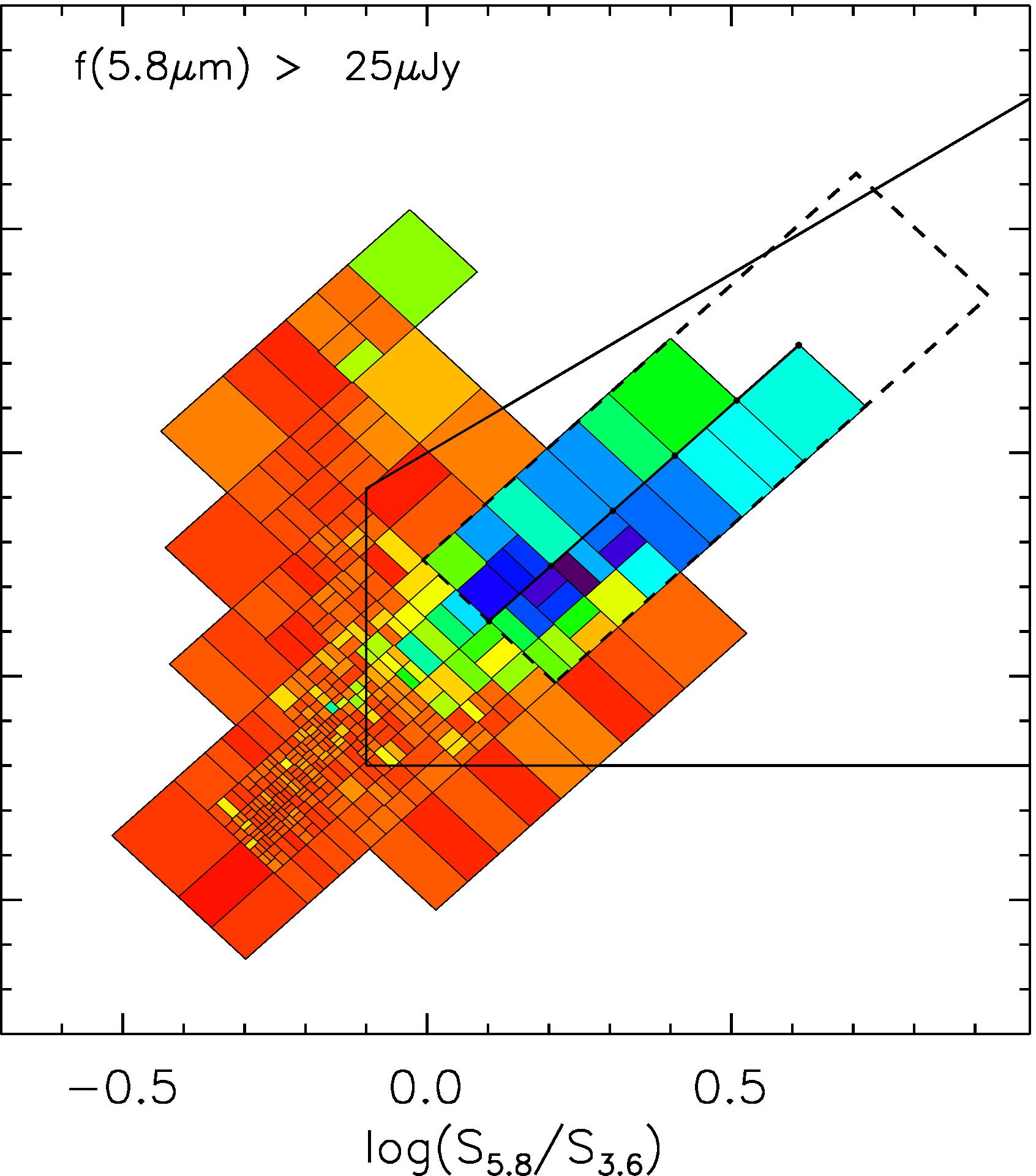} &
\includegraphics[angle=0,height=2.2in]{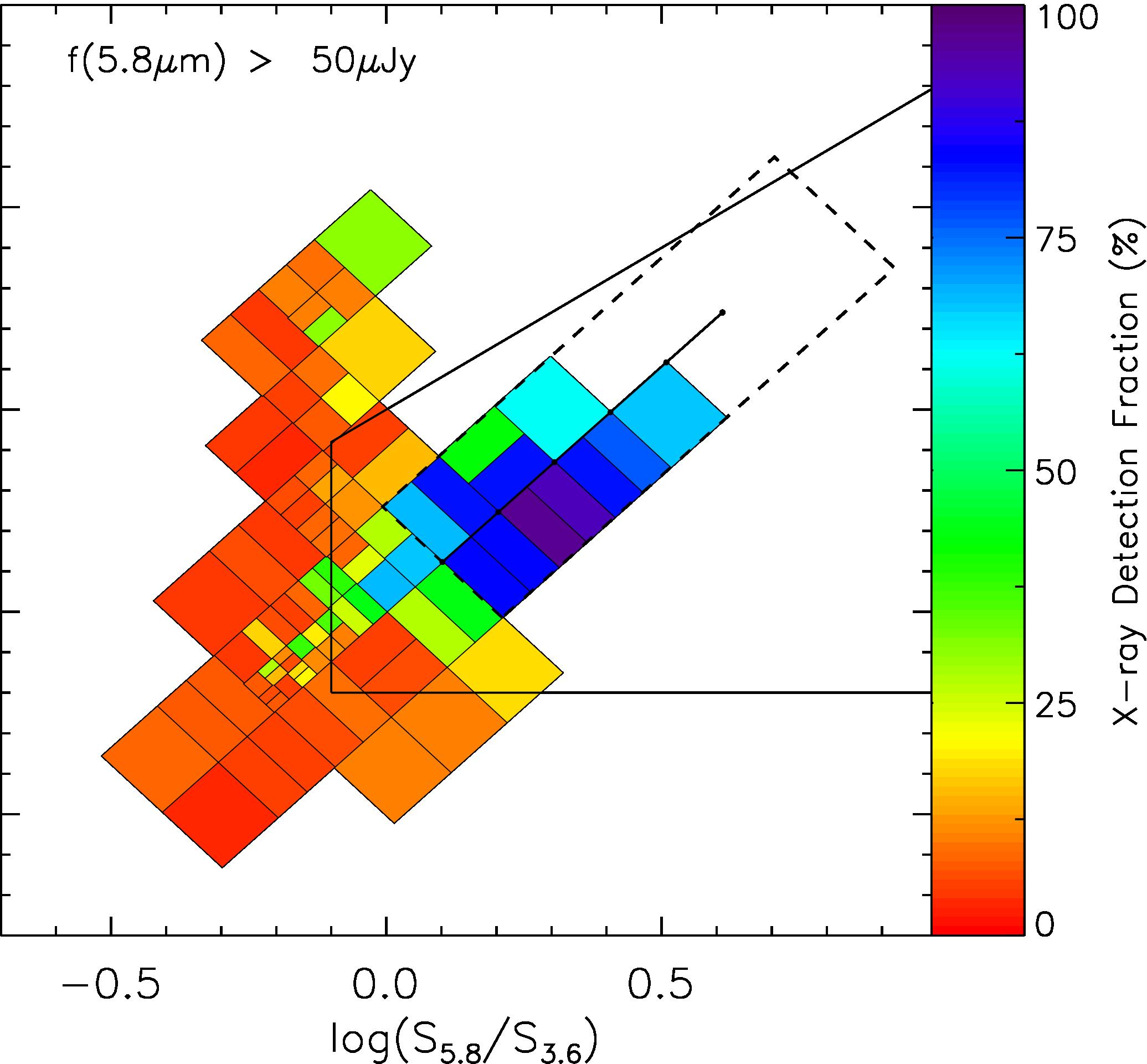}\\
\includegraphics[angle=0,height=2.2in]{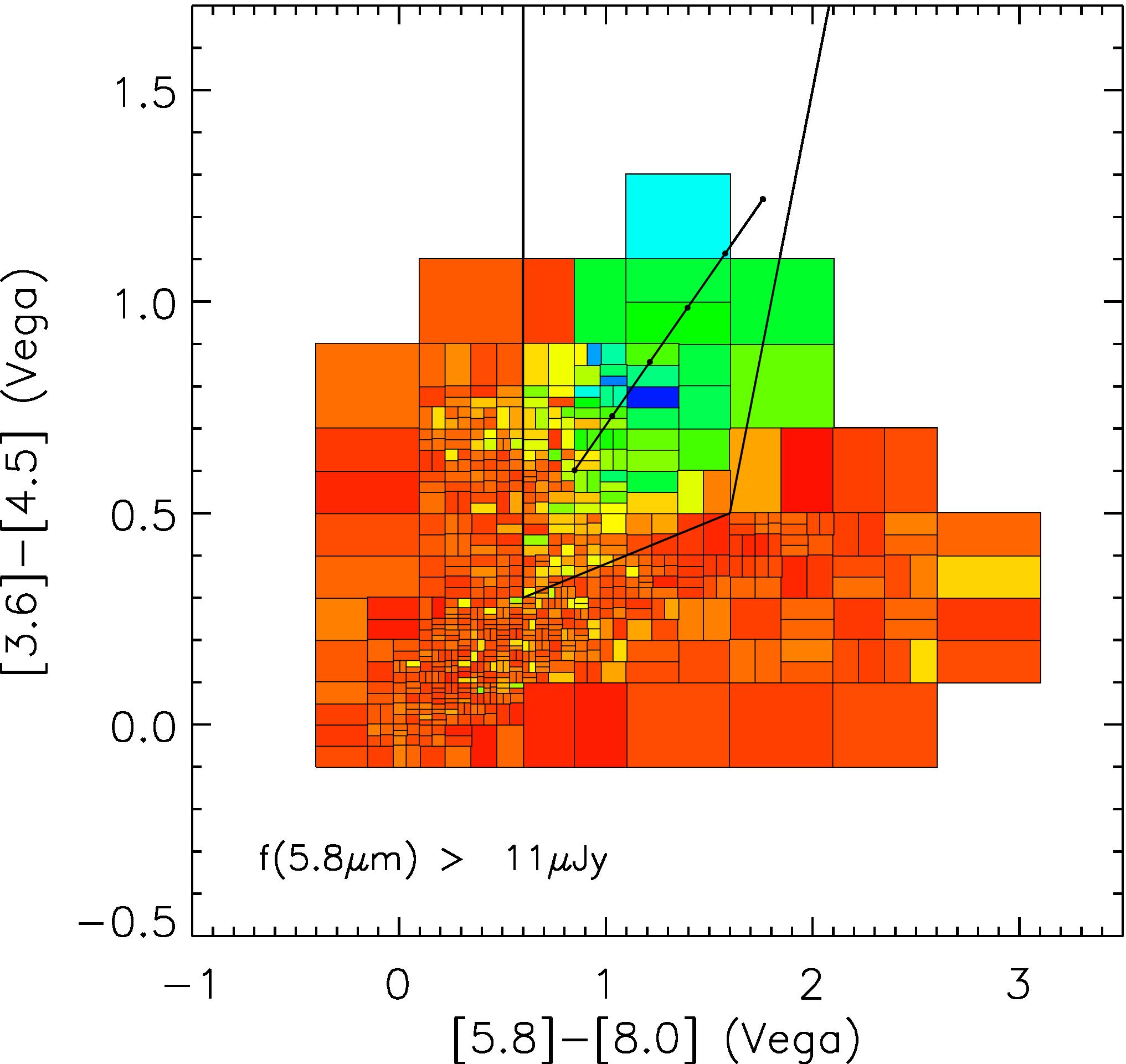} &
\includegraphics[angle=0,height=2.2in]{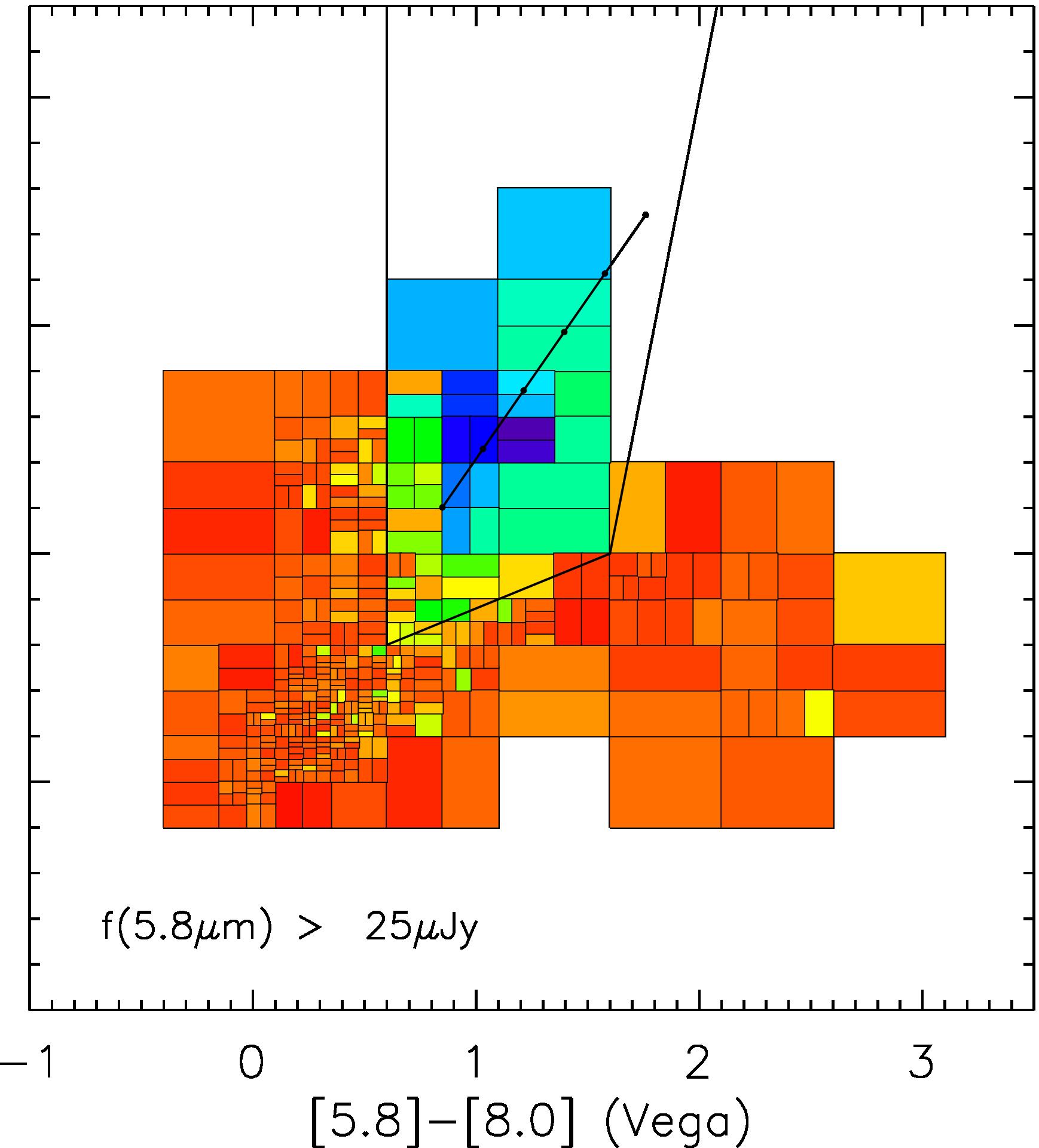} &
\includegraphics[angle=0,height=2.2in]{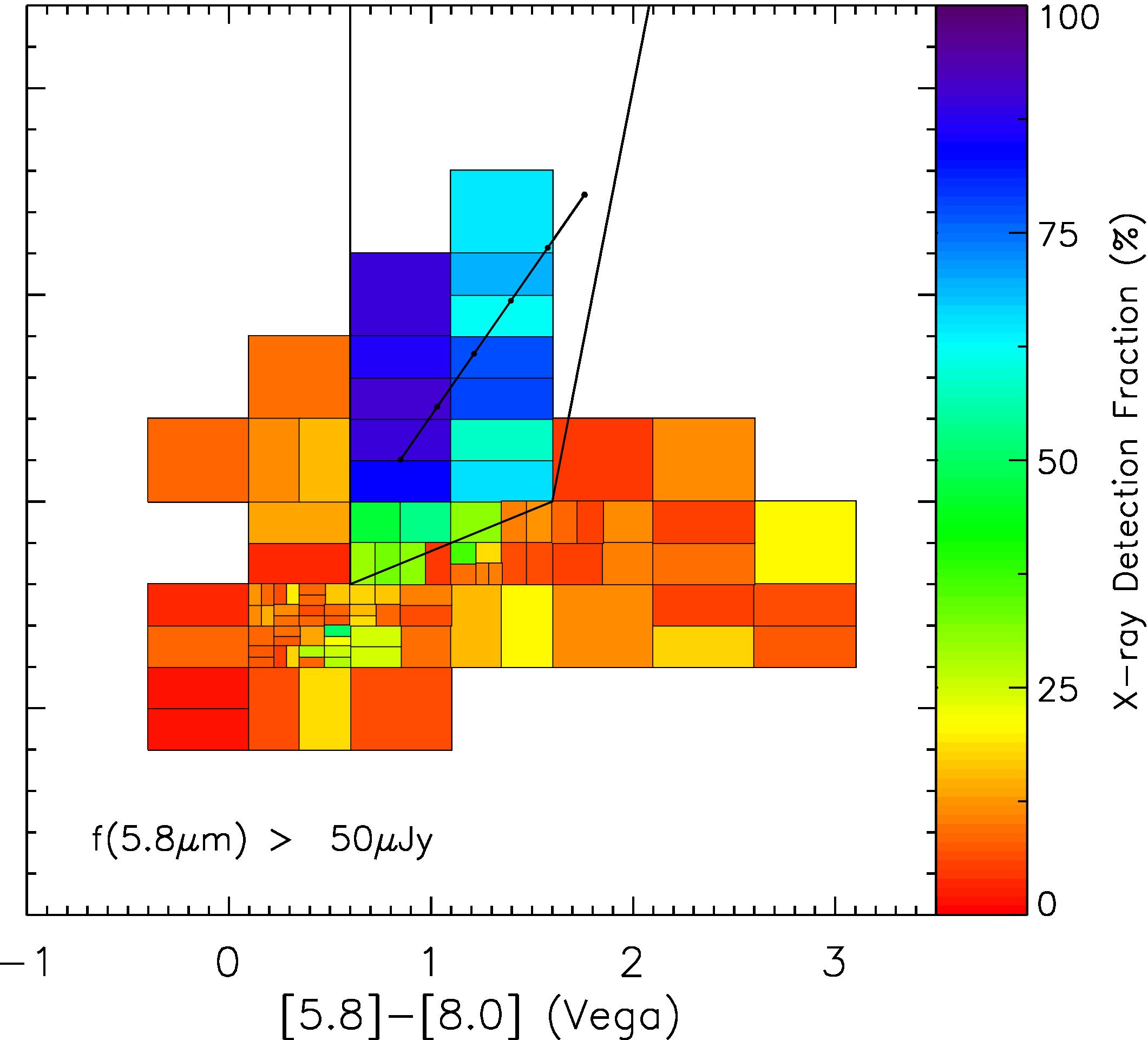} \\
\end{array}$
\caption{X-ray detection fraction for sources with at least 50 ks of \chandra\ coverage, for bins containing at least 10 sources. The top and bottom panels show \cite{lacy04,lacy07} and \cite{stern05} color space, respectively.  From left to right, the three rows show the X-ray detection fraction for all sources in our IRAC sample ($S_{\rm 5.8 \micron} > 11.3 \micron$), and for sources with $S_{\rm 5.8 \micron} > 25 \micron$  and $S_{\rm 5.8 \micron} > 50 \micron$. The dashed line in the top panels shows the $-3 \le \alpha \le -0.5$ power-law box defined in \S4.  At the full depth of COSMOS, the X-ray detection fraction is relatively constant at $\sim 50\%$ within the power-law box, but it drops to $\lsim 10\%$ across much of the remainder of IRAC color-space, including regions enclosed by the current AGN selection wedges.}
\end{figure*}

An optically thick torus could also contribute to the observed trend by masking the hot dust signature from obscured AGN. As discussed in \S2, however, the large NIR-MIR anisotropies predicted by models of smooth, geometrically and optically thick tori are not supported by recent observations, particularly when the intrinsic X-ray luminosity is used to normalize the MIR luminosity \citep{lutz04,gandhi09}.  However, studies that normalize the AGN's luminosity by the radio continuum find a factor of a few difference in MIR luminosity between unobscured and obscured AGN \citep{heckman95,haas08,leipski10}.  We return to this issue in \S9. 

\section{IRAC sample}

While the \xmm\ sample sheds light on the potential completeness of IRAC color selection to AGN of different luminosities and/or type, \xmm\ sources comprise only 4\% of the full sample of COSMOS IRAC sources.  In the sections that follow, we test our ability to cleanly separate the AGN-dominated IRAC sources from the remainder of the primarily star-formation-dominated IRAC population. 

\vspace*{0.5cm}
\subsection{X-ray Detection Fraction}

We plot in Figure 8 the \chandra\ X-ray detection fraction \citep[down to a limiting flux of $5.7 \times 10^{\-16}$~ergs~s$^{-1}$,][]{elvis09} for the 11324 IRAC sources with at least 50~ks of \chandra\ coverage, for which the median (mean) X-ray exposure is 148~ks (128~ks). To maximize the resolution in IRAC color space, we divide the IRAC sample into grids containing at least 10 sources.  The typical grid size is therefore inversely proportional to the density of IRAC sources in a given region of color space.  While the \chandra\ dataset only covers the central 0.9 deg$^2$ of the COSMOS field, its increased depth and higher resolution than the larger-area \xmm\ data allow us to probe more deeply the X-ray properties of the IRAC sample. 

At the full depth of the COSMOS IRAC survey, 42\% of the sources in the $\alpha \le -0.5$ power-law box are detected by \chandra, compared to only 5\% of the sources outside the power-law box (see Figure 8).  Within the power-law box, the X-ray detection fraction remains relatively constant down to a slope of $\alpha \sim -0.75$.   It drops somewhat, however, at the bluest extremes of the power-law box where we might expect contamination from high-redshift galaxies (see Figures 2, 3, and 4).  An additional cut in 8.0 to 4.5 \micron\ color may therefore be warranted, and will be explored in more detail in \S8.  

It is also clear from Figure 8 that at the moderate depth of the COSMOS IRAC data, significant fractions of the current AGN selection regions of \cite{lacy04,lacy07} and \cite{stern05} are characterized by low X-ray detection fractions indistinguishable from the typical values outside of the selection wedges.  These regions of color-space coincide with the regions where we expect contamination from low and high-redshift star-forming galaxies, and further motivate the need for new AGN selection criteria. 

To illustrate the effects of IRAC depth, we also plot in Figure 8 the X-ray detection fractions for limiting flux densities of 25 and 50 \microjy\ in the 5.8 \micron\ band (which effectively limits the detection of red, power-law like sources). For comparison, the IRAC data used by \cite{lacy04} and \cite{stern05} to define the current AGN selection regions have limiting 5.8 \micron\ depths of 60 \microjy\ and 76 \microjy, respectively.  As the IRAC data become progressively shallower, the X-ray detection fractions rise (to 64\% and 77\% in the power-law box at limiting flux densities of 25 and 50 \microjy, respectively) and the number of grids in the \cite{lacy04,lacy07} and \cite{stern05} AGN selection regions with low, star-formation-like X-ray detection fractions falls.  This trend reflects the increase in the AGN fraction with increasing MIR flux density \cite[e.g.,][]{brand06,treister06,donley08}.  While the likelihood of contamination by star-forming galaxies therefore decreases at higher IRAC flux densities, so too does the number of X-ray non-detected AGN candidates in the power-law box, from 570 at the full IRAC depth to 182 and 67 at limiting 5.8\micron\ flux densities of 25 and 50 \microjy, respectively.

\begin{centering}
\begin{figure*}
$\begin{array}{cccc}
\hspace*{0.8cm}
\includegraphics[angle=0,scale=0.4]{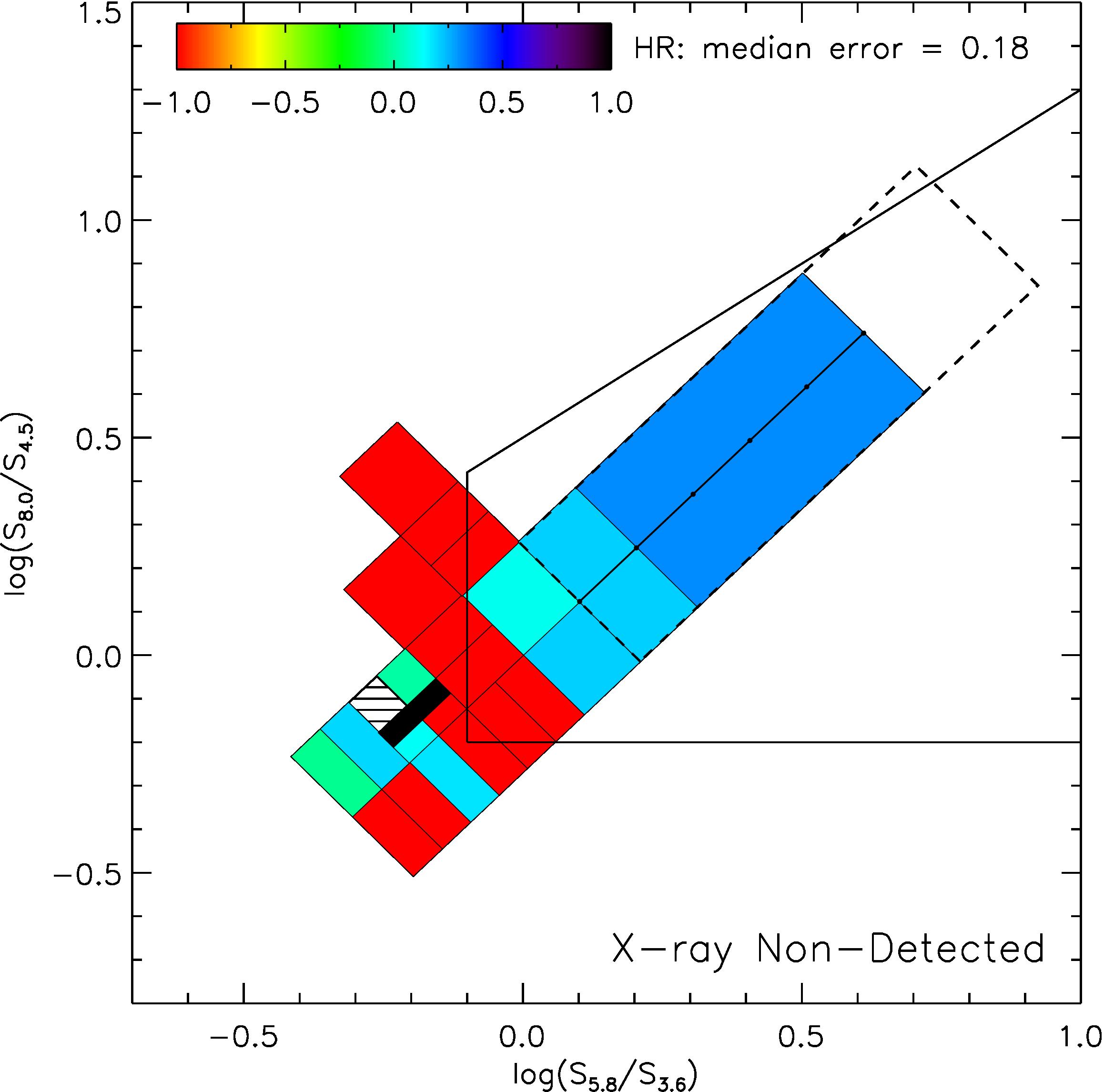}&
\includegraphics[angle=0,scale=0.4]{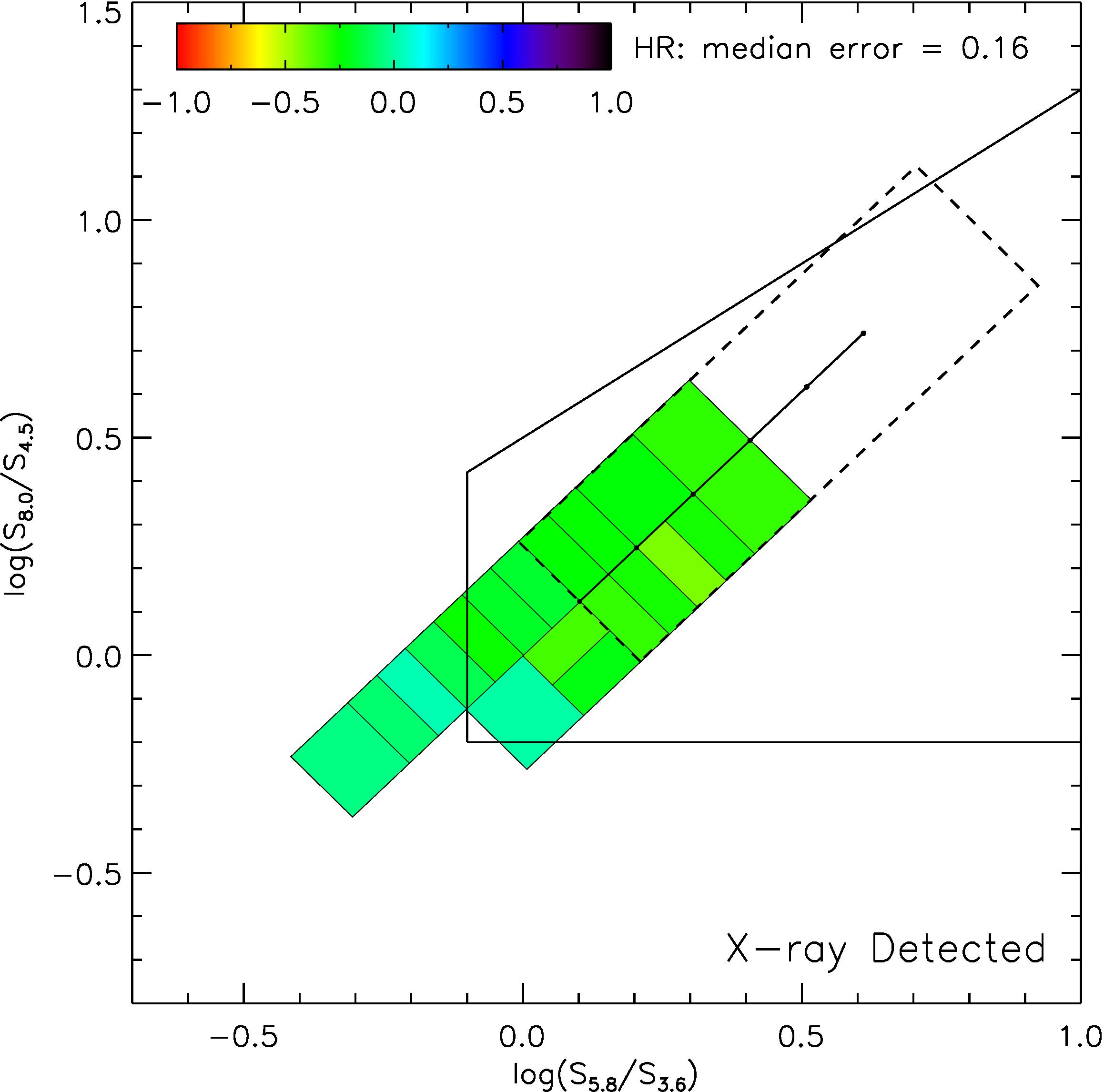}\\
\hspace*{0.8cm}
\includegraphics[angle=0,scale=0.4]{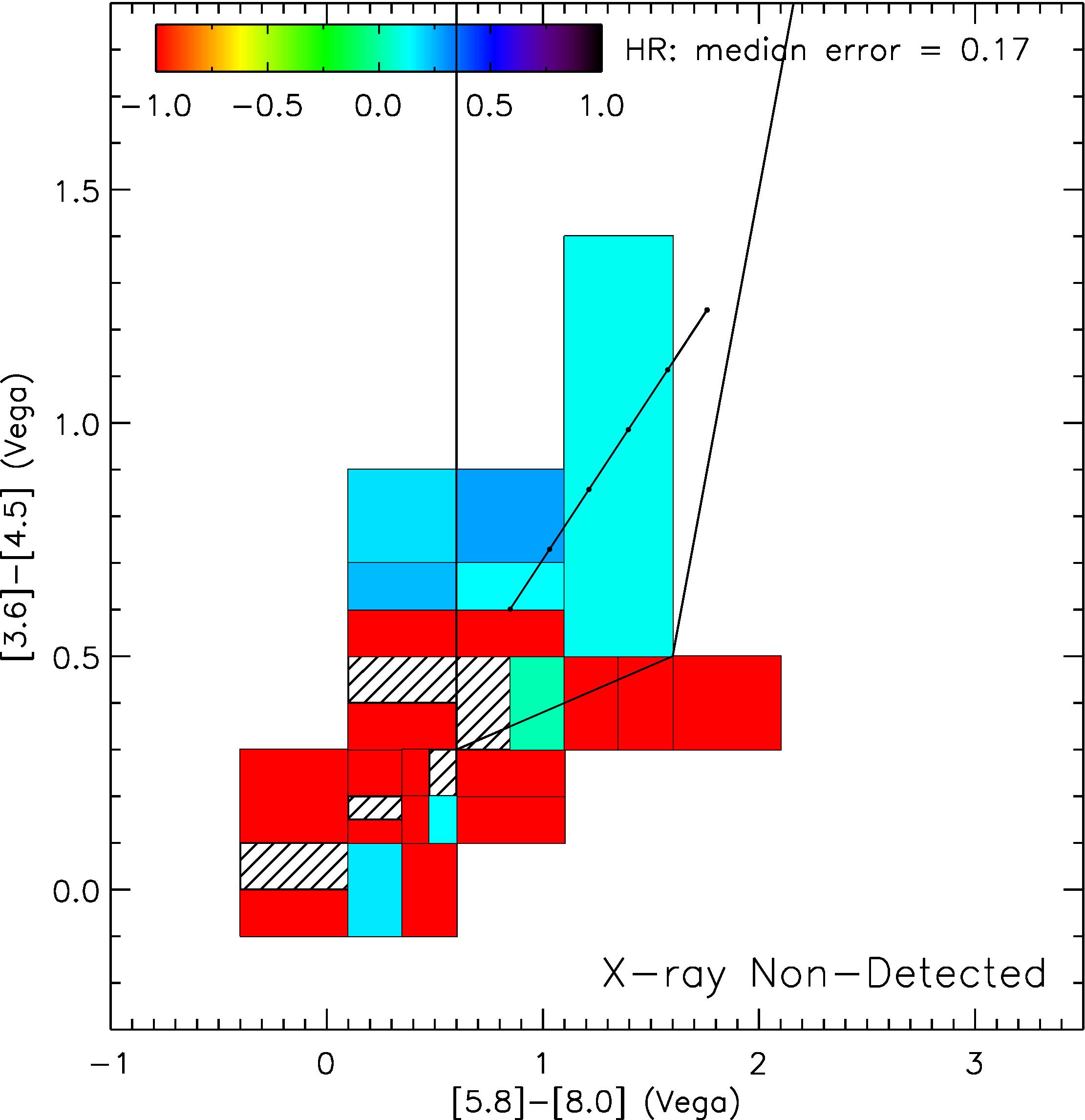}&
\includegraphics[angle=0,scale=0.4]{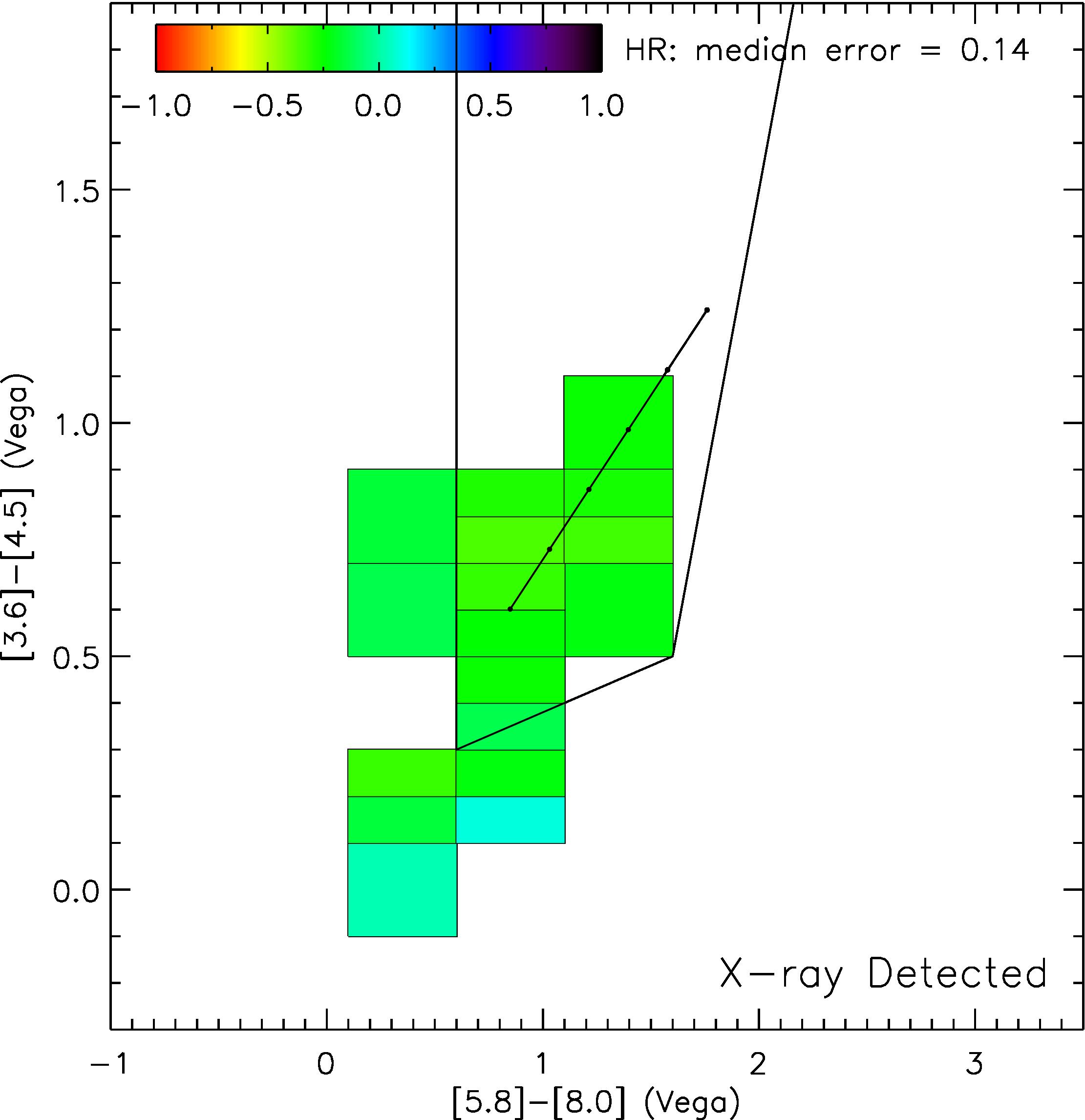}\\
\end{array}$
\caption{Hardness ratios (HR = $(H-S)/(H+S)$) of the stacked emission from X-ray non-detected (left) and X-ray detected (right) IRAC sources with at least 50 ks of \chandra\ coverage, where $H$ and $S$ represent the 2-8 keV and 0.5-2 keV ACIS count rates, respectively. The top and bottom panels show \cite{lacy04,lacy07} and \cite{stern05} color space, respectively.  The minimum number of sources per grid is 200 for the X-ray non-detected sources, and 25 for the X-ray detected sources. White (red in the online version) and black boxes indicate those grids with a detection in only the soft or hard bands, respectively.  Grids detected in neither band are shaded with hashed lines. The dashed line in the top panels shows the power-law box defined in \S4.  The hardest X-ray signal, consistent with emission from obscured to mildly Compton-thick AGN, comes from sources in the power-law box.}
\end{figure*}
\end{centering}

\vspace*{0.5cm}
\subsection{X-ray Stacking}

While the vast majority of IRAC sources lack X-ray counterparts, X-ray stacking can shed light on the nature of the X-ray non-detected population. Using v3.0beta of the program CSTACK\footnote{http://cstack.ucsd.edu/cstack/}, we stack the X-ray emission for grids in IRAC color space containing at least 200 X-ray undetected sources.  We exclude all sources that lie more than $8$\arcmin\ from the \chandra\ aim-point, and extract source counts within the smaller of either the position-dependent 90\% encircled-energy radius or the 5\arcsec\ inner background subtraction radius.  We then measure the background across the remaining 15\arcsec x15\arcsec\ image, after masking all known X-ray sources. 

We plot in Figure 9 the X-ray hardness ratios, HR, for grids detected to $\ge 3 \sigma$ in the soft and/or hard X-ray bands, and plot for comparison the stacked hardness ratios of the X-ray detected sources (HR = $(H-S)/(H+S)$, where $H$ and $S$ represent the 2-8 keV and 0.5-2 keV ACIS count rates, respectively). While low source densities limit the resolution along the power-law locus, we detect hard X-ray emission from all red ($\alpha \le 0$) power-law grids in \cite{lacy04,lacy07} color space. In contrast, the majority of grids outside of the power-law box are detected only in soft X-rays, and those with measurable hard X-ray emission lie in regions where we expect to find lower-luminosity AGN whose hosts dominate their MIR SEDs.

The hardest X-ray signal is observed for sources in the $\alpha < -1.0$ power-law box: HR$=0.31 \pm 0.13$, where the errors have been derived from the CSTACK bootstrap analysis.  For comparison, the stacked emission from X-ray detected sources in this region of color space is significantly softer: HR$=-0.31 \pm 0.13$.  At the typical photometric redshifts of the X-ray and non-X-ray sources, $z=1.8 \pm 0.7$ and $z=2.1 \pm 1.0$, respectively, these hardness ratios correspond to column densities of log~$N_{\rm H}$(cm$^{-2}$)$=22.4 \pm 0.4$ and log~$N_{\rm H}$(cm$^{-2}$)$=23.5 \pm 0.4$.

While 61\% of the X-ray sources have spectroscopic redshifts and the remainder have AGN-specific photometric redshifts, only 5\% of the X-ray non-detected sources in this region of color space have spectroscopic redshifts and their photometric redshifts have been calculated using only normal galaxy templates \citep{ilbert09}. To place an independent constraint on the typical redshift of the X-ray non-detected sources in the $\alpha < -1.0$ power-law box, we turn to their IRAC fluxes, which are on average  $3.1 \pm 0.3$ times fainter than those of the X-ray sample.   If we assume that this fairly uniform dimming across the four IRAC bands can be attributed solely to redshift effects (i.e., that the X-ray and non-X-ray power-law samples have similar intrinsic MIR luminosities), we estimate a typical redshift of $z=2.7$ and a typical column density of log~$N_{\rm H}$(cm$^{-2}$)$=23.7$ for the X-ray non-detected sources, slightly higher than the values quoted above. 

\subsection{X-ray/MIR Relation}
The observed X-ray to MIR luminosity ratio, a proxy for AGN obscuration, allows us to place an independent constraint on the column densities of the $\alpha \le -1.0$ power-law sources \citep{lutz04,maiolino07,gandhi09,park10}.  For the X-ray detected and non-detected AGN, we approximate the median 6.7 \micron\ rest-frame luminosity from the median power-law slope and observed 8.0 \micron\ IRAC flux density, and we calculate the rest-frame 2-10 keV luminosity from the stacked 0.5-2 keV (1.5-6 keV at $z\sim 2$) X-ray flux, assuming $\Gamma = 1.4$.  For the subsample of sources with MIPS 24 \micron\ (rest-frame 6.5 \micron) counterparts, this extrapolated 6.7 \micron\ rest-frame flux density is in excellent agreement with the median observed 24 \micron\ flux density.

Using the X-ray/MIR relation of \cite{maiolino07}, we measure a MIR-derived intrinsic X-ray luminosity of log~$L_{\rm{2-10keV}}$~(ergs~s$^{-1}$)$ \sim 43.8$ for the individually X-ray non-detected $\alpha \le -1.0$ power-law galaxies, $\sim 20$ times higher than observed in the stacked signal.  To achieve this degree of hard X-ray suppression, a marginally Compton-thick column density of log~$N_{\rm H}$(cm$^{-2}$)$ \sim 24$ is required, generally consistent with the HR-derived values above. In comparison, we see no significant offset between the observed and MIR-derived hard X-ray luminosities of the X-ray detected sources in this region of color space (log~$L_{\rm{2-10keV}}$(ergs~s$^{-1}$)$ \sim 44.1$), as expected for sources with low to moderate (log~$N_{\rm H}$(cm$^{-2}$)$ \le 23$) obscuring columns.

\section{High Redshift Galaxies}

The IRAC colors of purely star-forming galaxies generally avoid the power-law region of color space, at least out to $z \lsim 3$ (see Figure 2).  At $z>3$, however, only one IRAC channel samples the red side of the 1.6 \micron\ stellar bump, and by $z\sim5$, the IRAC bands fall entirely on the blue side of this feature.  In high-redshift galaxies with older stellar populations and/or measurable dust extinction, this stellar bump has the potential to mimic the power-law emission from AGN.  To determine whether high-redshift star-forming galaxies are an important source of contamination, we investigate below the IRAC colors of high-$z$ sources in COSMOS and in the deeper GOODS fields.  

While we could attempt to identify high-redshift candidates purely on the basis of their photometric redshifts, we are primarily interested in sources whose stellar bumps fall redward of the IRAC bands and whose fluxes are faint, precisely those sources for which photometric redshift fitting is most difficult.   To constrain the likelihood of contamination by high-$z$ star-forming galaxies, we therefore turn to spectroscopically-confirmed and color-selected high-$z$ samples: BzK-selected galaxies \citep{daddi04}, distant red galaxies \citep[DRGs,][]{franx03,vandokkum03}, Lyman-break galaxies \citep[LBGs, e.g.,][]{steidel03}, $z>3$ evolved and/or reddened galaxy candidates, and sub-mm galaxies (SMGs).  In the IRAC-selected sample of \cite{perezgonzalez08} drawn from the HDF-N, CDF-S, and Lockman Hole fields, 90\% of the IRAC sources at $z=3-3.5$ meet at least one of the BzK, DRG, or LBG criteria, as do $\sim 80\%$ of the sources at $z=3.5-4$.  

\subsection{BzK Galaxies}

The BzK selection technique of \cite{daddi04} was designed to identify star-forming (sBzK) and passive (pBzK) galaxies at $1.4 \lsim z \lsim 2.5$, although it also identifies luminous AGN generally removed from BzK-selected samples on the basis of their hard X-ray or MIR power-law emission \citep[see, e.g.,][]{sharp02,daddi07sf}. To constrain the IRAC colors of BzK-selected galaxies in COSMOS, we adopt the COSMOS BzK catalog of \cite{mccracken10}.

We plot in Figures 10a and 11a the IRAC colors and redshift distributions of the 14\% of sBzK galaxies that meet our IRAC detection criteria.  Based on the IRAC color distribution and the relative density of X-ray sources, we divide the IRAC-detected sBzK galaxies into two populations.  The dominant population (79\%) occupies the same region of IRAC color space used by \cite{huang09} to identify $1.5 < z < 3$ star-forming galaxies: log$(8.0 \micron/4.5 \micron) < 0.15$ (see Figure 4).  Only 11\% of these relatively blue $T_{\rm x}>50$~ks sources have a \chandra\ counterpart. 

\begin{figure*}
$\begin{array}{cccc}
\includegraphics[angle=0,scale=0.45]{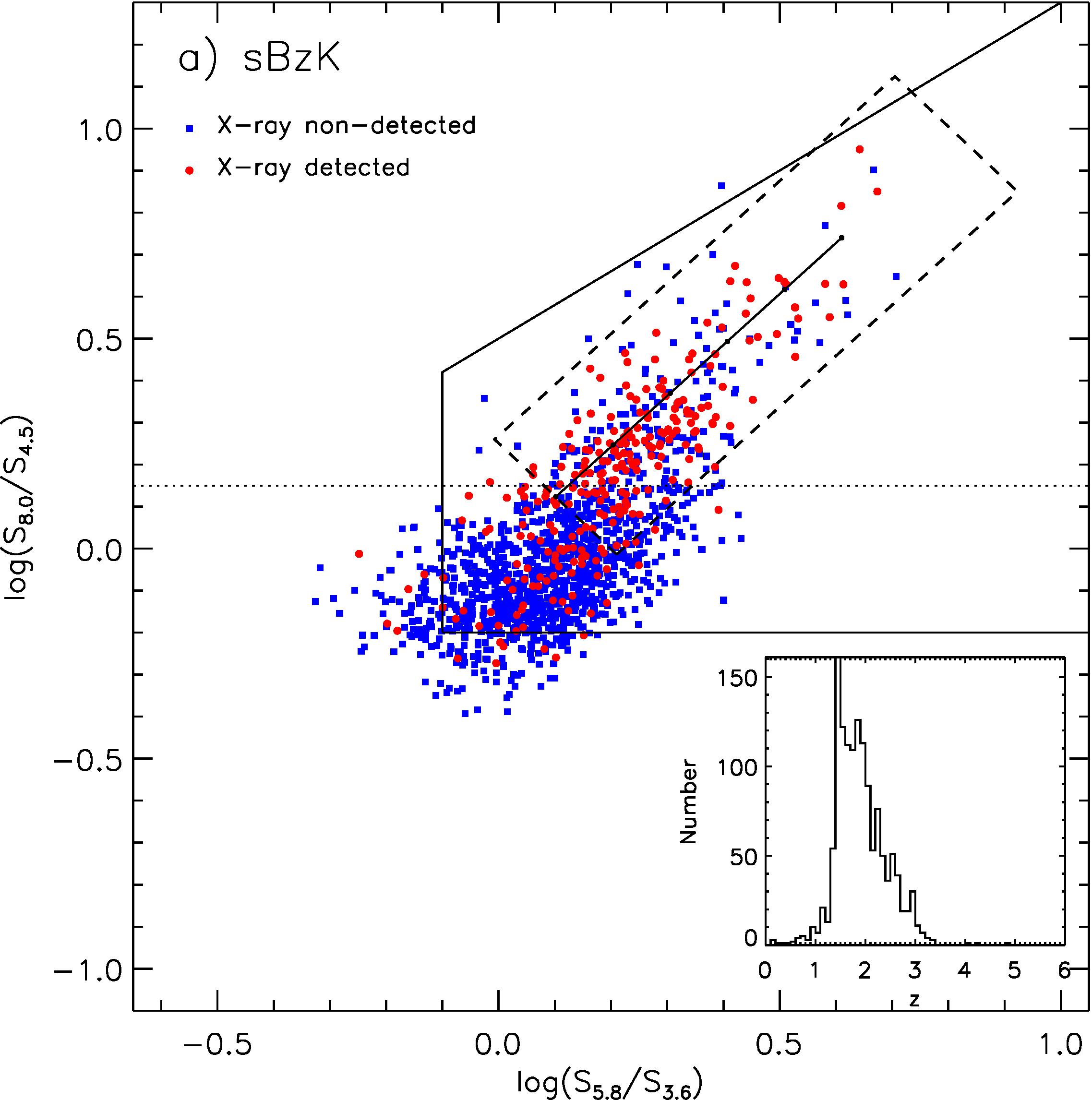} &
\includegraphics[angle=0,scale=0.45]{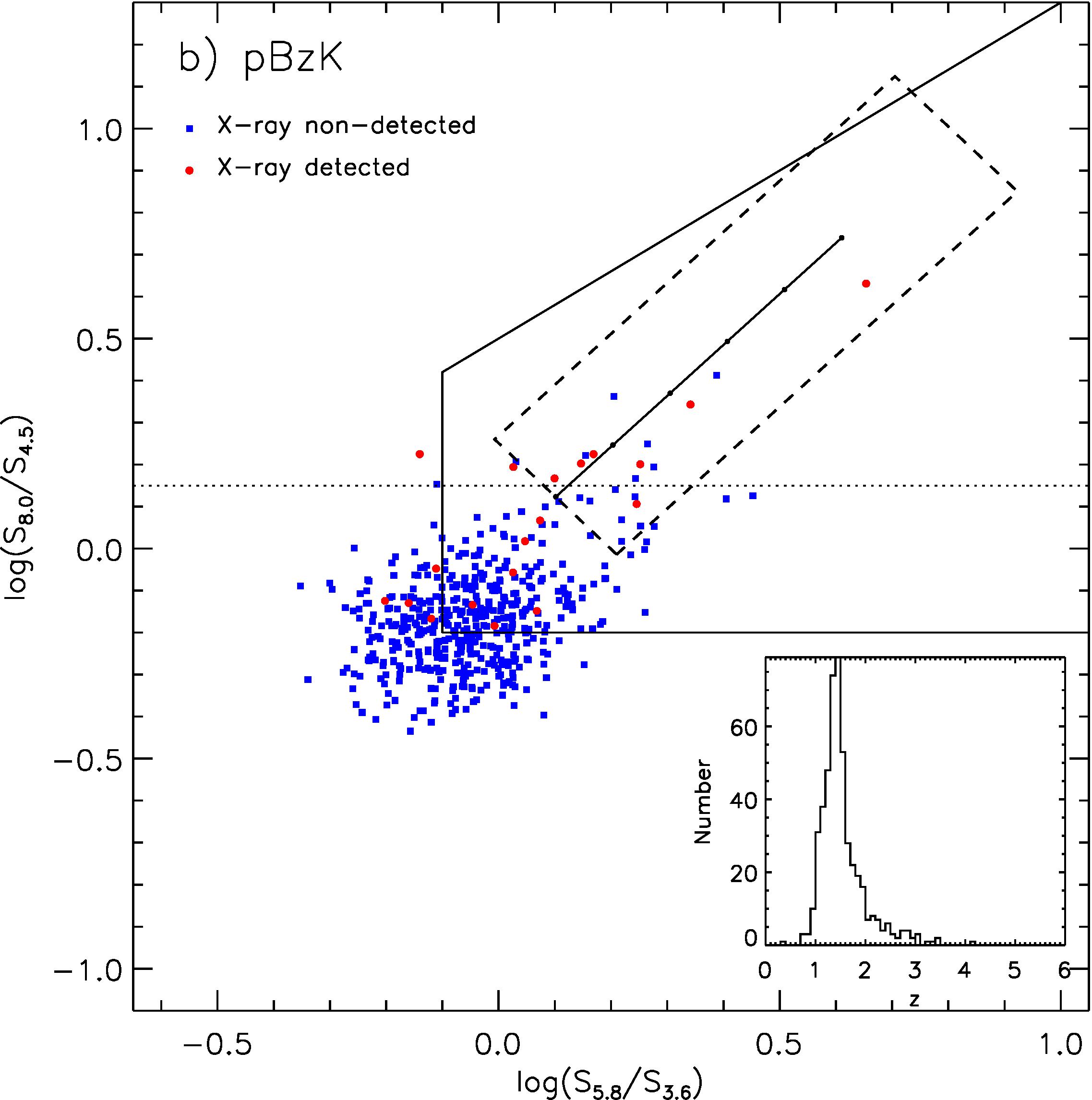} \\
\includegraphics[angle=0,scale=0.45]{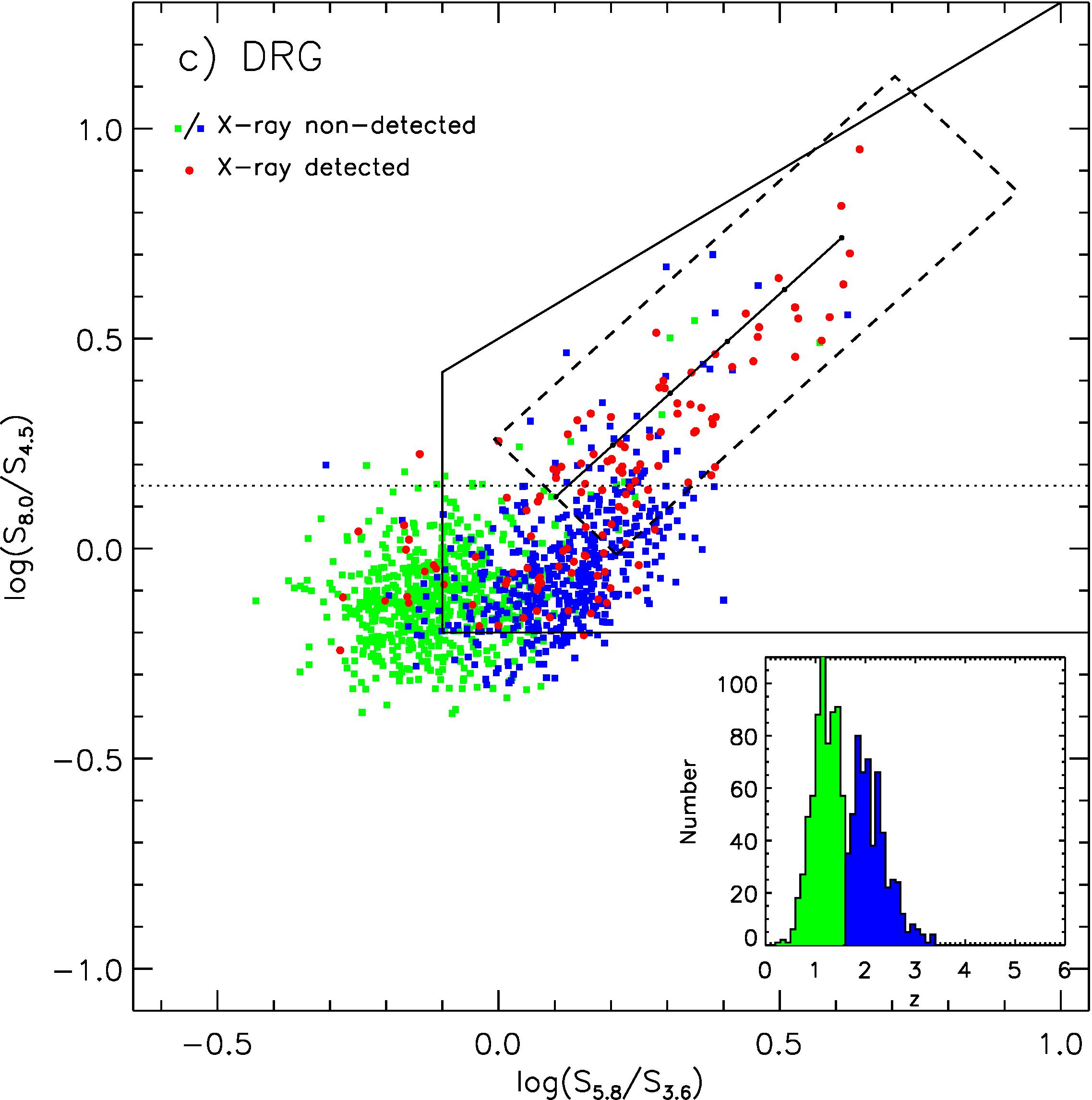} &
\includegraphics[angle=0,scale=0.45]{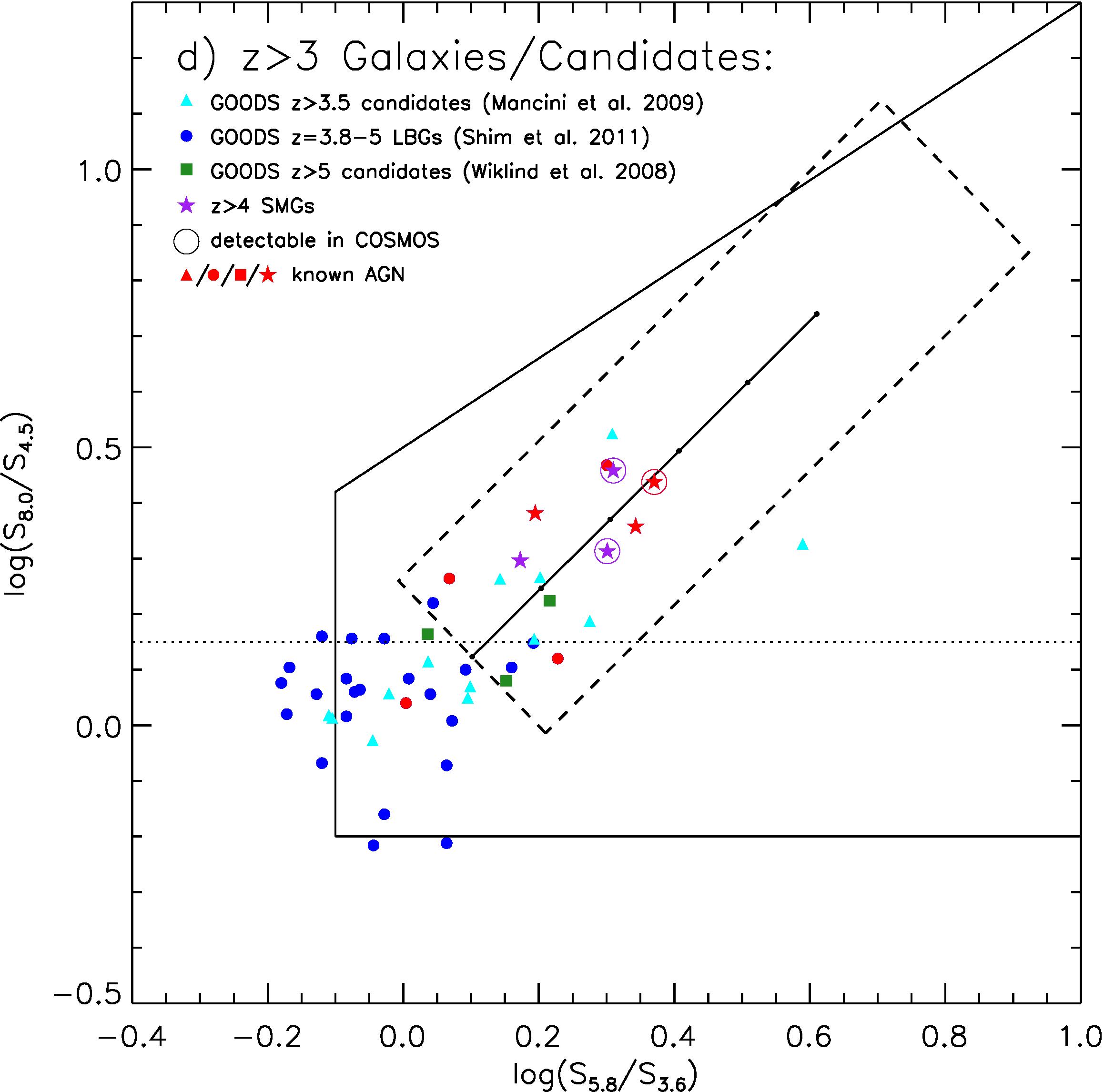} \\
\end{array}$
\caption{IRAC colors of high-$z$ galaxy populations with at least $50$~ks of \chandra\ coverage, in \cite{lacy04,lacy07} color space.  Black circles (red circles in the online version) indicate sources with \chandra\ counterparts, and the inset histograms give the redshift distributions of the high-$z$ populations. The dotted line shows the log$(8.0 \micron/4.5 \micron) < 0.15$ cut defined in \S8 to exclude high-redshift star-forming galaxies from the power-law box (dashed line). In panel (d), only the three sources indicated by large circles are bright enough to meet the COSMOS 5$\sigma$ limits in all four IRAC bands.  After applying the 8.0 \micron/4.5 \micron\ cut indicated by the dotted line, we therefore expect minimal contamination from high-redshift star-forming galaxies. }
\end{figure*}

\begin{figure*}
$\begin{array}{cccc}
\includegraphics[angle=0,scale=0.45]{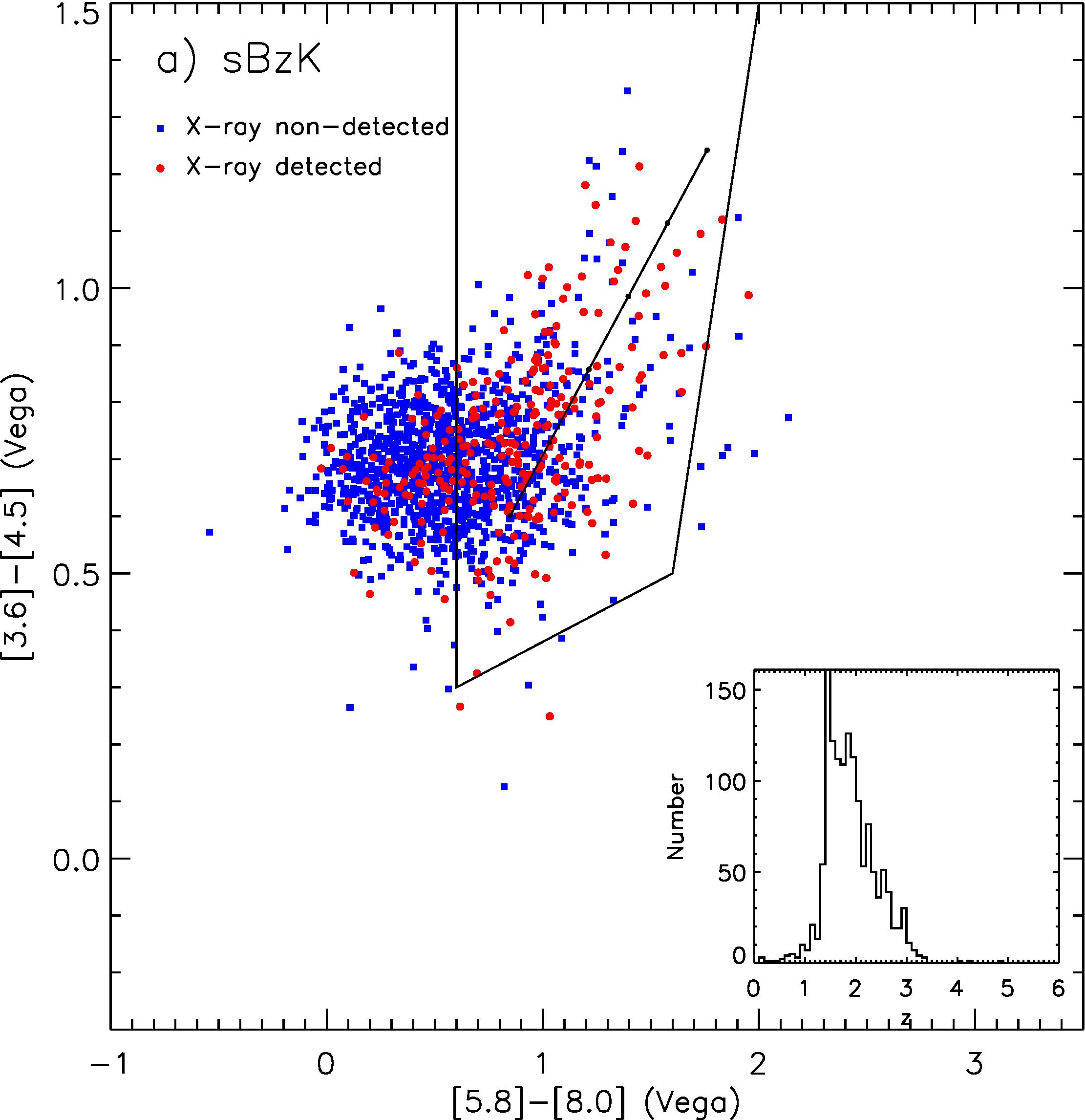} &
\includegraphics[angle=0,scale=0.45]{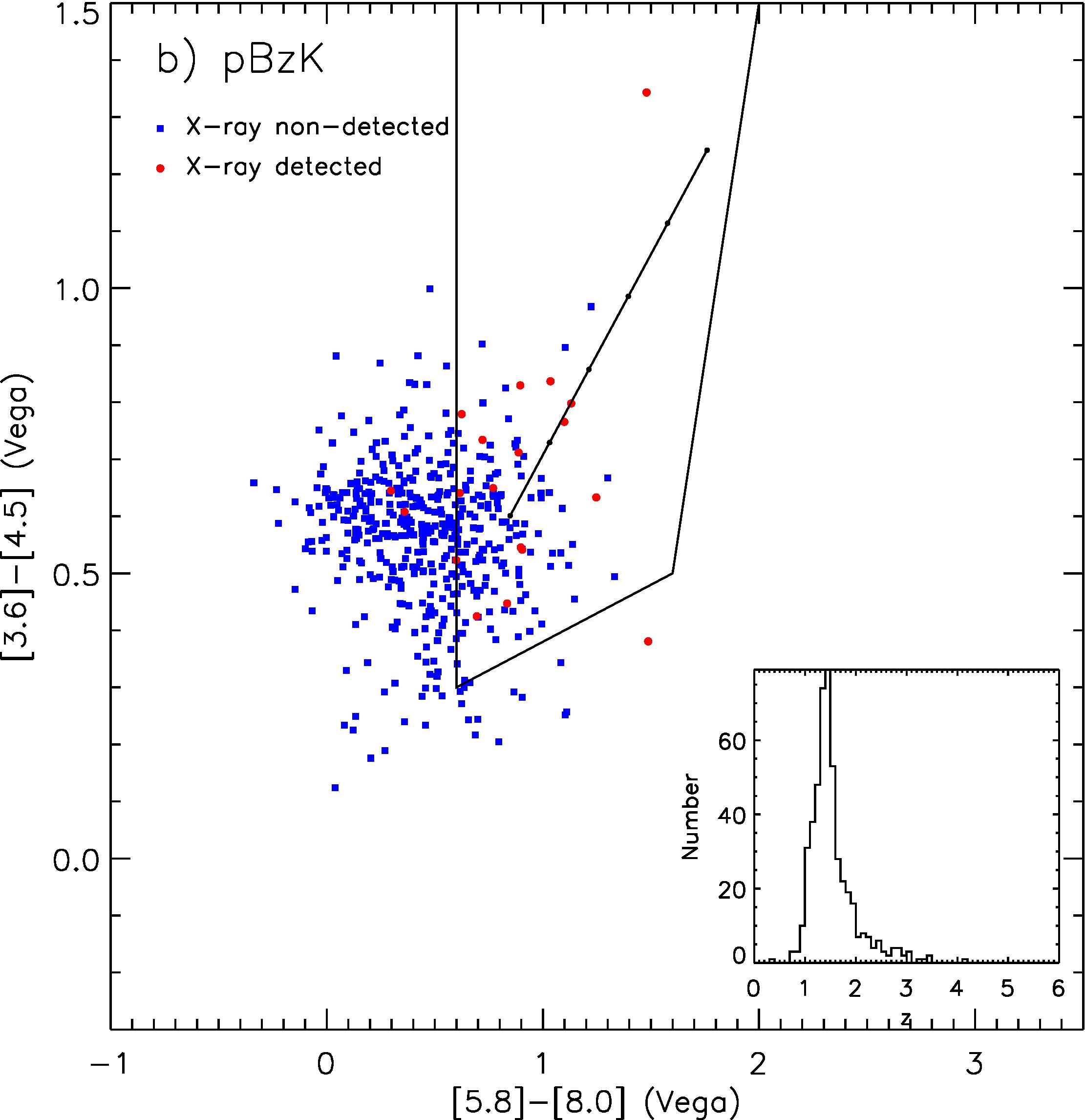} \\
\includegraphics[angle=0,scale=0.45]{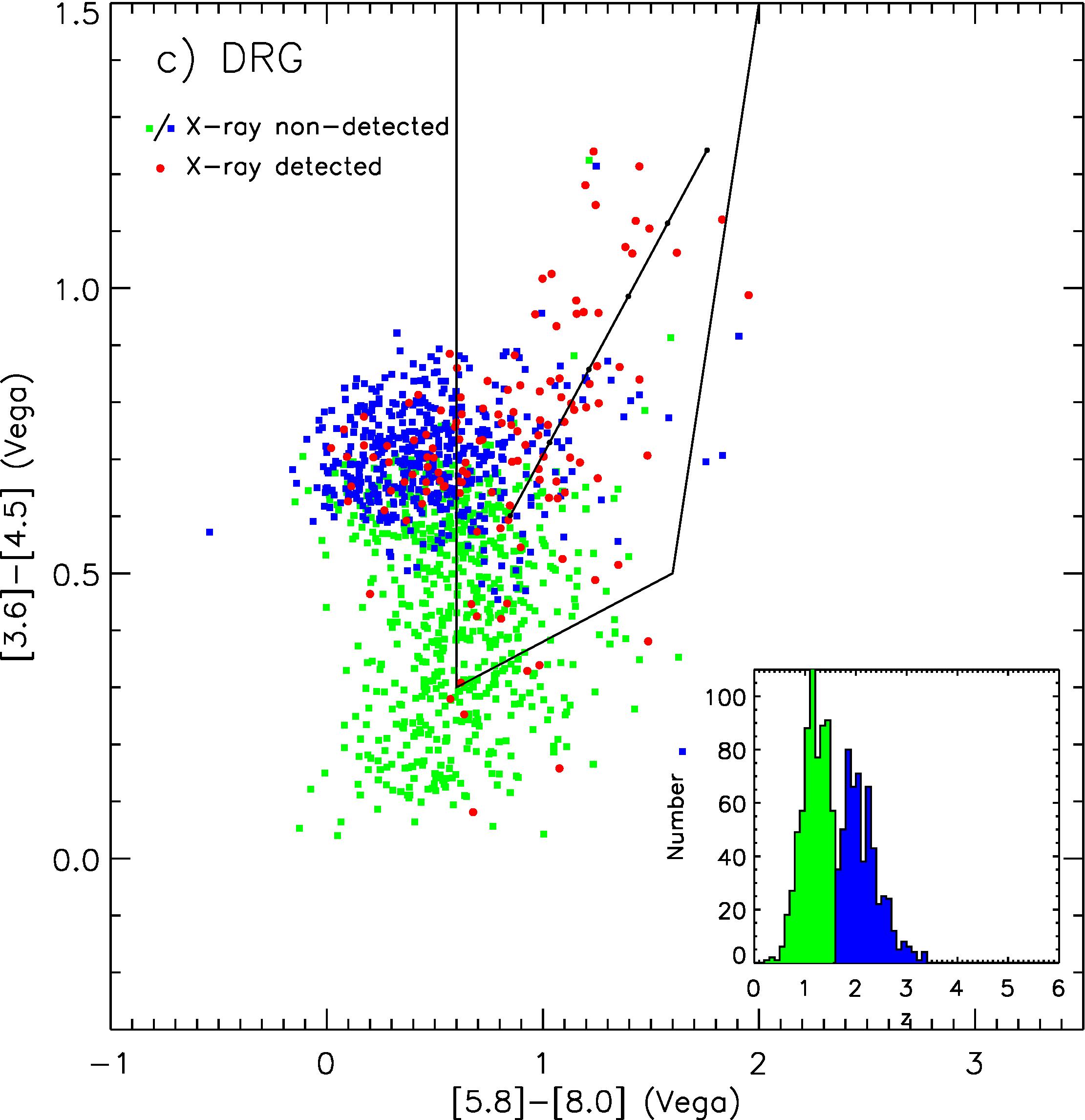} &
\includegraphics[angle=0,scale=0.45]{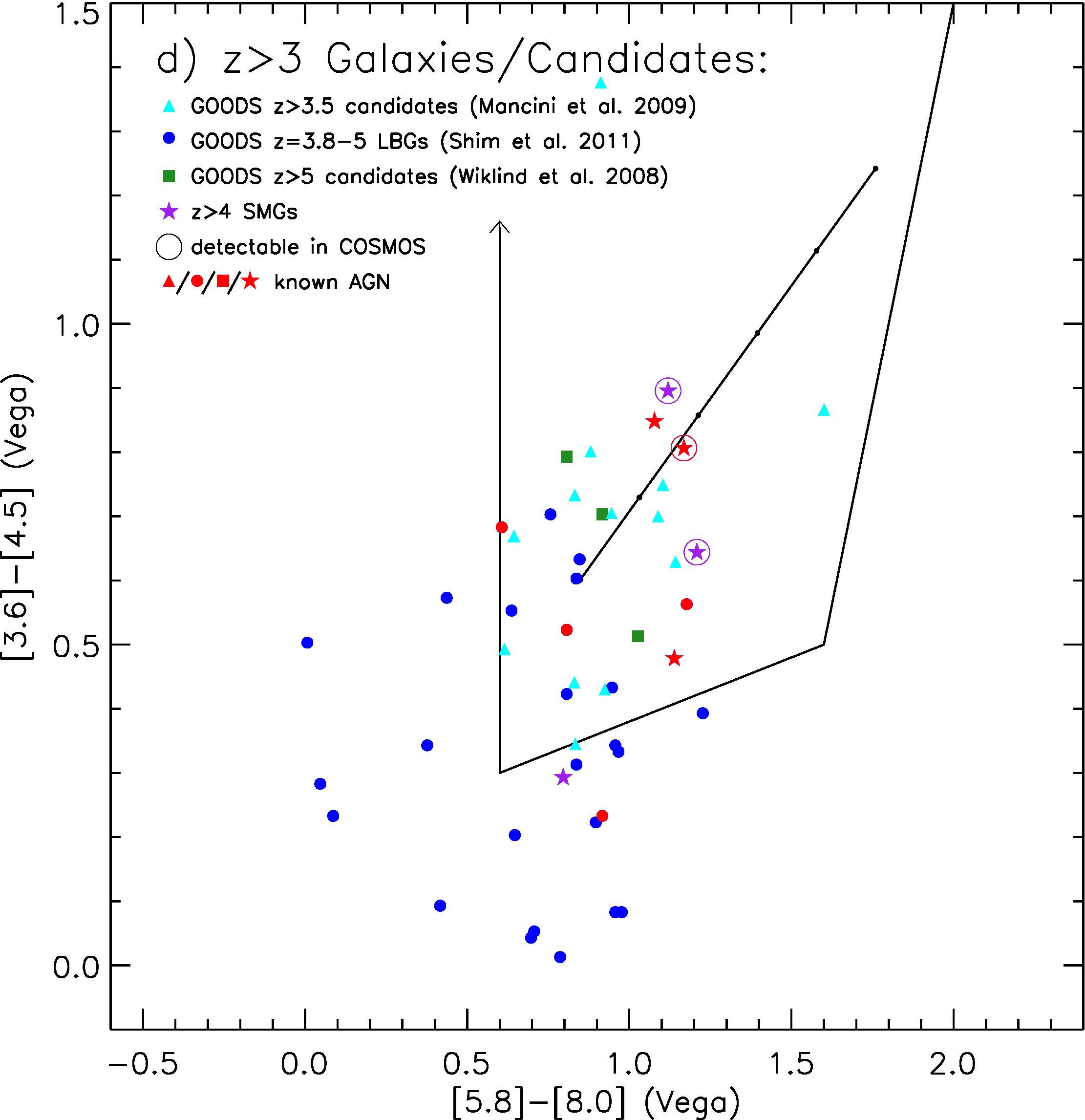} \\
\end{array}$
\caption{Same as Figure 10, but for \cite{stern05} color space.}
\end{figure*}

The population of sBzK sources with red IRAC colors (log$(8.0 \micron/4.5 \micron) \ge 0.15$) extends upwards along the power-law locus and has an X-ray detection fraction of 49\%, higher even than the median value in the power-law box (see \S7.1).  While these red sources are therefore strong AGN candidates, Figures 4 and 10 indicate that an additional color cut of log$(8.0 \micron/4.5 \micron) \ge 0.15$ should be applied to the AGN selection criteria to prevent contamination from high-redshift star-forming galaxies.  This cut, shown in Figure 10, rejects sources in the lower left corner of the power-law box, and is slightly bluer than the cut proposed by \cite{coppin10} to distinguish AGN and star-formation--dominated SMGs, log$(8.0 \micron/4.5 \micron) \ge 0.22$ \citep[see also][]{yun08}. 

We plot in Figures 10b and 11b the IRAC colors of the 31\% of pBzK galaxies detected to $\ge 5\sigma$ in the four IRAC bands. Nearly all (97\%) of the pBzK galaxies have IRAC colors blueward of the cut defined above, and only 4\% of these are X-ray sources.  Of the few sources with red IRAC colors, 50\% are again X-ray detected.

\subsection{DRGs}

The distant red galaxy (DRG) criterion, $(J-K)_{\rm AB} > 1.37$  or $(J-K)_{\rm Vega} > 2.3$, was designed to select $z=2-3.5$ galaxies with a strong Balmer or 4000\AA\ break and identifies 77\% of massive galaxies at $z=2-3$ \citep{vandokkum06}.  Using the UKIRT $J$ and CFHT $K_{\rm s}$ data from \cite{capak11} and \cite{mccracken10}, we identify $\sim 5000$ DRGs in COSMOS, 25\% of which lie above the $\ge 5 \sigma$ sensitivity limits in each of the IRAC bands.  Their IRAC colors and redshift distribution are shown in Figures 10c and 11c.  

While the DRG color cut was originally intended to select sources with a strong Balmer break at $z\sim 2-3.5$, it also recovers dusty star-forming galaxies and AGN at $z < 1.4$ \citep{conselice07,lane07}.  For clarity, we therefore divide the DRG sample in Figures 10c and 11c into low and high redshift components, adopting a cut near the minimum in our redshift distribution, $z=1.6$. 

The DRGs, like the BzK-selected galaxies, can be divided into two main populations using a cut of log$(8.0 \micron/4.5 \micron) = 0.15$.  The bluer sources, which comprise 91\% of the population, have an X-ray detection fraction of only 6\% and can again be excluded from the AGN selection region using the cut defined above.  The red sources extend along the power-law locus and have a far higher X-ray detection fraction of 53\% (or 56\% if we consider only the red sources that lie in the power-law box), indicative of AGN emission.  In the deeper GOODS-S field, \cite{papovich06} find X-ray or MIR evidence of AGN emission in 25\% of DRGs.  In COSMOS, if we take as AGN all X-ray detected DRGs as well as all red (log$(8.0 \micron/4.5 \micron) \ge 0.15$) X-ray non-detected DRGs in the power-law box, we conclude that at least 14\% of the IRAC-detected DRGs are AGN.

\subsection{LBGs}

The Lyman-break dropout technique remains the most successful method of identifying galaxies at $z\sim3-7$.  Of the photometrically-selected U, B, g, V, R, and i dropout candidates in COSMOS, however, fewer than $2\%$ are detected at or above the $5 \sigma$ sensitivity limits in all four IRAC bands.  As these bright sources are also those most likely to be low-redshift interlopers \cite[see, e.g.,][]{reddy08}, we limit our analysis of LBGs to sources that have been spectroscopically-confirmed. 

Only three spectroscopically-confirmed star-forming $z \ge 3$ LBGs in COSMOS are detected to $\ge 5 \sigma$ in all four IRAC bands, and two of these are in a close pair with blended IRAC photometry.  We therefore turn to the ultra-deep GOODS fields and adopt the sample of 74 spectroscopically-confirmed $z=3.8-5.0$ LBGs from \cite{shim11}, selected to have $\ge 5 \sigma$ detections in the 3.6 and 4.5 \micron\ IRAC bands.  We plot in Figures 10d and 11d the IRAC colors of the 27 LBGs detected to $\ge 3 \sigma$ in all four IRAC bands.  As none of the \cite{shim11} GOODS sources meet the COSMOS $5 \sigma$ sensitivity cuts in all in four IRAC bands, we do not expect LBGs to be a significant source of contamination at the depth of COSMOS.  Nonetheless, their IRAC properties shed light on the potential contamination by high-redshift star-forming galaxies in surveys with deeper IRAC data. 

Despite their median redshift of $z=4.2$, the LBGs have relatively blue IRAC colors.  Of the 27 LBGs, only six fall within the power-law box, and half of these are X-ray--detected AGN.  This behavior can be attributed to the requirement that LBGs have a bright UV continuum (which in turn biases the selection towards galaxies with young stellar populations, low extinctions, and thus blue UV-optical rest-frame colors) and to the bright H$\alpha$ emission observed in 70\% of the \cite{shim11} sources, which leads to an excess of 3.6 \micron\ emission relative to the best-fit stellar continuum.  If no H$\alpha$ were present, however, the \cite{shim11} sources would simply shift to the right in \cite{lacy04,lacy07} color space and continue to be excluded by the $8.0 \micron/4.5 \micron$ cut defined above. For a high-redshift galaxy to display red colors in the observed IRAC bands, it must be more heavily reddened or have an older stellar population than the typical LBG.  

\subsection{Evolved and Reddened $z>3$ Candidates}
Several attempts have been made to select reddened and/or evolved galaxies at $z>3$ \citep{yan06,rodighiero07,wiklind08,mancini09,marchesini10}.   The high 24 \micron\ detection fraction for these sources, however, points to likely contamination by star-forming galaxies at $z=2-3$ or to a dominant obscured AGN component at high redshift \cite[see, e.g.,][]{dunlop07,marchesini10}.  We therefore plot in Figures 10d and 11d only the ``no-MIPS'' $z>3.5$ IRAC-selected and $z>5$ Balmer-break--selected high-$z$ candidates from the GOODS samples of \cite{mancini09}  and \cite{wiklind08}, and further restrict the \cite{mancini09} sample to those sources lacking a $z\le 3$ redshift solution within the 90\% confidence interval. 

None of the high-$z$ evolved and/or reddened galaxy candidates plotted in Figures 10d and 11d meet the COSMOS $5 \sigma$ sensitivity limits in all four IRAC bands, so we again expect little to no contamination from such galaxies in COSMOS.  Furthermore, only 5 of the 13 $z>3.5$ candidates from \cite{mancini09} lie within the power-law box, primarily at colors blueward of $\alpha \sim -1.0$.   (The reddest \cite{mancini09} source in the power-law box is the $z_{\rm phot} = 3.93$ SMG GN1200.5/AzGN06.)  Likewise, two of the three $z \ge 5$ Balmer-break candidates of \cite{wiklind08} fall in the power-law box, though again at  $\alpha \ge -1.0$.   The relatively blue continua of these galaxies, compared to the redshifted tracks of local templates, can likely be attributed to the  bluer UV continua and lower dust extinctions of high redshift galaxies \citep{bouwens09}.  Even in the deep GOODS field, we therefore expect only minimal contamination from typical high-redshift galaxies, and only in the bluest regions of the power-law box. 

\begin{figure*}
$\begin{array}{cccc}
\includegraphics[angle=0,scale=0.46]{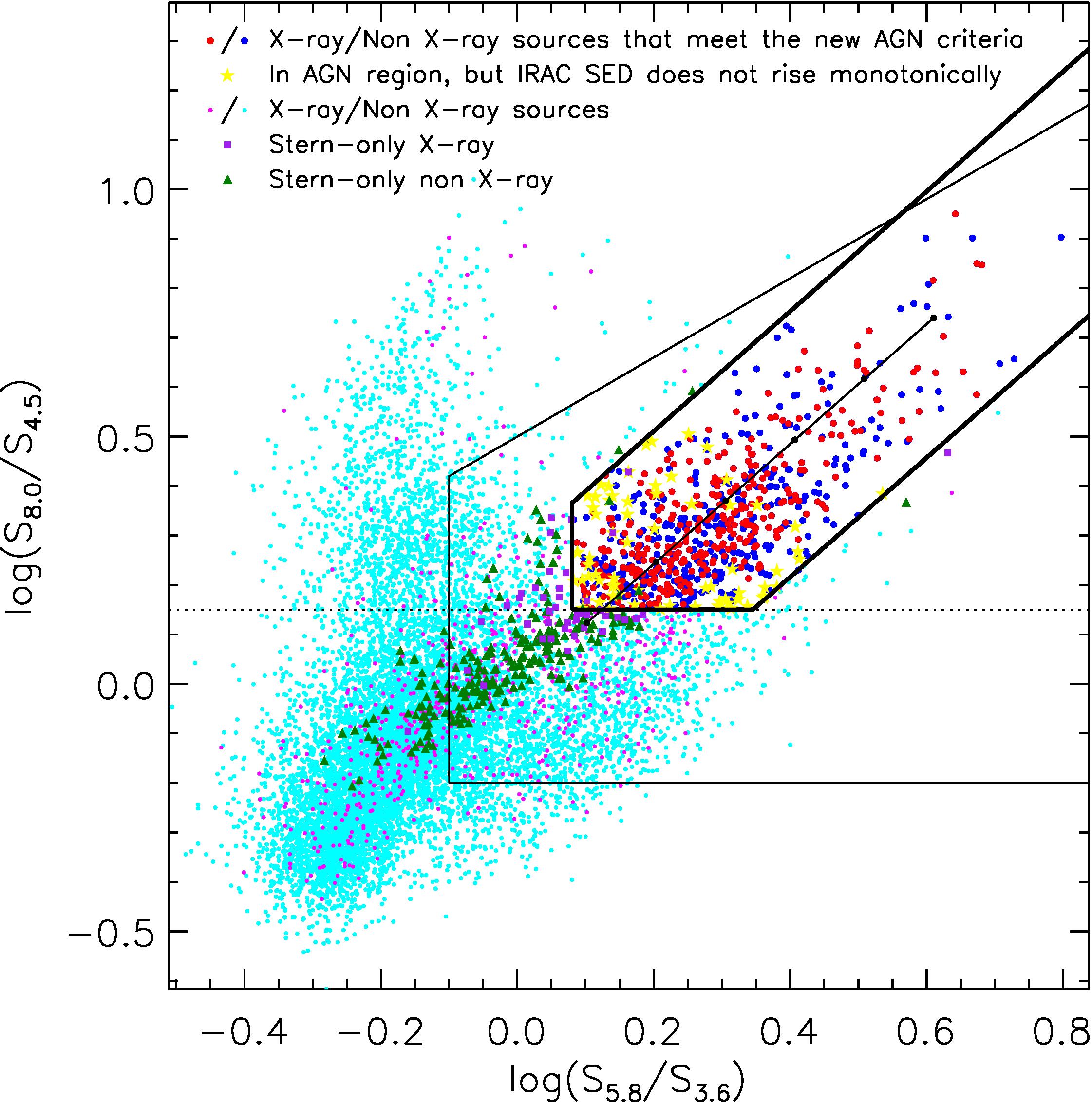} &
\includegraphics[angle=0,scale=0.46]{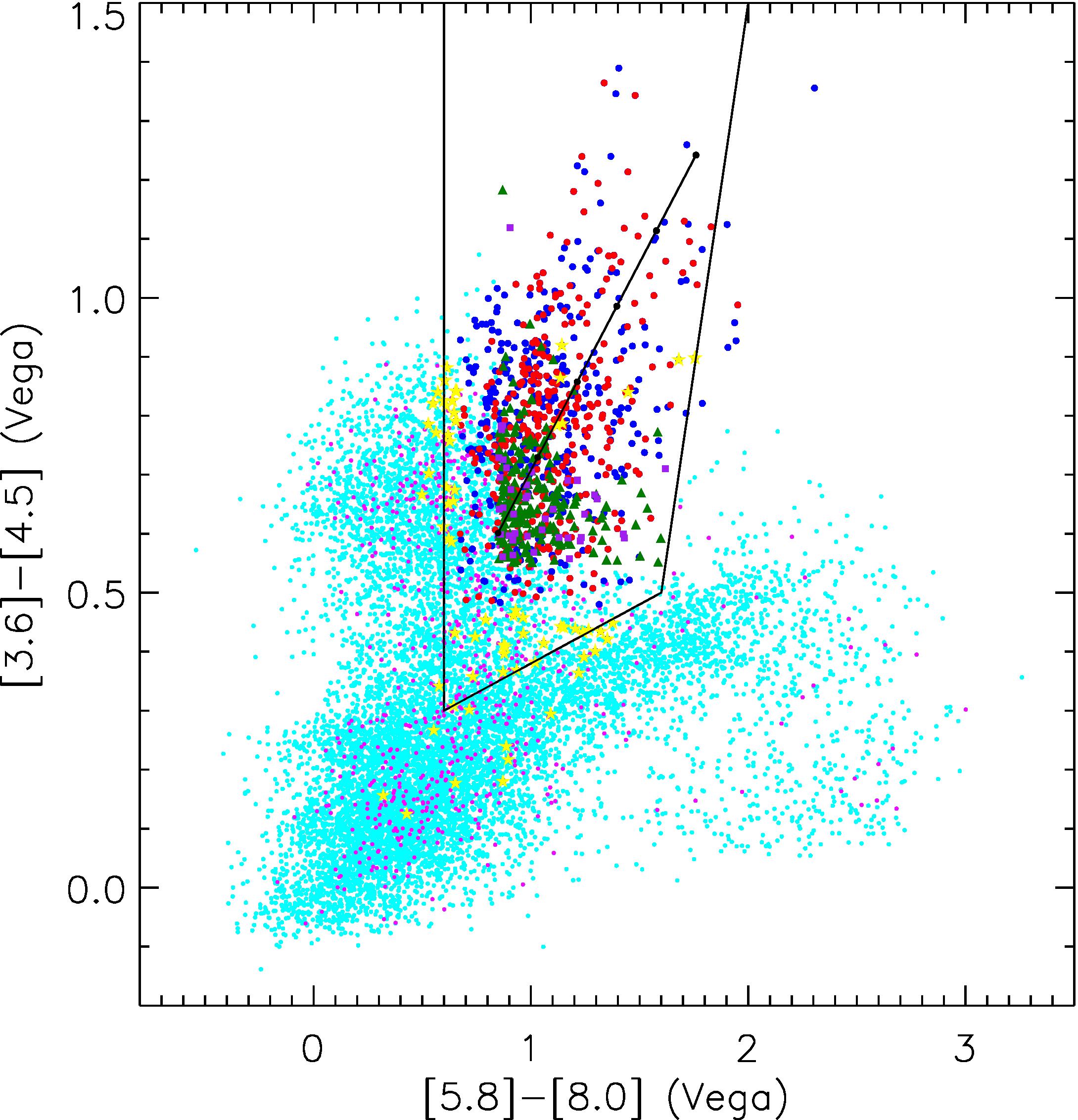} \\
\end{array}$
\caption{New AGN selection criteria (thick solid lines).  The full COSMOS IRAC sample with $T_{x} \ge 50$~ks are plotted as small points, and sources that meet our new criteria are shown by large filled circles.   Sources plotted in black (or red/magenta in the online version) have X-ray counterparts, and sources plotted in grey (or blue/cyan in the online version) do not.  Large (yellow) stars represent sources that lie within the revised wedge in \cite{lacy04,lacy07} color space but that are excluded from our final selection because their IRAC SEDs do not rise monotonically. The (purple) squares and (green) triangles represent the X-ray and non-X-ray sources, respectively, that fall well within the \cite{stern05} AGN selection region but that are excluded by our new AGN selection criteria (see \S9.2.3).}
\end{figure*}

\subsection{SMGs}

While dust extinction appears to be less prevalent in the early universe,  heavily dust obscured luminous star-forming galaxies have been selected as sub-mm galaxies (SMGs) out to $z\sim5$ \citep{capak08,schinnerer08,daddi09a,daddi09b,coppin09highz,knudsen10,riechers10,smolcic11}.   At moderate redshifts of $1.4 < z < 2.6$, more than $80\%$ of bright radio-selected SMGs meet one or more of the BzK, DRG, or LBG (e.g, BM/BX) criteria \citep{reddy05}.  Their IRAC colors are therefore coincident with the BzK and DRG star-forming populations shown in Figures 10a, 11a, 10b, and 11b (see also Figure 4).   At $z \gsim 4$, however, UV-NIR color cuts are unlikely to identify the dust-obscured and IR-luminous SMGs most likely to enter the AGN selection region \citep[though see][]{mancini09}.

In COSMOS, six spectroscopically-confirmed $z\gsim4$ SMGs have been identified over 0.2 deg$^2$.  Of these, only AzTEC-5 \citep[][Capak et al. 2011, in prep.]{marchesini10} is detected to $\ge 5 \sigma$ in all four IRAC bands, although AzTEC1 \citep{smolcic11} falls just below our cuts.   We plot in Figures 10d and 11d the IRAC colors of these two high-$z$ SMGs, as well as the four (of seven) additional spectroscopically-confirmed $z>4$ SMGs with high-$\sigma$ IRAC counterparts:  the GOODS-N sources GN20 and GN20.1 \citep{daddi09a}, the EGS source LESS J033229.4 \citep{coppin09highz,gilli11}, and the Abell 2218 source SMM J163555.5 \citep{knudsen10}. 

All of the $z\ge 4$ SMGs fall in the power-law box at $-1 < \alpha < -2$, although only three are bright enough to be included (or nearly included) in our COSMOS IRAC sample.  At least half of the $z\ge4$ SMGs, however, are known AGN (GN20.2 is a radio-bright AGN \citep{daddi09a}, LESS J033229.4 is a Compton-thick AGN detected in the 4Ms \chandra\ data \citep{gilli11}, and the IRS spectrum of GN20 indicates a considerable AGN contribution to the rest-frame 6 \micron\ emission (D. Riechers et al. 2011, in prep.)), and all but one have a bright MIPS counterpart that points to a likely AGN contribution.  We therefore expect little to no contamination from predominantly star-forming SMGs at moderate to high redshifts.

\section{Revised IRAC Criteria}

We take as a starting point for the revised IRAC selection criteria the $\alpha \le -0.5$ power-law box defined in \S4. Not only does this box enclose the vast majority of sources that would be identified as power-law galaxies \citep{aah06,donley07,park10}, but it also tightly encloses the templates of AGN-dominated sources (see Figure 2) and the region of color space with a high ($\sim 50\%$) X-ray detection fraction (see Figure 8). 

To prevent contamination from high-redshift ($z\ge2$) galaxies bright enough to be included in the COSMOS IRAC sample, we then impose a cut of log$(8.0 \micron/4.5 \micron) \ge 0.15$, defined in \S8.  While this high$-z$ cut removes the majority of grids in the power-law box with low X-ray detection fractions (see Figure 8), the leftmost corner of the power-law box also has a lower than average X-ray detection fraction, presumably due to contamination by low-redshift star-forming galaxies.  To exclude these sources, we impose a vertical cut that coincides with the intersection of the $\alpha \le -0.5$ power-law box and the high-$z$ cut discussed above: log$(5.8 \micron/3.6 \micron) \ge 0.08$.

We plot in Figure 12 the IRAC colors of the sources that meet these criteria.  While all of the new AGN candidates fall by definition within the original \cite{lacy04,lacy07} AGN selection wedge, 9\% have IRAC SEDs that do not rise monotonically, but that decrease between 3.6 and 4.5 \micron\ or 5.8 and 8.0 \micron\ (or, in rare cases, between 4.5 and 5.8 \micron), placing them on the outskirts of the \cite{stern05} AGN selection region (of these, 38\% formally fall outside of the \cite{stern05} AGN wedge).  This behavior is rare among the XMM AGN in our selection region, 97\% of which have monotonically rising IRAC SEDs (a fraction that rises to 99.6\% if we consider only luminous QSOs with log $L_{\rm 2-10 keV} $(ergs~s$^{-1}$)$\ge 44$), and may be due to the 1.6 \micron\ stellar bump passing through the IRAC bandpasses. This interpretation is consistent with the median redshifts, $z=0.7$ and $z=2.2$,  and low $T_{\rm x} > 50$~ks X-ray detection fractions, 17\% and 30\%, of the sources with blue 3.6/4.5 \micron\ and 5.8/8.0 \micron\ colors, respectively.  We therefore exclude from our final selection criteria all sources with non-monotonically-rising IRAC SEDs, plotted as yellow stars in Figure 12.  Of the low redshift sources previously excluded by the vertical cut in \cite{lacy04,lacy07} color space, 81\% would also be excluded by this criterion. 

The final revised AGN selection criteria are as follows, where $\wedge$ is the logical ``AND'' operator: 

\begin{eqnarray}
& x = \rm{log_{10}}\left(\frac{f_{5.8\mu m}}{f_{3.6\mu m}}\right), \hspace*{0.2cm} y = \rm{log_{10}}\left(\frac{f_{8.0\mu m}}{f_{4.5\mu m}}\right) \\
 &\hspace*{0.1cm} x \ge 0.08 \hspace*{0.2cm} \wedge \hspace*{0.2cm} y \ge 0.15 \hspace*{0.2cm}  \\
\nonumber & \wedge \hspace*{0.2cm} y \ge (1.21\times x) - 0.27 \hspace*{0.2cm}\wedge \hspace*{0.2cm} y \le (1.21\times x) + 0.27  \\
\nonumber & \wedge \hspace*{0.2cm} f_{4.5\mu m} > f_{3.6\mu m} \hspace*{0.2cm}  \wedge \hspace*{0.2cm} f_{5.8\mu m} > f_{4.5\mu m} \hspace*{0.2cm} \wedge \hspace*{0.2cm} f_{8.0\mu m} > f_{5.8\mu m} 
\end{eqnarray}

\vspace*{0.2cm}

\noindent These new criteria identify 1506 AGN candidates in COSMOS, only 38\% of which have \xmm\ or \chandra\ counterparts. In regions of deep \chandra\ coverage ($T_{\rm x} =50-160$~ks), the X-ray detection fraction is 52\%. 

Repeating the X-ray stacking analysis for the final AGN candidate sample confirms the results of \S7.2.  The X-ray detected AGN that meet our IRAC criteria lie at $z=1.7 \pm 0.8$ and have typical stacked hardness ratios and column densities of HR$=-0.30 \pm 0.11$ and log~$N_{\rm H}$(cm$^{-2}$)$=22.4\pm 0.4$.  In comparison, the X-ray non-detected AGN candidates with $T_{\rm x} > 50$~ks (median $T_{\rm x} = 125$~ks) have slightly higher photometric redshifts: $z_{\rm phot}=2.2 \pm 0.9$.   While this population lacks individual X-ray counterparts, X-ray stacking leads to $\sim6\sigma$ detections in both the soft and hard X-ray bands with HR$=0.29 \pm 0.13$ or log~$N_{\rm H}$(cm$^{-2}$)$=23.5 \pm 0.4$.  IRAC selection therefore appears to successfully recover large samples of both unobscured to moderately obscured X-ray--detected AGN, as well as heavily obscured, high-redshift AGN missed by deep X-ray surveys. 

\begin{figure*}
$\begin{array}{cc}
\hspace*{0.5cm}
\includegraphics[angle=0,height=3.2in]{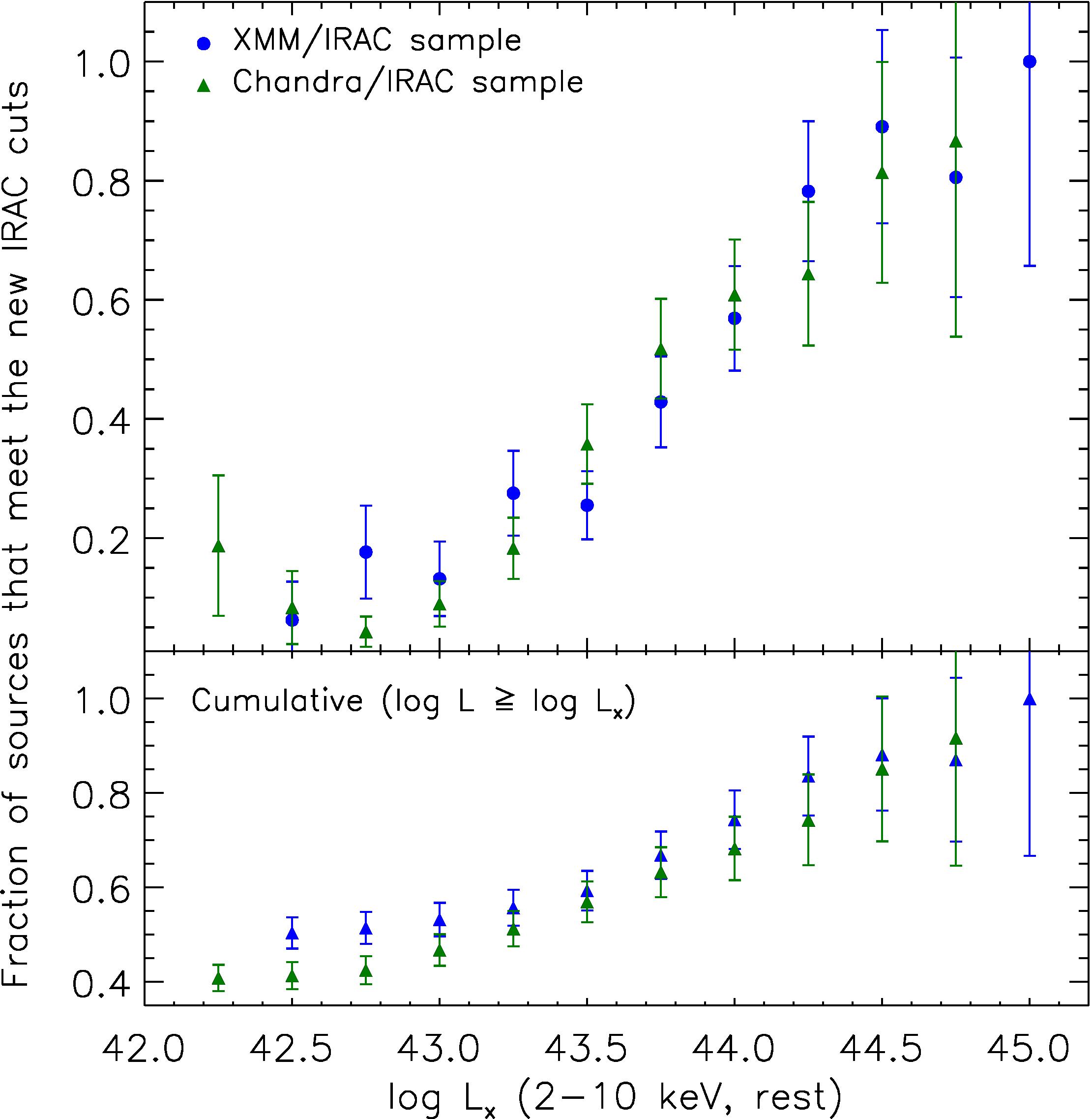} &
\includegraphics[angle=0,height=3.2in]{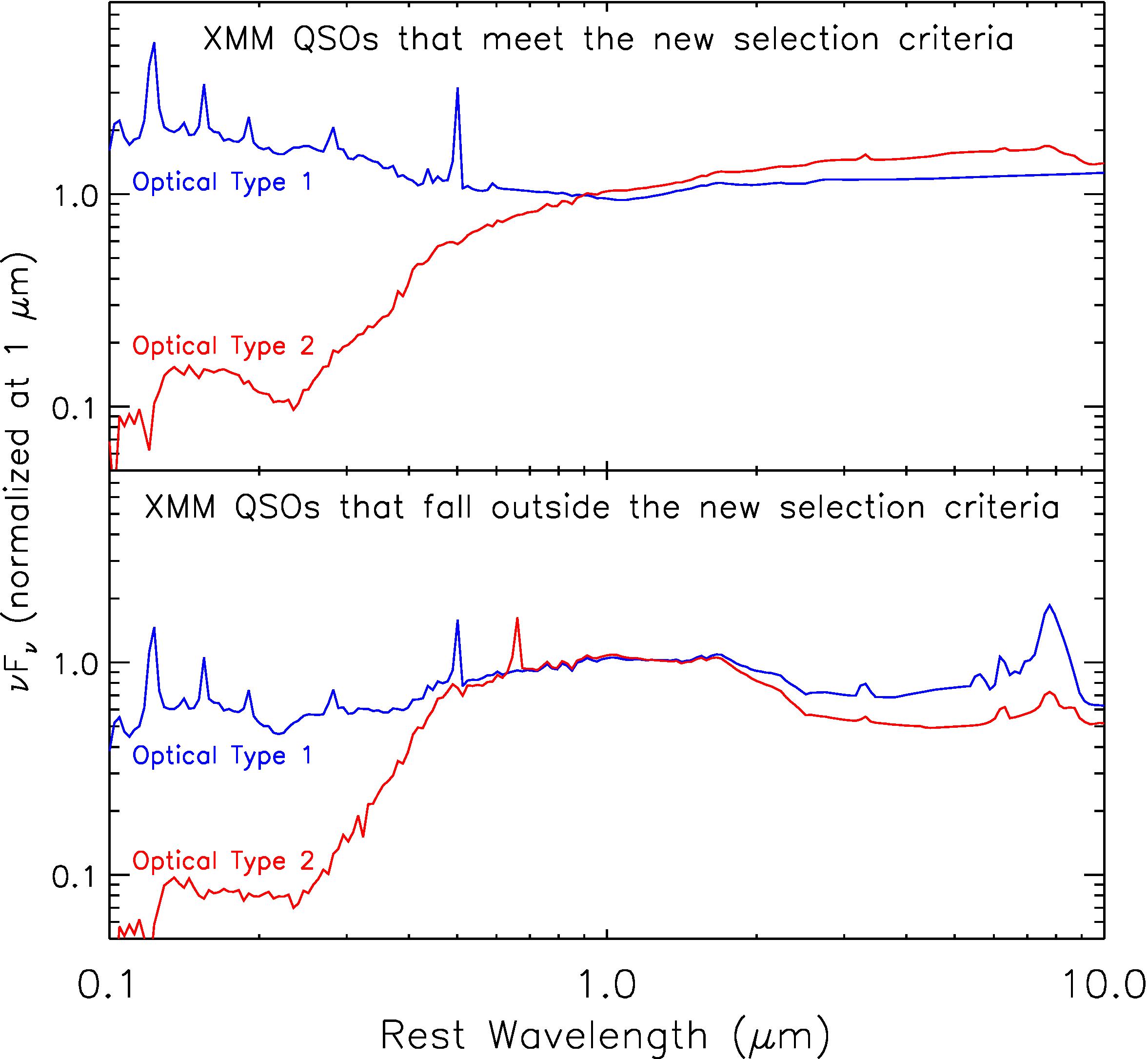} 
\end{array}$
\caption{Left: Fraction of hard X-ray and IRAC-detected \xmm\ and \chandra\ sources that meet the new IRAC cuts, as a function of intrinsic 2-10 keV X-ray luminosity.  While few low-luminosity AGN meet our criteria, the IRAC selection criteria recovers $75\%$ of \xmm\ QSOs and $69\%$ of \chandra\ QSOs with log $L_{\rm 2-10 keV} $(ergs~s$^{-1}$)$\ge 44$. Right: Best fits to the median SEDs of QSO-luminosity \xmm\ sources that do or do not meet our selection criteria.  Both the Type 1 and Type 2 QSOs missed by the IRAC criteria have slightly redder UV-optical continua and prominent 1.6 \micron\ stellar bumps, indicating that IRAC selection is most likely to miss AGN with luminous hosts, particularly when the AGN emission is itself obscured.}
\end{figure*}

\subsection{Reliability}

The criteria defined above have been designed to reject the majority of low- and high-redshift star-forming contaminants that enter the \cite{lacy04,lacy07} and \cite{stern05} AGN selections wedges (see Sections 2, 7, 8, and 9).  While approximately half of the AGN candidates selected by these criteria are not individually detected in deep ($T_{\rm x} = 50-160$~ks) \chandra\ data, their stacked X-rays are consistent with emission from heavily obscured AGN.  However, 32\% of the $T_{\rm x}>50$~ks X-ray undetected IR AGN candidates have $z_{\rm phot} \ge 2.7$, the redshift at which the templates of purely star-forming local LIRGS and ULIRGS from \cite{rieke09} begin to enter our new selection region.  To determine if these sources are star-forming contaminants, we repeat the X-ray stacking analysis for three equally-populated redshift bins: $z=0-1.8$, $z=1.8-2.7$, and $z=2.7-4.8$.

The stacked emission from the $z>2.7$ IR AGN candidates is detected to $3.9\sigma$ in the 0.5-2.0 keV band and to $4.0\sigma$ in the 2-8 keV band with an observed hardness ratio of $HR = 0.16 \pm 0.18$ (or log~$N_{\rm H}$(cm$^{-2}$)$=23.6 \pm 0.2$ at the mean redshift of $z=3.2$).  As the effective X-ray photon index, $\Gamma_{\rm eff} = 0.67^{+0.35}_{-0.36}$, is harder than even the hardest star-forming galaxy \cite[$\Gamma_{\rm eff}=1-2$, e.g.,][]{lehmer08}, a significant fraction of these sources must be obscured AGN\footnote{We use v4.2 of the Portable, Interactive Multi-Mission Simulator (PIMMS) to estimate $\Gamma_{\rm eff}$ using the Galactic column density of $2.7\times10^{20}$~cm$^{-2}$ \citep{elvis09} and the Chandra Cycle 8 effective area curves.}.  For comparison, the low redshift bin is detected to $2.4\sigma$ in the soft band and to $4.1\sigma$ in the hard band with $HR=0.50\pm0.23$ or $\Gamma_{\rm eff} = -0.04^{+0.51}_{-0.55}$ (log~$N_{\rm H}$(cm$^{-2}$)$=23.2 \pm 0.4$ at $z=1.2$), and the medium redshift bin is detected to $3.2\sigma$ in the soft band and to $2.7\sigma$ in the hard band with $HR=0.26\pm0.27$ or $\Gamma_{\rm eff} = 0.49^{+0.51}_{-0.58}$ (log~$N_{\rm H}$(cm$^{-2}$)$=23.5 \pm 0.3$ at $z=2.2$). 

Many of the high-$z$ AGN candidates, however, show a clear infrared excess relative to even the reddest star-forming template.  To better isolate the sources most likely to be star-forming contaminants, we develop the following empirical criteria to identify high-$z$ sources whose IRAC photometry can plausibly be fit by the \cite{rieke09} star-forming templates:

\begin{eqnarray}
& \rm{log_{10}}\left(\frac{f_{8.0\mu m}}{f_{3.6\mu m}}\right) < 
\begin{cases}
0.39\times z - 0.69 & \text{if $z=2.7-3.1$} \\
0.18\times z - 0.04 & \text{if $z=3.1-4.2$} \\
0.06\times z + 0.47 & \text{if $z=4.2-5.0$} 
\end{cases}\\
& {\rm and}\hspace*{0.2cm} {{f_{5.8\mu m}/f_{3.6\mu m}}\over{f_{8.0\mu m}/f_{4.5\mu m}}} \ge 0.95
\end{eqnarray}

\noindent The first criterion identifies sources whose 8.0 to 3.6 \micron\ flux ratio is not significantly redder than the reddest \cite{rieke09} LIRG/ULIRG template at a given redshift, and the second criteria ensures that the curvature of the IRAC photometry is roughly consistent with the redshifted 1.6 \micron\ stellar bump.   Approximately 50\% of the $z>2.7$ X-ray non-detected IR AGN candidates (or 15\% of the full sample of X-ray non-detected AGN) meet these criteria and therefore have IRAC photometry potentially consistent with star-formation at their assumed redshift (though nearly all show an infrared excess relative to the bluest of the \cite{rieke09} templates).   X-ray stacking of this subsample, however, again returns significant ($>3 \sigma$) detections in both the soft and hard X-ray bands with $HR=0.33 \pm 0.22$ or $\Gamma_{\rm eff} = 0.35^{+0.44}_{-0.50}$, indicating that any contamination by high redshift star-forming galaxies is minimal.  Instead, these sources appear to be either high redshift AGN whose underlying hosts are bluer than the reddest \cite{rieke09} template, or lower redshift AGN whose photometric redshifts have been artificially inflated by the fact that they were estimated using only star-forming templates, which are consistent with red, power-law like IRAC emission only at high-$z$.   Interestingly, we do not detect a significant stacked signal for the 50\% of $z>2.7$ AGN candidates with extremely red IRAC colors, perhaps because these sources lie at genuinely higher redshifts or are more heavily obscured. 

\subsection{Completeness}

To quantify the completeness of the new selection criteria, we plot in Figure 13a the fraction of hard X-ray and IRAC-detected \xmm\ and \chandra\ sources that fall in our new selection region as a function of their intrinsic 2-10 keV luminosities.  The completeness of our new selection criteria is a strong function of AGN luminosity, as expected.  At log~$L_{\rm x}$(ergs~s$^{-1}$)$ < 43$, fewer than 20\% of AGN meet our criteria due to dilution of the AGN continuum by the host galaxy and/or to the disappearance of the hot dust torus at low Eddington accretion rate \citep{trump11b}. At QSO luminosities of log~$L_{\rm x}$(ergs~s$^{-1}$)$ \ge 44$, however, 75\% of the \xmm\ AGN and 68\% of the \chandra\ AGN fall in the revised IRAC selection region.

\subsubsection{Nature of the QSOs missed by our selection criteria}

We plot in Figure 13b the best fits to the median SEDs of the \xmm\ QSOs (e.g., log~$L_{\rm x}$(ergs~s$^{-1}$)$ \ge 44$) that fall inside and outside of our selection region, calculated as described in \S5.3.  The \xmm-selected QSOs that meet our criteria have a median redshift of $z=1.8$, a median luminosity of log~$L_{\rm x}$(ergs~s$^{-1}$)$ = 44.4$, a median column density of log~$N_{\rm H}$(cm$^{-2}$)$=22.3$ (for the 82\% of sources with a measurable column), and X-ray and optical Type 2 fractions of 53\% and 28\%, respectively.  The Type 1 and Type 2 SEDs are remarkably similar at $\lambda \gsim 1 \micron$, and they show no sign of host galaxy emission. 

In comparison, the \xmm-selected QSOs (again defined to have log~$L_{\rm x}$(ergs~s$^{-1}$)$ \ge 44$) that fall outside of our selection region have a median redshift of $z=1.6$, a median luminosity of log~$L_{\rm x}$(ergs~s$^{-1}$)$ = 44.2$, a median column density of log~$N_{\rm H}$(cm$^{-2}$)$=22.6$ (for the 88\% of sources with a measurable column), and X-ray and optical Type 2 fractions of 74\% and 54\%, respectively.  Their SEDs are characterized by slightly redder UV-optical continua and by noticeable 1.6 \micron\ stellar bumps, even in the case of the Type 1 QSOs.  The QSOs missed by IRAC selection therefore appear to be somewhat more heavily obscured, lower-luminosity AGN whose host galaxies contribute a larger relative fraction of their optical-NIR flux. 

\subsubsection{Completeness to Heavily Obscured AGN}

To better constrain our completeness to heavily obscured AGN, we turn to the luminous ($f_{\rm 24} \ge 700$ \microjy) dust-obscured galaxies (DOGs) of \cite{donley10}, selected on the basis of their high 24 \micron\ to R-band flux ratios.  All but one of these AGN-dominated sources at $z\sim 2$ have QSO-like luminosities (e.g., log~$L_{\rm x}$(ergs~s$^{-1}$)$ \ge 44$), and as many as 80\% may be Compton-thick.  Eight of the 11 QSO-luminosity DOGs with good IRAC data meet our new IRAC criteria, with one additional source falling just beyond our selection region. 

The two DOGs that fall well outside of our selection region (IRBG10 and IRBG13) are the brightest radio sources in the sample, and both show a distinct curvature in their IRAC SEDs, with an excess of 3.6 and 8.0 \micron\ emission and/or a deficit of 4.5 and 5.8 \micron\ emission \citep[see Figure 1 of ][]{donley10}.  \cite{haas08} and \cite{leipski10} observe similar SEDs among their sample of $z>1$ 178 MHz-selected double-lobed radio galaxies (e.g., radio-selected Type 2 AGN), which they attribute to the combined effects of AGN obscuration and host-galaxy emission. It is perhaps not surprising, then, that while the new IRAC selection criteria recover all of the Type 1 3CRR radio quasars of \cite{haas08}, they recover only 33\% of the $z=1.3 \pm 0.2$ narrow-line (e.g., Type 2) 3CRR radio galaxies, whose low-frequency selection and complete optical identification should result in an obscuration-independent sample of luminous radio-loud AGN\footnote{The $z>1$ 3CRR radio galaxies discussed here lie at higher redshifts and have higher radio luminosities ($\nu L_{\rm \nu} $(178 MHz, rest)$ > 10^{44}$~ergs~s$^{-1}$) than the intrinsically MIR-weak and LINER-like narrow-line radio galaxies of \cite{ogle06}, which may lack a hot dust torus.}.  

To extend this test to higher redshift, we adopt the Spitzer high-$z$ radio galaxy (SHzRG) sample of \cite{seymour07} and \cite{debreuck10}.  Of the $z = 2.2 \pm 0.8$ Type 2 SHzRGs with good IRAC photometry (approximately half of the full SHzRG sample), 61\% lie within our AGN selection region. The AGN that are missed tend to have the highest star-forming contributions to their 1.6 \micron\ rest-frame flux and the lowest radio core fractions, indicating that they are preferentially viewed in edge-on, potentially heavily obscured orientations.  While the SHzRG and $z>1$ 3CRR radio galaxy samples lie at different redshifts, they have similar rest-frame radio luminosities.  The higher recovery fraction of the SHzRGs is therefore likely due to the more heterogeneous selection of this sample, whose larger range of core fractions (e.g. orientations) relative to the 3CRR sample suggests a bias towards less heavily obscured radio galaxies.

The X-ray stacking results from Sections 7 and 9 and the \cite{donley10} DOG sample illustrate that a large fraction of IR-luminous, potentially Compton-thick AGN can be recovered by our IRAC criteria.  Nonetheless, many to most radio-selected Type 2 luminous AGN lie outside of our selection region, depending on how the radio sample is selected.  This difference may stem from the more luminous host galaxies of radio-loud AGN \citep[e.g.,][]{lacy00}.  While obscuration in AGN far more luminous than their hosts will lead to a reddening of the IRAC SED similar to that seen in the top panel of Figure 13b, obscuration in lower luminosity AGN or those with particularly luminous early-type hosts will more quickly reveal the stellar continuum, resulting in the curved NIR-MIR SEDs shown in the bottom panel of Figure 13b and observed by \cite{donley10}, \cite{haas08}, and \cite{leipski10}.  IRAC selection will therefore preferentially miss AGN with more luminous hosts, particularly when the AGN emission is itself obscured.

\subsubsection{\cite{stern05}-selected Sources}

By definition, the AGN candidates occupy a well-defined region of \cite{lacy04,lacy07} IRAC color space (though we select only 17\% of the sources that fall in the \cite{lacy04,lacy07} wedge).    
While nearly all of our new AGN candidates also lie within the \cite{stern05} AGN selection wedge, they comprise only 28\% of the sources in the \cite{stern05} wedge and cannot be cleanly separated from the remaining galaxies in this representation of IRAC color space.  To investigate the properties of the sources that do not meet our selection criteria, but that lie well within the \cite{stern05} wedge in regions populated by our AGN candidates, we plot in Figure 12 the IRAC colors of all $T_{\rm x} \ge 50$~ks sources that do not meet our criteria but that have [3.6]-[4.5] (Vega) $>$ 0.55 and [5.8]-[8.0] (Vega) $>$ 0.85.  

\begin{figure}
\epsscale{1.15}
\plotone{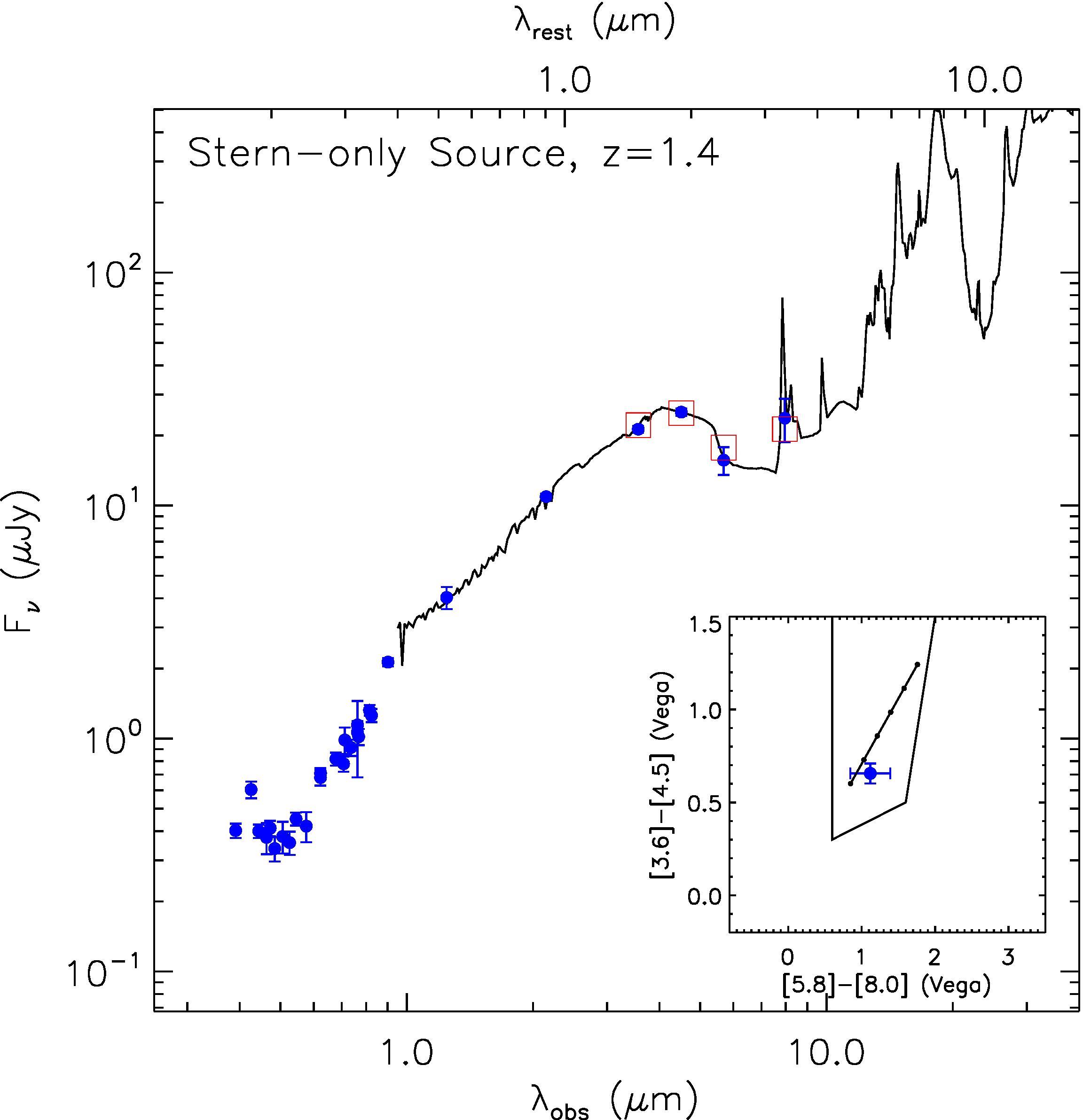}
\caption{SED of a ``Stern-only'' source at $z_{\rm phot}=1.4$, with the \cite{rieke09} template of the purely star-forming ULIRG IRAS 12112+0305 overplotted (after scaling to match the observed 4.5 \micron\ flux density).  Circles give the source photometry, and squares give the photometry of the template convolved with the IRAC bandpasses.  The apparently red color of this source in \cite{stern05} color space can be attributed solely to star-forming features (e.g., the 1.6 \micron\ stellar bump and the 3.3 \micron\ PAH emission feature). }
\end{figure}

This ``Stern-only'' sample, which comprises 33\% of the sources in this region of IRAC color space (and 17\% of the \cite{stern05}-selected sources that do not meet our new criteria), has a median redshift of $z=1.5$, an X-ray detection fraction of only 14\%, and median 5.8 and 8.0 \micron\ flux densities ($f_{\rm 5.8 \micron} = 15$~\microjy, $f_{\rm 8.0 \micron} = 23$~\microjy) that are $2-3$ times fainter than our AGN candidates in this region of color space (though no such offset is seen at 3.6 and 4.5 \micron).  

While these sources have red 3.6 to 4.5 \micron\ and 5.8 to 8.0 \micron\ colors, their colors in \cite{lacy04} color space tend to be quite blue (see Figure 12).  The reason for this behavior is illustrated in Figure 14.  While our new AGN selection criteria require sources to have monotonically rising IRAC SEDs, only 15\% of the ``Stern-only'' sample display this behavior.  Instead, their SEDs tend to rise between 3.6 and 4.5 \micron, turn over between 4.5 and 5.8 \micron, and then rise again towards long wavelengths.  As shown in Figure 14, at the sample's median redshift of $z=1.5$, this behavior can be attributed to the 3.6 and 4.5 \micron\ bands sampling the rising side of the 1.6 \micron\ stellar bump, the 5.8 \micron\ band falling redward of the bump, and the 8.0 \micron\ band sampling the bright MIR emission from the 3.3 \micron\ PAH feature and warm star-formation heated dust.   Moderate redshift star-forming galaxies like the one plotted in Figure 14 (which were excluded from the original \cite{stern05} sample by their shallow flux limit of $S_{\rm 8\mu m} \ge 76$\microjy) will therefore mimic power-law AGN when adjacent wavebands are used to determine their IRAC colors, but can effectively be separated from AGN using a longer wavelength baseline or a requirement that the SED rise monotonically.

\section{Summary}

IRAC selection provides a powerful tool for identifying luminous AGN.  However, the AGN selection wedges currently in use \cite[e.g.,][]{lacy04,stern05} are heavily contaminated by normal star-forming galaxies in deep IRAC data. Using the large samples of luminous AGN and high-redshift star-forming galaxies in COSMOS, we redefine the IRAC AGN selection criteria for use in deep IRAC surveys.  The new cuts, presented in \S9, incorporate the best aspects of the current AGN selection wedges and of infrared power-law selection \citep{aah06,donley07,park10}, while improving on both.  They are designed to be both highly complete and reliable, and effectively exclude high-redshift star-forming galaxies selected via the BzK, DRG, LBG, and SMG criteria down to IRAC flux limits of 0.9, 1.7, 11.3, and 14.6~\microjy\ in the 3.6, 4.5, 5.8, and 8.0 \micron\ bands, respectively.  

The completeness of the new selection criteria is highly luminosity dependent. While fewer than 20\% of Seyfert galaxies with log $L_{\rm 2-10 keV} $(ergs~s$^{-1}$)$\le 43$ meet the IRAC selection criteria, the new cuts recover $75\%$ of hard X-ray detected AGN with QSO-luminosities of log $L_{\rm 2-10 keV} $(ergs~s$^{-1}$)$\ge 44$.  Despite this bias towards luminous AGN, however, only 38\% of the IRAC-selected AGN candidates in COSMOS would be identified as AGN in the X-ray, a fraction that rises to 52\% in regions with deep \chandra\ data ($T_{\rm x} = 50-160$~ks).  

X-ray stacking of the individually X-ray non-detected AGN candidates leads to $\ge 6\sigma$ detections in both the hard and soft X-ray bands, with a large implied column density of log~$N_{\rm H}$ (cm$^{-2}) \sim 23.5 \pm 0.4$.  In comparison, the hard X-ray selected AGN have a typical column density of only log~$N_{\rm H}$(cm$^{-2}$)$=22.4 \pm 0.4$.  While some X-ray non-detected AGN are likely to be missed in the X-ray because of their higher typical redshifts ($z\sim2.2$ compared to $z\sim1.7$ for the X-ray AGN), heavily obscured to mildly Compton-thick obscuration appears to be primarily responsible for driving these intrinsically luminous AGN below the current X-ray flux limits.

The selection criteria defined here provide a reliable method for identifying highly complete samples of luminous unobscured \textit{and} obscured AGN with high-quality (e.g., $\gsim 5 \sigma$) counterparts in the four IRAC bands, a condition met by 88\% of the \xmm-COSMOS sample.  However, IRAC selection cannot effectively identify low-luminosity AGN with host-dominated MIR SEDs, and also appears to be incomplete to luminous heavily obscured AGN with particularly bright hosts (e.g., Type 2 radio galaxies).  X-ray and radio selection therefore remain important tools for identifying unobscured to moderately obscured low-luminosity AGN and obscured radio-loud AGN, respectively.  While the upcoming $>10$~keV NuSTAR X-ray mission \citep{harrison10} will begin to probe the population of heavily obscured Seyfert-luminosity AGN at low to moderate redshift \citep[$z \lsim 0.4$, e.g.,][]{ballantyne11}, new methods such as composite SED fitting (e.g. A. Del Moro 2012, in prep.) will be required to identify the distant radio-quiet and heavily obscured Seyfert-luminosity AGN missed by current X-ray, radio, and MIR-based selection techniques.

\acknowledgements

We thank Ranga Chary, Harry Ferguson, Joanna Kuraszkiewicz, Mark Lacy, Daniel Stern, Belinda Wilkes, Steve Willner, and the anonymous referee for helpful discussions and comments that improved this paper.  J. L. D. is supported by the Giacconi Fellowship at STScI.  This work is partially supported by CONACyT Apoyo 83564, DGAPA UNAM Grant PAPIIT IN110209 and NASA ADP Grant NNX07AT02G. 

\appendix

At high redshift, the observed X-ray bands sample progressively harder X-ray emission less sensitive to intrinsic obscuration.  As a result, low column densities become poorly constrained and are often overestimated \citep[see also][]{ueda03,akylas06}.  To illustrate and quantify this bias, we use a Monte Carlo simulation to study the effect of observational errors on the measured column densities of X-ray detected AGN (but do not consider here any additional biases in the Type 2 fraction introduced by X-ray detectability as a function of redshift and/or obscuration). 

We start with the intrinsic column density distribution of QSO-luminosity AGN (log $L_{\rm x} $(ergs~s$^{-1}$)$\sim 44$) from \cite{ueda03}, which we parameterize as 500 discrete columns.  For redshifts of $z=0.5-3.0$, we then calculate the hard and soft X-ray fluxes that would be observed for each of the input columns, and vary these ideal fluxes using random Gaussian errors scaled by the observed errors on the X-ray fluxes of the \xmm-COSMOS sample. Using these randomized fluxes, we then remeasure the column density using the method described in \S5.1.  The results are shown in Figure 15a.  While low column densities are more poorly constrained than high column densities even at low-redshift, very few unobscured AGN (log~$N_{\rm H}$(cm$^{-2}$)$ < 22$) will be misclassified as obscured AGN (log~$N_{\rm H}$(cm$^{-2}$)$ \ge 22$) at $z\lsim0.5$.  However, as the redshift increases, progressively larger fractions of intrinsically unobscured AGN will scatter into the obscured region of Figure 15a. Because heavily obscured AGN are far less likely to be misclassified as unobscured, this observational scatter leads to an apparent increase in the Type 2 fraction of high redshift X-ray detected AGN.

To estimate the effect of this bias on the measured X-ray Type 2 fraction of the hard X-ray detected \xmm-COSMOS sample, 89\% of which have log~$N_{\rm H}$(cm$^{-2}$)$ \le 23$, we first truncate the \cite{ueda03} column density distribution at log~$N_{\rm H}$(cm$^{-2}$)$ = 23$. The resulting Type 2 fractions recovered from the Monte Carlo simulation are shown in Figure 15b, as is the intrinsic value of 25\%. For the assumed column density distribution, the recovered X-ray Type 2 fraction is overestimated by only 4\% at at $z=1$, the approximate redshift of the bluest \xmm\ sources in Figure 7.  However, at $z=2$, the typical redshift of the reddest \xmm\ sources, the observed Type 2 fraction is likely to be overestimated by $\sim 40\%$.  This overestimation of the X-ray Type 2 fraction at high redshift contributes to the apparent discrepancies between X-ray and optically-classified AGN (see \S5 and \S6).  

\begin{figure*}
$\begin{array}{cc}
\hspace*{1cm}
\includegraphics[angle=0,height=3.2in]{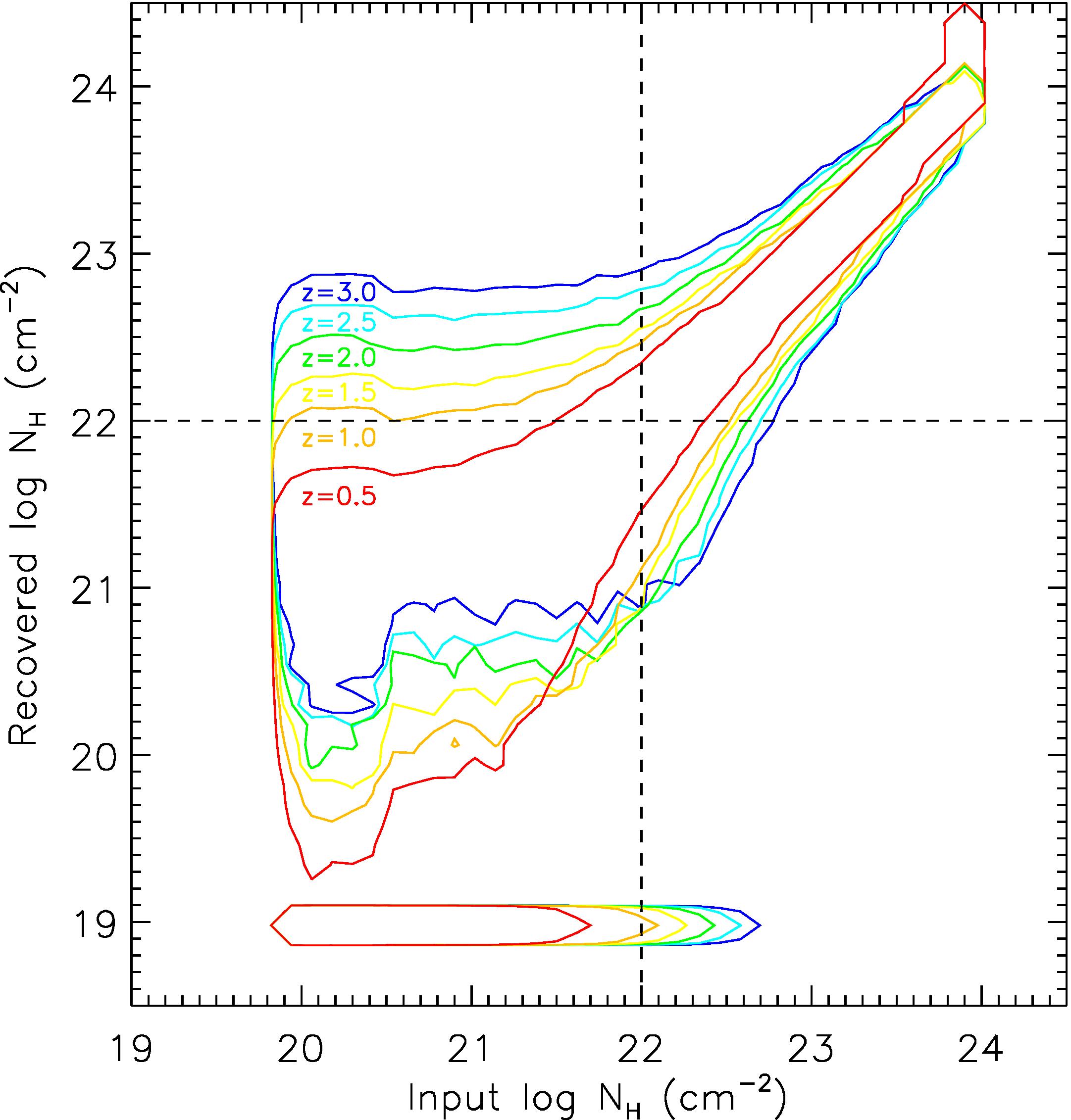} &
\includegraphics[angle=0,height=3.2in]{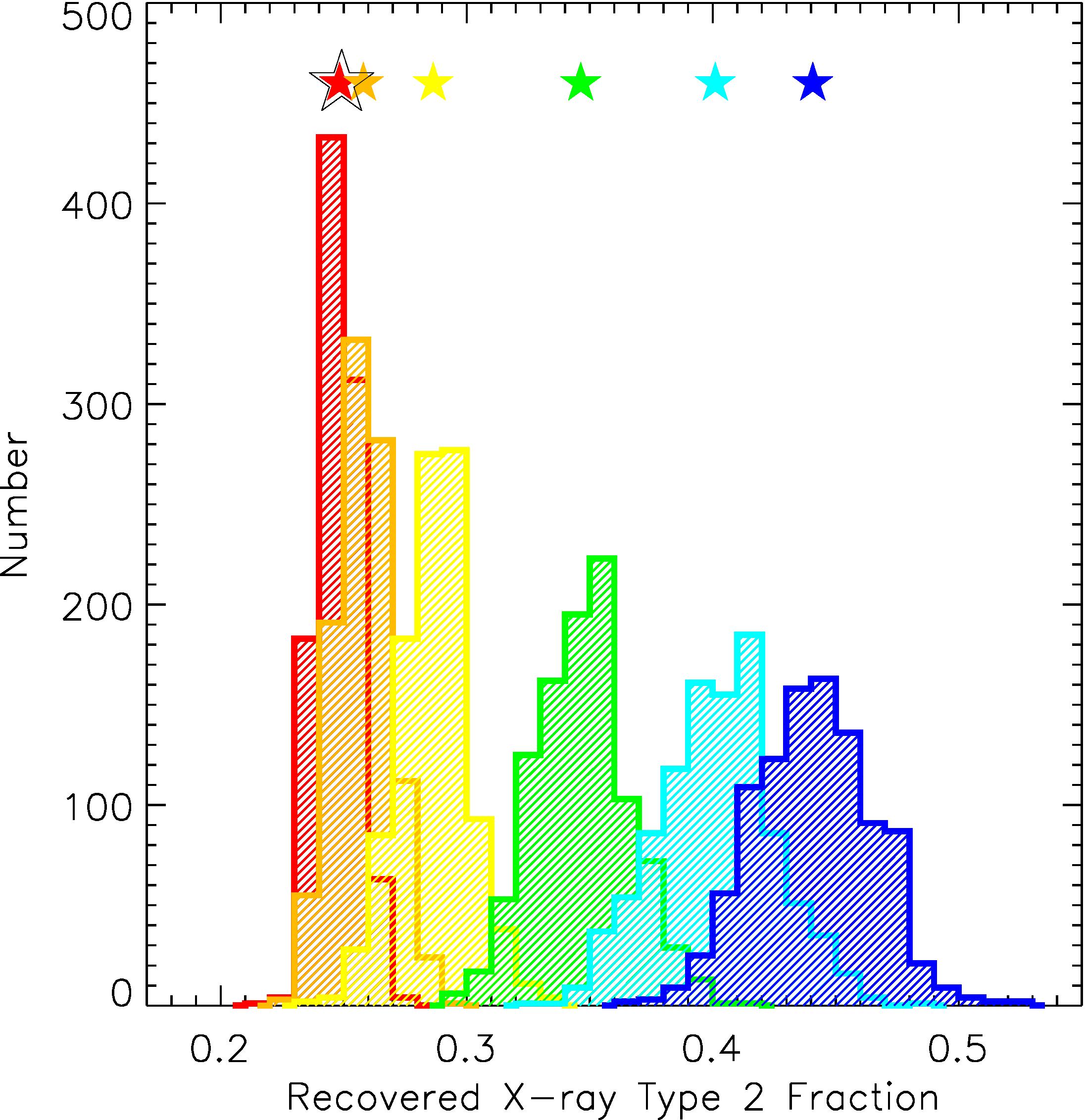} 
\end{array}$
\caption{left: Contours representing the agreement between model column densities and those recovered from our Monte Carlo simulation for $z=0.5-3.0$.  Sources with unmeasurably low columns are assigned a value of log~$N_{\rm H} = 19$. right: Recovered Type 2 fractions for the same range of redshifts, assuming the column density distribution of \cite{ueda03} truncated at log~$N_{\rm H} = 23$.  The intrinsic Type 2 fraction of 25\% is given by an open star, and the median recovered Type 2 fractions are given by filled stars. Because low column densities have little effect on high X-ray energies, both the measured column densities and the X-ray Type 2 fractions will be overestimated at high redshift.}
\end{figure*}

\end{document}